\documentclass[twocolumn]{aastex63}
\usepackage{graphicx}
\usepackage{wrapfig}
\usepackage{natbib}
\usepackage{color}
\graphicspath{{./figures/}} 
\usepackage{mathtools}
\usepackage{epstopdf}
\usepackage{hyperref} 
\def    \apjl  		{\rm {ApJL}}

\def    \apjl  		{\rm {ApJL}}

\def	\cm		{\,{\rm {cm}}}
\def	\K		{\,{\rm K}}
\def	\g		{\,{\rm {g}}}
\def	\mum	{\,{\mu \rm{m}}}
\def	\erg		{\,{\rm {erg}}}
\def	\s		{\,{\rm {s}}}

\def \bea {\begin{eqnarray}}
\def \ena {\end{eqnarray}}

\begin{document}
\shorttitle{Rotational disruption and extinction of AGN}
\title{Modelling Grain Rotational Disruption by Radiative Torques and Extinction of Active Galactic Nuclei}
\author{Nguyen Chau Giang}
\affil{Korea Astronomy and Space Science Institute, Daejeon 34055, Republic of Korea; \href{mailto:thiemhoang@kasi.re.kr}{thiemhoang@kasi.re.kr}}
\affil{Korea University of Science and Technology, Daejeon 34113, Republic of Korea}
\affiliation{University of Science and Technology of Hanoi, VAST, 18 Hoang Quoc Viet, Vietnam}

\author{Thiem Hoang}
\affil{Korea Astronomy and Space Science Institute, Daejeon 34055, Republic of Korea; \href{mailto:thiemhoang@kasi.re.kr}{thiemhoang@kasi.re.kr}}
\affil{Korea University of Science and Technology, Daejeon 34113, Republic of Korea}

\begin{abstract}
Extinction curves observed toward individual Active Galactic Nuclei (AGN) usually show a steep rise toward far-ultraviolet (FUV) wavelengths and can be described by the Small Magellanic Cloud (SMC)-like dust model. This feature suggests the dominance of small dust grains of size $a\leq 0.1\mum$ in the local environment of AGN, but the origin of such small grains is unclear. In this paper, we aim to explain this observed feature by applying the RAdiative Torque Disruption (RATD) to model the extinction of AGN radiation from FUV to mid-infrared (MIR) wavelengths. We find that in the intense radiation field of AGN, large composite grains of size $a \geq 0.1\mum$ are significantly disrupted to smaller sizes by RATD up to $d_{\rm RATD} > 100$ pc in the polar direction and $d_{\rm RATD} \sim 10$ pc in the torus region. Consequently, optical-MIR extinction decreases, whereas FUV-near-ultravilet (NUV) extinction increases, producing a steep far-UV rise extinction curve.  The resulting total-to-selective visual extinction ratio thus significantly drops to $R_{\rm V} < 3.1$ with decreasing distances to AGN center due to the enhancement of small grains. The dependence of $R_{\rm V}$ with the efficiency of RATD will help us to study the dust properties in the AGN environment via photometric observations. In addition, we suggest that the combination of the strength between RATD and other dust destruction mechanisms which are responsible for destroying very small grains of $a \leq 0.05\mum$ is the key for explaining the dichotomy observed 'SMC' and 'gray' extinction curve toward many AGN.

\end{abstract}
\keywords{active galactic nuclei: dust, extinction}

\section{Introduction}\label{sec:intro}
The unified model (\citealt{Antonucci_93}, \citealt{Urry_95}) of Active Galactic Nuclei (AGN) posits that all radio-quiet AGN are intrinsically the same, and the difference in their observational properties arises from the inclination effects due to the presence of an optically thick dusty torus around the nuclei. When observing AGN from an inclined angle (i.e., not perpendicular to the disk), the ultraviolet (UV)-optical radiation from the accretion disk and the line emission from the broad-line region (BLR) are absorbed by the dusty torus. Thus, the UV-optical signal from these regions is absent in the observed spectrum of AGN (i.e., type 2 AGN). In contrast, when looking AGN in the perpendicular direction to the disk (face-on), the  UV-optical radiation emitting from the central region is not significantly attenuated due to the small presence of dust grains along the polar direction (i.e., type 1 AGN).

The absorption of UV-optical radiation heats dust grains to high temperatures, followed by their re-emission in infrared (IR) with a peak at $\sim3-30 \mum$ (\citealt{Ree69}). The broad and excess emission in the mid-IR range, which is called 'infrared bump', is usually observed in the sub-arcsecond scale of numerous AGN and successfully explained by the thermal dust emission from the circumnuclear region (\citealt{Barvainis87}, \citealt{Fritz_06}, \citealt{Nenkova08b}, \citealt{Sieben15}). Recently, the high-resolution interferometric observations in mid-IR reveal the significant fraction ($ \geq 50-80\%$) of mid-IR emission in the extended region along the polar direction of many AGN (\citealt{Honig_12}, \citealt{Honig_13}, \citealt{Tristram_14}, \citealt{Lopez_16}). Moreover, mid-IR observations with single-dish telescopes at a resolution of $\sim0.3" $ have detected the thermal emission from warm grains on the scales of 100 pc out to $\sim$ kpc along the polar direction (\citealt{Asmus14}, \citealt{Asmus16}). The plausible explanation for the large-scale polar dust is due to radiation pressure on dust (see e.g., \citealt{Honig_10}). Unfortunately, the physical properties of dust grains (size distribution, shape, and composition) in the torus and polar region are not well constrained.
 
Photometric observations of individual AGN usually report the 'red tail' in its spectrum energy distribution (\citealt{Webster_1995}, \citealt{Brotherton_2001}, \citealt{Richards_2002}), which can be explained by the extinction of dust in the host galaxies (\citealt{Richards_2003}). The observed extinction curves of these AGN exhibit a steep rise in the far-ultraviolet (FUV) range and can be described by the Small Magellanic Clouds (SMC)-like dust model. For example, \cite{Richards_2003} found that 273 over 4576 quasars in the Sloan Digital Sky Survey (SDSS), appear to be redder and can be explained by the SMC-like extinction curve. Similarly, \cite{Hopkins_2004} reported the similar feature for 9655 quasars observed by SDSS and 1887 quasars observed by Two Micron All Sky Survey (2MASS). In addition, several studies of the dust reddening of AGN report the low values of color excess and visual extinction per gas column density, $E(B-V)/N_{\rm H}$ and $A_{\rm V}/N_{\rm H}$, compared to the standard values produced by interstellar grains (\citealt{Reichert}, \citealt{Granato}, \citealt{Maioline}). The origin of such low values is unclear.
 
The steep far-UV rise in the extinction curves requires an enhancement of small grains of size $a\leq 0.1\mum$ in the local environment of AGN compared to the standard interstellar dust model (\citealt{Mathis77}, \citealt{Wein01}). This feature suggests the modification of dust by the intense radiation field of AGN. Thermal sublimation of AGN dust was first studied by \cite{Laor_93} for the polar cone and \cite{Barvainis87} for the torus region. They found that polar and torus  grains can be heated to the sublimation temperatures and evaporate near the center of AGN. However, this mechanism is more efficient in destroying small grains than large grains, and it converts small grains directly to the gas phase, which reduces the overall extinction. Dust destruction by Coulomb explosion induced by energetic photons (e.g., X-ray and extreme UV, \citealt{Draine79}) was studied by \cite{Weingartner_2006} and recently by \cite{TazakiRyo} for AGN. This process also works effectively for very small grains because they can only accommodate a small amount of charge. Lately, \cite{Tazaki_20} proposed a new scenario of drift-induced sputtering that can remove sub-micron grains in the polar direction. However, this mechanism is not effective enough to enhance small grains as required to reproduce a steep far-UV rise extinction curve. Thus, the origin of such a dominance of small grains in AGN is still unclear.

Recently, \cite{Hoang_2019} introduced a new dust destruction mechanism of dust in intense radiation fields, namely RAdiative Torque Disruption (RATD). The mechanism is based on the fact that centrifugal stress within a dust grain that is spun up to extremely fast rotation by radiative torques (\citealt{Draine96}; \citealt{Laza07}; \citealt{Hoang09}) can exceed the maximum tensile strength of the material. As a result, rapidly spinning grains can be disrupted into small fragments (see \citealt{Hoang:2020} for a review). As shown in \cite{Hoang_2019}, large grains of $a \geq 0.1 \mum$ exposed to a strong radiation field can be quickly disrupted into smaller ones even at large distances from the source. We expect that RATD can explain the dominance of small grains around AGN. Thus, the goal of this paper is to quantify the effect of RATD and model the observed extinction curves in the presence of RATD. 
 
The structure of the paper is as follows. In Section \ref{sec:RATD}, we present the radiation field of AGN and briefly describe the RATD mechanism that determines the grain size distribution. The physical model of the polar cone and the dusty torus, and the radiative transfer modeling, are presented in Sections \ref{sec:model} and \ref{sec:radiative_transfer}, respectively. We present our numerical results for the dust disruption and the corresponding extinction curves in Sections \ref{sec:adisr} and \ref{sec:Photometry}. An extended discussion and a summary of our main findings are presented in Sections \ref{sec:dicuss} and \ref{sec:sum}, respectively.

 \section{The RATD Mechanism for AGN}\label{sec:RATD}

\subsection{Spectral energy distribution of AGN}
  The spectral luminosity, $L_{\lambda}$, of an unobscured AGN can be described by a power law function with different slopes for different ranges of wavelength as follows (\citealt{Stalevski_2012}):
\bea 
\lambda L_{\lambda}= \alpha \left\{
\begin{array}{l l}    
    0.158 ~ (\lambda/1\mum)^{1.2} ~ ~ ~ {\rm  if~ } 0.001 \mum \leq \lambda \leq 0.01 \mum\\
    6.3\times10^{-4} ~ (\lambda/1\mum) ~  {\rm ~ if~ } 0.01 \mum < \lambda \leq 0.1\mum \\
    2\times10^{-4} ~ (\lambda/1\mum)^{-0.5} ~ ~ {\rm ~ if~ } 0.1 \mum < \lambda \leq 5\mum\\
    0.011 ~ (\lambda/1\mum)^{-3} ~ ~ ~ ~ ~ ~ ~ ~ ~ {\rm ~ if~ } 5 \mum < \lambda \leq 50 \mum
\end{array}\right\},
\label{eq:u_AGN}
\ena

where $\alpha$ is a normalization constant, which is determined by the bolometric luminosity. We consider an AGN with a constant bolometric luminosity of $L_{\rm bol} = 10^{46}\erg\s^{-1}$ and disregard their time variability. 

To model the grain disruption by RATD in AGN environments, we only consider photons with wavelengths from $\lambda_{1} = 0.1\mum$ to $\lambda_{2} = 20\mum$, where the lower limit is determined by the Lyman limit (i.e., photons of $\lambda< 0.1\mum$ are entirely absorbed by hydrogen), and the upper limit is chosen such that above this wavelength the RAT efficiency is negligible for interstellar grains of size $a\lesssim 1\mum$. The strength of the radiation field is defined as $U=u_{\rm rad}/u_{\rm ISRF}$, where $u_{\rm rad}$ is the energy density of the assumed radiation spectrum (Equation \ref{eq:u_AGN}), and $u_{\rm ISRF}=8.64\times10^{-13} \erg \cm^{-3}$ is the energy density of the average interstellar radiation field (ISRF) in the solar neighborhood \citep{Mathis83}.

The mean wavelength of the radiation field considered for RATD is given by:
\bea
\bar{\lambda} = \frac{\int_{\lambda_{1}}^{\lambda_{2}} \lambda L_{\rm \lambda} d \lambda}{\int_{\lambda_{1}}^{\lambda_{2}} L_{\lambda} d \lambda},
\ena 
 and we get $\bar{\lambda} = 0.876 \mum$ with the usage of $L_{\lambda}$ from Equation (\ref{eq:u_AGN}).

  \subsection{The RATD mechanism}
An irregular dust grain exposed to an anisotropic radiation field experiences radiative torques due to differential scattering and absorption of left-handed and right-handed photons (\citealt{Draine96}, \citealt{Laza07}). In an intense radiation field, the grain can be spun up to extremely fast rotation so that it is disrupted into small fragments when the centrifugal stress exceeds the maximum tensile strength of the grain material (\citealt{Hoang_2019}). This mechanism is named RAdiative Torque Disruption mechanism (RATD, see \citealt{Hoang_2019} for details). Here, we briefly summarize the main formulae for reference.
 
Let $a$ be the effective size of an irregular grain, which is defined as the radius of an equivalent spherical grain with the same volume as the irregular grain. The angular velocity of the irregular grain spun-up by RATs is obtained by solving the equation of motion (\citealt{Hoang_2019}): 
\bea
\frac{I d\omega}{dt} = \Gamma_{\rm RAT}-\frac{I\omega}{\tau_{\rm damp}},
\label{eq:eq_motion}
\ena
where $I=8 \pi \rho a^{5}/15$ is the grain inertia moment with $\rho$ the grain mass density, $\tau_{\rm damp}$ is the characteristic timescale of grain rotational damping (see \citealt{Hoang_2019} for details). The first term of the right-hand side $\Gamma_{\rm RAT}$ is the radiative torque arising from the interaction of the anisotropic radiation field of AGN with grain of size $a$, which equals:
\bea
\Gamma_{\rm RAT} &=& \int_{\lambda_{\rm 1}}^{\lambda_{\rm 2}} \Gamma_{\rm \lambda} d\lambda \nonumber \\
&=& \int_{\rm \lambda_{\rm 1}}^{\rm \lambda_{\rm 2}} \pi a^{2} \gamma_{\rm rad} u_{\rm \lambda,0} e^{-\tau_{\rm \lambda}} \left(\frac{\lambda}{2 \pi}\right) Q_{\rm \lambda} d\lambda.
\ena
In above equation, $u_{\rm \lambda,0} = L_{\rm \lambda,0}/(4 \pi c d^{2})$ is the intrinsic energy density at wavelength $\lambda$ with intrinsic spectral luminosity given by Equation (\ref{eq:u_AGN}), $\tau_{\rm \lambda}$ is the optical depth, and $\gamma_{\rm rad}$ is the anisotropy degree of the radiation field ($0 \leq \gamma_{\rm rad} \leq 1$). Here, we adopt $\gamma_{\rm rad} = 1$ for the unidirectional radiation field of AGN. $Q_{\rm \lambda}$ is the RAT efficiency at wavelength $\lambda$, which is (\citealt{Hoang08}, \citealt{Hoang14}, \citealt{Hoang_2019}):
\bea
Q_{\lambda}=\left\{
\begin{array}{l l}	
     \approx 2.33 \left(\frac{\lambda}{a} \right)^{-3} & {\rm for~} a \lesssim \lambda/1.8\\
     \approx 0.4 & {\rm for~} a > \lambda/1.8
\end{array}\right\}. \label{eq:Q_lambda}
\ena 
In the constant radiation field, the grain of size $a$ will be spun-up continuously by a constant radiative torque, $\Gamma_{\rm RAT}$, and achieve a maximum angular speed of $\omega_{\rm RAT}$ after about a damping timescale. An analytical formula for $\omega_{\rm RAT}$ as follows:
\bea
  \omega_{\rm RAT} = \frac{\Gamma_{\rm RAT} ~  \tau_{\rm damp}}{I} .
\label{eq:w_terminal}
\ena

 \begin{figure}[htb!]
            \includegraphics[width=0.45\textwidth]{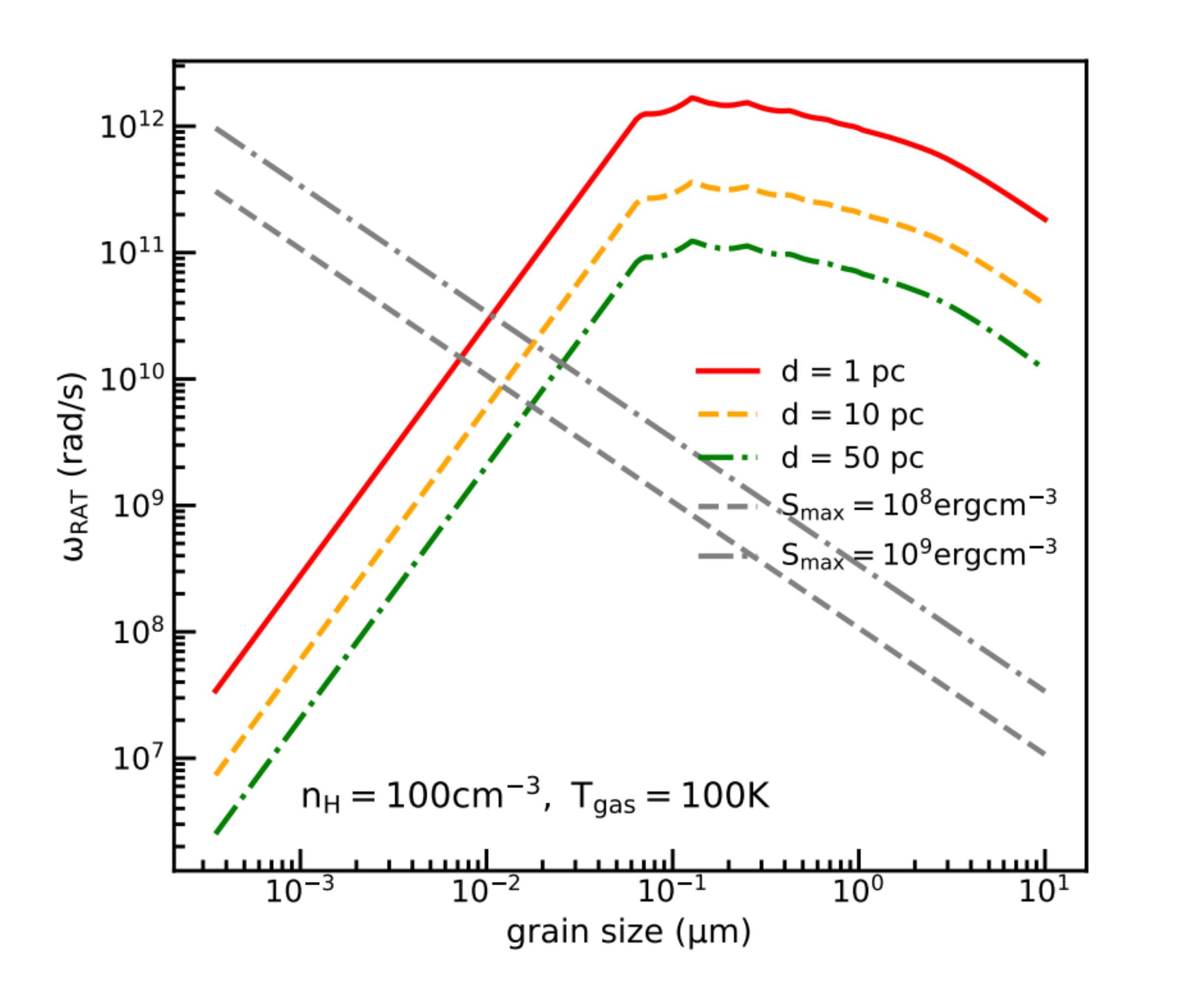}
\caption{Maximum angular speed of grains spun-up by RATs due to AGN radiation as a function of the grain size for different distances $d$ to the central source. The gray dashed lines show the disruption speed for grains with $S_{\rm max}=10^{8}$ and $10^{9}\erg\cm^{-3}$. Optically thin environment with the gas density of $n_{\rm H}=100 \cm^{-3}$ and the gas temperature of $T_{\rm gas}=100$ K are assumed. The intersection of $\omega_{\rm RAT}$ and $\omega_{\rm disr}$ determines the disruption size.}
\label{fig:omega_sat}
\end{figure}

Figure \ref{fig:omega_sat} shows the variation of the maximum angular speed due to RATs, $\omega_{\rm RAT}$, with the grain size from $a=0.001 \mum$ to $10 \mum$ at different distances to the center of AGN in the absence of dust reddening, with a gas density of $n_{\rm H} = 100\cm^{-3}$ and gas temperature of $T_{\rm gas} = 100$ K. At the same distance, $\omega_{\rm RAT}$ increases with the grain size to the peak at $a \sim 0.2 - 0.5\mum$ then decreases for larger grains. The reason is as follows: for grains of $a\leq 0.2\mum$, smaller grains receive less energy due to smaller cross section $\pi a^{2}$ and smaller RAT efficiency $Q_{\rm \lambda}$ from the considered spectrum (see Equation \ref{eq:Q_lambda}). Thus, they are spun up to lower $\omega_{\rm RAT}$ compared with larger grains. On the contrary, grains of $a\geq 0.2\mum$ nearly experience the same RAT efficiency that larger grains will be spun-up to lower $\omega_{\rm RAT}$ due to larger mass. For example, at $d = 10$ pc, grains of $a = 0.001 \mum$, $0.01 \mum$ and $a = 0.1 \mum$ are spun up to $\omega_{\rm RAT} \sim 3\times10^{7}~\rm rad ~ s^{-1}$, $\sim 10^{10} ~ \rm rad ~ s^{-1}$ and $3\times10^{11} ~\rm rad ~s^{-1}$ , respectively. This value decreases to $\omega_{\rm RAT} \sim 10^{10} ~\rm rad ~ s^{-1}$ for micron-sized grains of $a \geq 1 \mum$. In addition, grains which are far from the center of AGN receive lower radiation flux and rotate slower.

The rapidly spinning grain will be disrupted into small fragments when its centrifugal stress $S=\rho a^{2}\omega^{2}/4$ exceeds the maximum tensile strength of the grain material, $S_{\rm max}$. The value of $S_{\rm max}$ depends on the grain material, internal structure, and perhaps the grain size, i.e., large grains usually have porous structure with lower value of $S_{\rm max}$ compared with small grains with compact structure. It can vary from $S_{\rm max}=10^{11}\erg\cm^{-3}$ for ideal materials, i.e., diamond (\citealt{Burke74}; \citealt{Draine79}) to $S_{\max}\sim 10^{9}-10^{10}\erg\cm^{-3}$ for polycrystalline bulk solid (\citealt{Hoang_2019}) and $S_{\max}\sim 10^{6}-10^{8}\erg\cm^{-3}$ for composite grains (\citealt{Hoang2019}). In this paper, we take $S_{\rm max}=10^{8}\erg\cm^{-3}$ as a typical value.

The critical angular speed at which rotational disruption occurs is obtained by setting $S$ equal to $S_{\rm max}$, which yields:
\bea
\omega_{\rm disr} &=& \frac{2}{a} \left(\frac{S_{\rm max}}{\rho}\right)^{1/2}.  
\label{eq:w_disr}
 \ena
 
Then, the grain disruption size, i.e., $a_{\rm disr}$, at which all larger grains are fragmented to smaller pieces by RATD, can be calculated by setting $\omega_{\rm disr}=\omega_{\rm RAT}$.  
In Figure \ref{fig:omega_sat}, the gray dashed and dashed-dot line present the dependence of the critical angular speed $\omega_{\rm disr}$ with the grain size. One can see that for the same internal strength, small grains are required to be spun up to higher angular speed in order to be disrupted by RATD. Thus, at the same distance, large grains are easier to be disrupted by RATD due to its higher $\omega_{\rm RAT}$ and lower $\omega_{\rm disr}$, i.e., the right-hand side of the intersection where $\omega_{\rm RAT} = \omega_{\rm disr}$. For example, at $d = 10$ pc for grains with $S_{\rm max} = 10^{8}\erg\cm^{-3}$, all large grains of $a \geq 0.02\mum$ are destroyed by RATD, while smaller grains can survive. Grains at large distances to AGN are less affected by RATD due to lower radiation energy density, i.e., lower $\omega_{\rm RAT}$. Similarly, grains with higher $S_{\rm max}$ are more difficult to be broken due to its higher disruption threshold $\omega_{\rm disr}$ (see Equation \ref{eq:w_disr} and dashed-dot gray line).
 
Note that the decrease of $\omega_{\rm RAT}$ with grain sizes for large grains of $a\geq 0.1\mum$ will set an upper limit at which all larger grains are not affected by RATD, which is denoted by $a_{\rm disr, max}$. From Figure \ref{fig:omega_sat}, one can see that micron-sized grains are disrupted totally by RATD in the optically thin environments. However, in the dense regions, the strong attenuation of the radiation field may reduce the value of $a_{\rm disr, max}$ that let more micron-sized grains survive (see \citealt{Hoang_202020} for a detailed explanation).
 
 \section{A physical model of AGN}\label{sec:model}
  \begin{figure}[t]
      \includegraphics[width=0.5\textwidth]{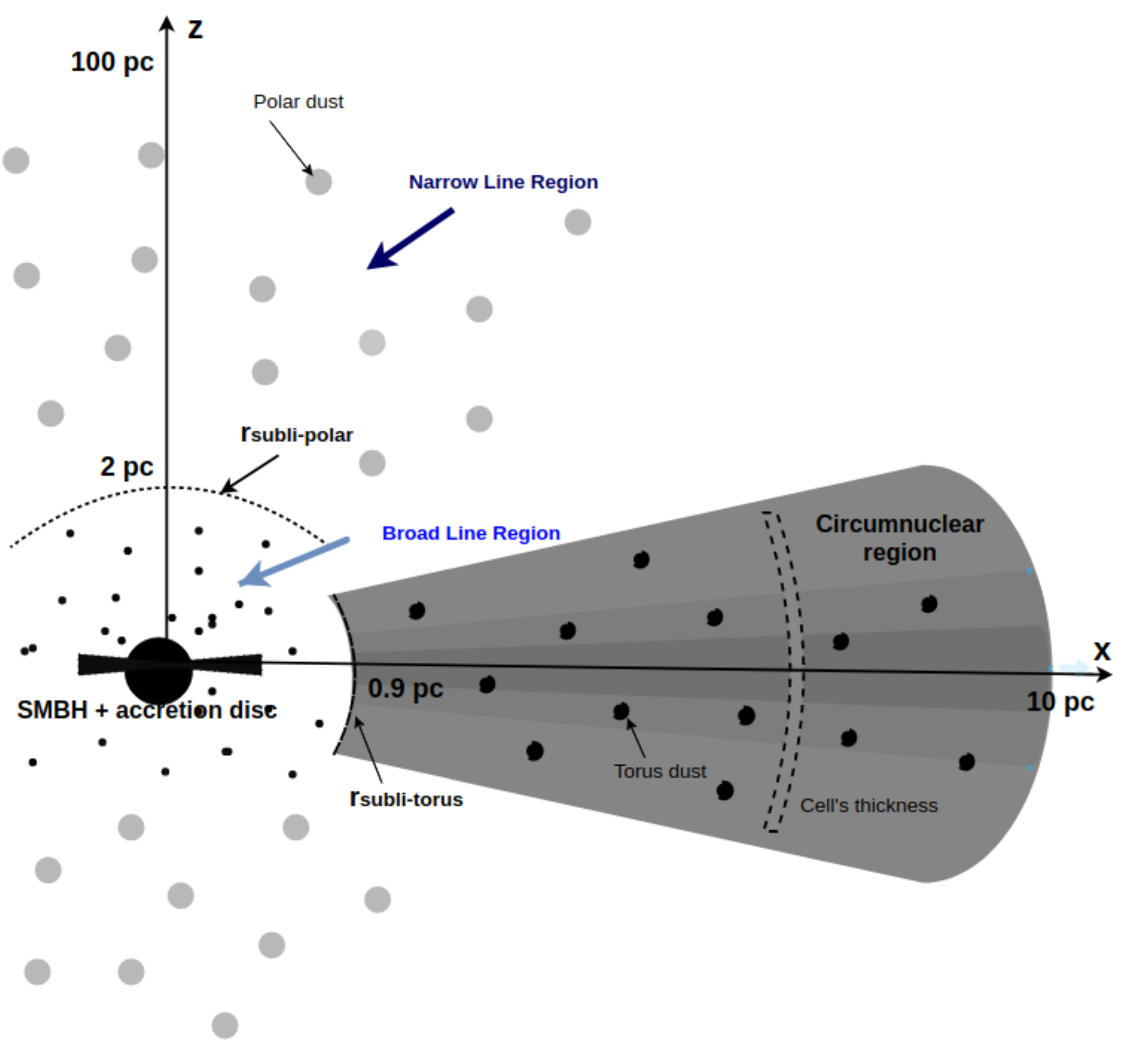}
\caption{A two-dimensional illustration of an axisymmetric structure of AGN. The central engine including the SMBH and an accretion disk (black circle and small black flared disk) is covered by a dusty toroidal region on the equatorial plane and an ionized region (BLR, NLR) extending along the polar direction. We consider that the torus region has the 'flared disc' geometry (\citealt{Fritz_06}).} The inner boundary of the polar cone and torus are determined by the sublimation distances $r_{\rm subli-polar}$ and $r_{\rm subli-torus}$, beyond which grains are not sublimated by AGN radiation. To model the dust destruction by RATs, we divide the polar cone and torus into spherical cells of the same thickness, $d_{\rm cell}$ (dashed donut shape). The distance of each cell toward the center of AGN is denoted by $r$, and $\Theta$ is the observed angle.
\label{fig:AGN}
\end{figure} 

  Figure \ref{fig:AGN} shows a schematic illustration of the AGN structure. The central engine, including a supermassive black hole (SMBH) and an accretion disk in the center of AGN, is presented by a black circle and a black flared disk in the center of the figure.  This region is covered by the dusty toroidal structure on the equatorial plane (gray region) and the scattering region, i.e., BLR and NLR, along the polar direction. The exact geometry of torus region has not yet been constrained. Therefore, in this paper, we consider a simple but realistic geometry of a 'flared disc' to be the shape of torus region (\citealt{Fritz_06}, \citealt{Efsta_95}). The innermost locations of dust in the polar cone and torus toward the AGN center are determined by the sublimation distances $r_{\rm subli-polar}$ and $r_{\rm subli-torus}$, respectively, which constitute sublimation fronts. Thus, grains can survive in the intense radiation field of AGN beyond the sublimation fronts.  
\subsection{Polar cone}
The presence of dust along the polar direction can arise from a magneto-hydrodynamical wind from the central engine of AGN or radiation pressure. Close to the center, grains are quickly heated to the high temperature and sublimated to the gas phase (\citealt{Laor_93}). The radius at which the grain temperature equals the sublimation threshold is called the sublimation distance, $d_{\rm subli-polar}$. The exact value of $d_{\rm subli-polar}$ depends on the size and composition of grains. In general, small grains are destroyed to larger distances due to their low heat capacity. Silicate grains are destroyed more efficiently than carbonaceous grains due to their lower sublimation temperature (see \citealt{Laor_93} for details). For example, from Figure 8 in \cite{Laor_93}, polar silicate grains of size $a \geq 0.005\mum$ can start to survive around AGN with $L_{\rm bol} = 10^{45}\erg~s^{-1}$ beyond the outer boundary of BLR of $d \sim 4$ pc, while it is $d \sim 1$ pc for carbonaceous grains of size $a \geq 0.005\mum$. Thus, we adopt the average value of $r_{\rm subli-polar} = 2$ pc to be the sublimation distance of both silicate and carbonaceous grains in the polar direction.

The gas number density of the polar region can be described by a power law (\citealt{Laor_93}):
\begin{equation}
    n_{\rm H, polar}(r) = n_{\rm H}^{0} \left(\frac{r}{r_{\rm subli-polar}}\right)^{\gamma_{\rm polar}},
    \label{eq:profile_polar}
\end{equation}
where $n_{\rm H}^{0}$ is the gas density at $r=r_{\rm subli-polar}$, and $\gamma_{\rm polar}$ is the power-law index of the distribution. The value of $\gamma_{\rm polar}$ varies from $\gamma_{\rm polar} = 0$ (corresponding to a uniform distribution) to $\gamma_{\rm polar} = -2$ (\citealt{Lyu_18}).

Gas in the polar region is strongly ionized due to the intense UV radiation of AGN. The ionization parameter is defined by:
\bea 
U_{\rm ion} &=& \frac{n_{\rm \gamma}(\geq 13.6~\rm eV)}{n_{\rm H}} \nonumber \\
&=& \frac{1}{n_{\rm H}} \int_{\rm 13.6~\rm eV}^{1~\rm keV} \frac{u_{\rm \lambda}}{E_{\rm \lambda}} d\lambda,
\label{eq:U_ion}
\ena
where $n_{\rm \gamma}(\geq 13.6~\rm eV)$ is the number density of ionizing photons of energy of $13.6 ~\rm eV \leq E_{\rm \lambda} \leq 1~\rm keV$ with $E_{\rm \lambda} = hc/\lambda$, and $u_{\rm \lambda}$ is the spectral energy density given by Equation (\ref{eq:u_AGN}). We can neglect the reduction of ionizing photons by neutral hydrogen due to its negligible fraction (i.e., $n({\rm H}^{+})\sim n_{\rm H}$). 

The gas temperature $T_{\rm gas}$ in the polar cone is thus determined by the balance between photoionization heating and radiative cooling, which follows (\citealt{Sazonov_04}, \citealt{Sazonov_05}, \citealt{Weingartner_2006}):
 \bea 
T_{\rm gas}= \left\{
\begin{array}{l l}    
     10^{4} ~\rm K ~~& {\rm if~} U_{\rm ion} \leq 2  \\
      5\times10^{3}U_{\rm ion} ~\rm K ~~& {\rm if~} 2 < U_{\rm ion} \leq 4\times 10^{3} \\
     2\times10^{7} ~\rm K ~~ & {\rm if~} U_{\rm ion} > 4\times10^{3}
\end{array}\right\}.
\label{eq:Tgas}
\ena

\subsection{Dusty torus}
Similar to the polar cone, grains in the torus region start to survive beyond the sublimation distance where the dust temperature drops below $T_{\rm subli} \sim 1500$ K. From the study of \cite{Honig_10}, we adopt the sublimation distance of $r_{\rm subli-torus}= 0.9 \sqrt(L_{\rm bol,46})$ (pc) with $L_{\rm bol,46} = L_{\rm bol}/(10^{46}\erg s^{-1})$. The torus structure and the morphology of grains in the circumnuclear region are still not constrained. Initially, \cite{Krolik_1988} suggested that grains might concentrate into a clumpy structure to avoid being destroyed by the intense radiation field of AGN. However, the difficulty in performing the clumpy model appeals to the simple case of the smooth distribution with different vertical and horizontal profiles (\citealt{Efsta_95}, \citealt{Manske_98}, \citealt{Fritz_06}). Nevertheless, the smooth distribution cannot explain the depletion of silicate absorption feature at $9.7\mum$ in type 1 AGN and the broad observed far-IR emission compared to the modeling (\citealt{Granato}, \citealt{gratano}). \cite{Rowan} suggested solving these problems by assuming the concentration of grains in the clumpy structure and this idea was confirmed by \cite{Nenkova2002}.  The clumpy model is also supported by observations of the inhomogeneous circumnuclear region (\citealt{Tristram2007}). The detailed model of the clumpy structure is then developed to solve the radiative transfer and compare with observations (e.g.,  \citealt{Nenkova2002}, \citealt{Nenkova08a}, \citealt{Nenkova08b}, \citealt{Dullemond05} (two-dimension), \citealt{Honig06}, \cite{Stalevski_2012}, \cite{Sieben15} (three-dimension)). 

In this paper, our main goal is to explore the effect of rotation disruption on polar and torus dust grains and to understand how it changes the dust extinction of AGN. Therefore, we adopt the simple smooth dust distribution model in the flared disk taken from \cite{Fritz_06}, and modeling with a more realistic clumpy AGN structure will be addressed in our future study. 

We consider a spherical coordinate system ($r, \Theta$, $\phi = 0^{\circ}$) in the x-z plane. The gas number density in the torus (see Figure \ref{fig:AGN}) can be described by (\citealt{Fritz_06}): 
\bea
n_{\rm H, torus}(r,\Theta) =  n_{\rm H}^{0} \left(\frac{r}{r_{\rm subli-torus}}\right)^{\gamma_{\rm torus}} e^{\beta \cos(\Theta)},
\label{eq:profile_torus}
\ena
where $\gamma_{\rm torus}$ and $\beta$ are the power-law index of the gas density profile along the radial direction $r$ and polar direction, (i.e., the observed angle $\Theta$), respectively, and $n_{\rm H}^{0}$ is the gas density at the sublimation front of $r = r_{\rm subli-torus}$. 

In the dense torus, dust and gas are essentially in thermal equilibrium (see other studies for dense regions like protoplanetary disks (\citealt{Tung2020}) and dense molecular clouds  (\citealt{Hoang2021})). The gas temperature can be obtained by the equilibrium temperature of grains, which is given by (\citealt{Draine11}):
\bea
T_{\rm dust} \approx 16.4U^{1/6} \K,
\label{eq:Tgas_torus}
\ena
where $U$ is the radiation field strength.
 
\section{Radiative transfer modelling} \label{sec:radiative_transfer}

\subsection{Grain model}\label{sec:grain_model}
Radiation emitting from AGN is attenuated due to scattering and absorption by surrounding dust. To calculate the dust extinction, we adopt a popular ISM mixed-dust model consisting of astronomical silicate and carbonaceous grains (see \citealt{Li_2001}, \citealt{Wein01}, \citealt{Drain07}). We assume that grains in AGN follow a power-law size distribution with a slope $\alpha$:
\bea
\frac{dn}{da}^{j} = C^{j} n_{\rm H} a^{\alpha},
\label{eq:grain_size}
\ena
where $C^{j}$ is the normalization constant which is determined by the dust-to-gas mass ratio $\eta$:
\bea 
\eta = \frac{4\pi}{3 m_{\rm H}}\frac{(a_{\rm max}^{4+\alpha} - a_{\rm min}^{4+\alpha})}{4+\alpha}  \sum_{j = sil,carb} C^{\rm j} \rho^{\rm j},
\label{eq:eta}
\ena
where $\rho^{j}$ is the mass density of component $j$, we take $\rho_{\rm sil} = 3.5\g\cm^{-3}$, $\rho_{\rm carb}=2.2\g\cm^{-3}$ and $C_{\rm sil}/C_{\rm carb} = 1.12$ (\citealt{Draine_84}, \citealt{Laor_93}). Without RATD, we consider the range of grain sizes from $a_{\rm min}=  3.5 ~ \rm \AA$ to $a_{\rm max}=10 \mum$ with $\alpha = -3.5$ (\citealt{Mathis77}). When increasing the maximum grain size to $10\mum$, the total dust mass will be higher than the standard dust model in our Galaxy. Thus, the original value of normalization constant $C_{\rm sil}$ and $C_{\rm carb}$ from \cite{Mathis77} must be reduced by a factor of 7 to keep the Galactic standard dust-to-gas mass ratio of $\eta = 0.01$. The new values are $C_{\rm sil} = 1.16\times10^{-26}\cm^{2.5}$ and $C_{\rm carb} = 1.036\times10^{-26}\cm^{2.5}$, for silicate and carbonaceous grains, respectively.

When we account for the effect of RATD, the grain sizes from $a_{\rm disr}$ to $a_{\rm disr,max}$ will be fragmented to smaller sizes, resulting in the two separated grain size distributions. The first one is the range of grain sizes of $ a_{\rm disr, max} - a_{\rm max}$ that still follows the MRN distribution with $\alpha = -3.5$. The second one is the grain sizes from $a_{\rm min}$ to $a_{\rm disr}$ that follows a steeper distribution with a new slope $\alpha$ due to the enhancement of small grains by RATD. By assuming the normalization constant of silicate and carbonaceous grains $C_{\rm sil}$ and $C_{\rm carb}$ to be a constant, the new grain size distribution of small grains can be found from the conservation of dust mass, which follows (see \citealt{Giang_2020} for details):
\bea
\label{eq:slope_new}
\frac{a_{\rm disr,max}^{4-3.5} - a_{\rm min}^{4-3.5}}{4-3.5}=\frac{a_{\rm disr}^{4+\alpha} - a_{\rm min}^{4+\alpha}}{4+\alpha}.
\ena

\subsection{Effect of dust reddening in the presence of RATD} \label{sec:dust_redding}
The emergent spectral energy density ($u_{\lambda}$) of AGN radiation after traveling through a dusty cloud of thickness $d$ is given by
\bea
u_{\rm \lambda} = u_{\rm \lambda, 0} e^{-\tau_{\rm \lambda}},
\label{eq:u_rad_reddening}
\ena
where $u_{\rm \lambda, 0} = L_{\rm \lambda, 0}/(4 \pi c d^{2})$ is the intrinsic spectral energy density given by Equation (\ref{eq:u_AGN}), and $\tau_{\rm \lambda}$ is the optical depth induced by intervening grains.

In the absence of RATD, AGN radiation is attenuated by all grains from $a_{\rm min}$ to $a_{\rm max}$ following the MRN distribution. When RATD is included, AGN radiation will be attenuated by two dust populations following different size distributions as described in Section \ref{sec:grain_model}. The extinction cross section per H by dust at distance $r$ at wavelength $\lambda$ in units of $\cm^{2}/H$ is then calculated as (\citealt{Hoang13}):
\bea
\sigma_{\rm ext}(\lambda, r) &=& \sum_{j=\rm sil,carb} \int Q_{\rm ext}^{j}(a) \pi a^{2} \frac{dn^{j}}{da}(r) da  \nonumber\\
&=& \sum_{j =\rm sil,carb} \Bigg(\int_{a_{\rm min}}^{a_{\rm disr}(r)} Q_{\rm ext}^{j}(a) \pi a^{2} a^{\alpha(r)} da \nonumber\\ 
&+& \int_{a_{\rm disr, max}(r)}^{a_{\rm max}} Q_{\rm ext}^{j}(a) \pi a^{2} a^{-3.5} da \Bigg) C^{j}, ~~\label{eq:sigma_ext}
\ena
where $Q_{\rm ext}$ is the extinction efficiency, $C^{j}$ is the normalization constant for silicate and carbonaceous grains, $a_{\rm disr}(r)$, $a_{\rm disr, max}(r)$, and $\alpha(r)$ are the grain disruption size, the maximum grain disruption size, and the new slope of the distribution of small grain size at distance $r$. We assume that grains of sizes between $a_{\rm disr}$ and $a_{\rm disr,max}$ are totally disrupted into grains smaller than $a_{\rm disr}$. For sub-micron grains, we adopt $Q_{\rm ext}$ calculated for the oblate spheroidal shape with an axial ratio of $2$ in \cite{Hoang13}. For micron grains, we calculated $Q_{\rm ext}$ for the same shape using DDSCAT (\citealt{Draine:1994}). The optical depth induced by dust in the cloud of thickness $d$ is:
\bea
\tau_{\lambda}(d) &=& \int_{0}^{d} \sigma_{\rm ext}(\lambda, r) n_{\rm H}(r) dr,  \nonumber\\
 ~~\label{eq:Aext}
\ena
where $n_{\rm H}(r)$ is the gas density at distance $r$, which is described by Equation (\ref{eq:profile_polar}) for the polar cone and Equation (\ref{eq:profile_torus}) for the torus region. 

To numerically model the dust reddening effect, we divide the smooth dusty polar cone and torus (see Section \ref{sec:model}) into a number of  thin cells of the same thickness of $d_{\rm cell}=0.05$ pc (see the dashed donut shape in Figure \ref{fig:AGN}). The thickness of the cell is chosen such that the change of the grain size distribution within the cell is negligible. With this choice, in the $i^{\rm th}$ cell, we can calculate the optical depth produced by grains in the cell $\Delta \tau_{\rm \lambda,i}$ (using Equation \ref{eq:sigma_ext} and Equation \ref{eq:Aext} with d replaced by $d_{\rm cell}$). Thus, the total optical depth produced by dust in the cloud of thickness $d$ containing $n$ dusty cells is then obtained as:
\bea
    \tau_{\rm \lambda,n} = \sum_{i=1}^{n} \Delta \tau_{\rm \lambda, i},
    \label{eq:tau_n}
\ena
where $i=1$ denotes the first cell at the sublimation front, and $i=n$ denotes the last cell at distance $d$.


\section{Modelling dust disruption in AGN environment}  \label{sec:adisr}
  \begin{figure}[t]
      \includegraphics[width=0.45\textwidth]{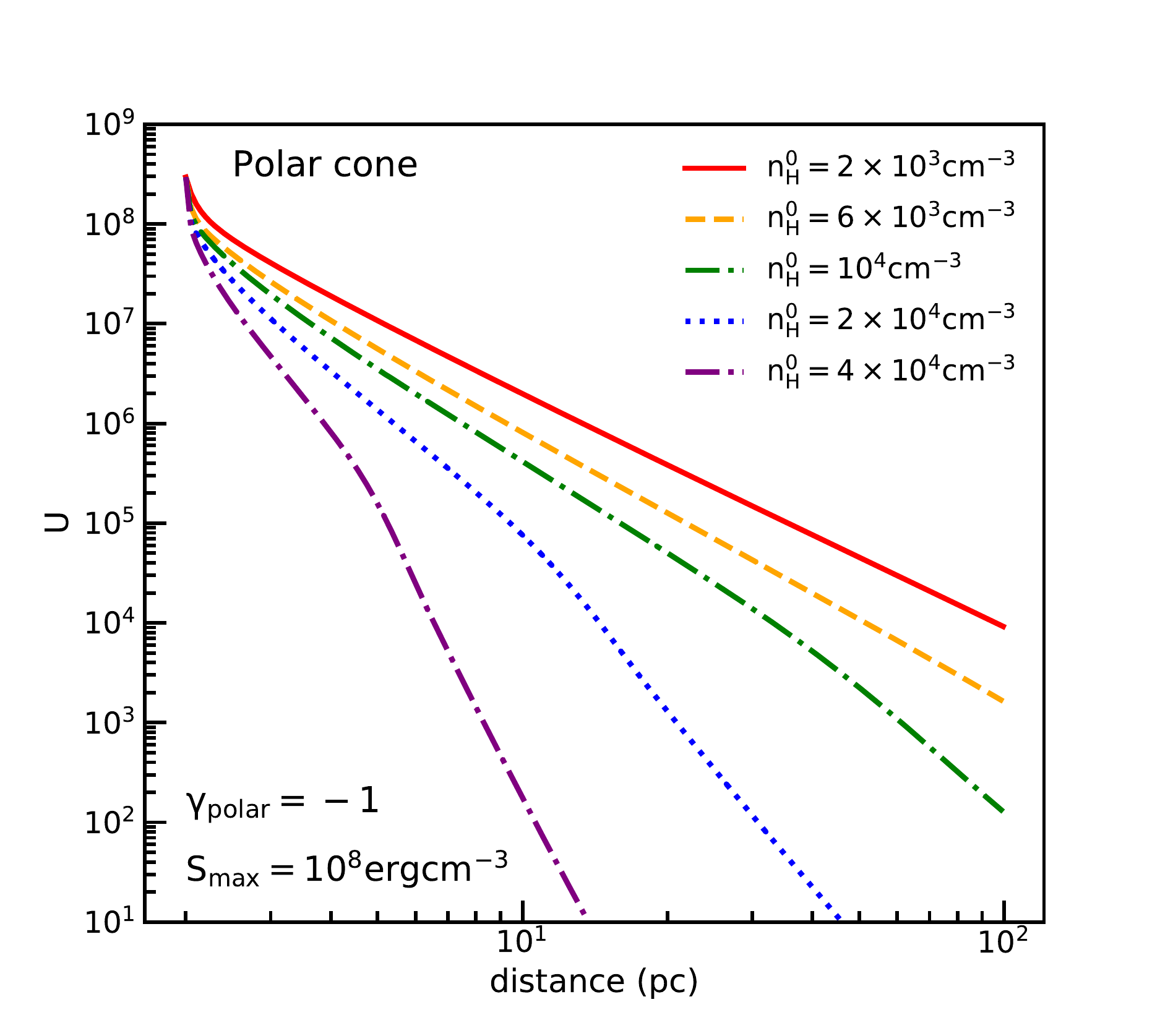}
      \includegraphics[width=0.45\textwidth]{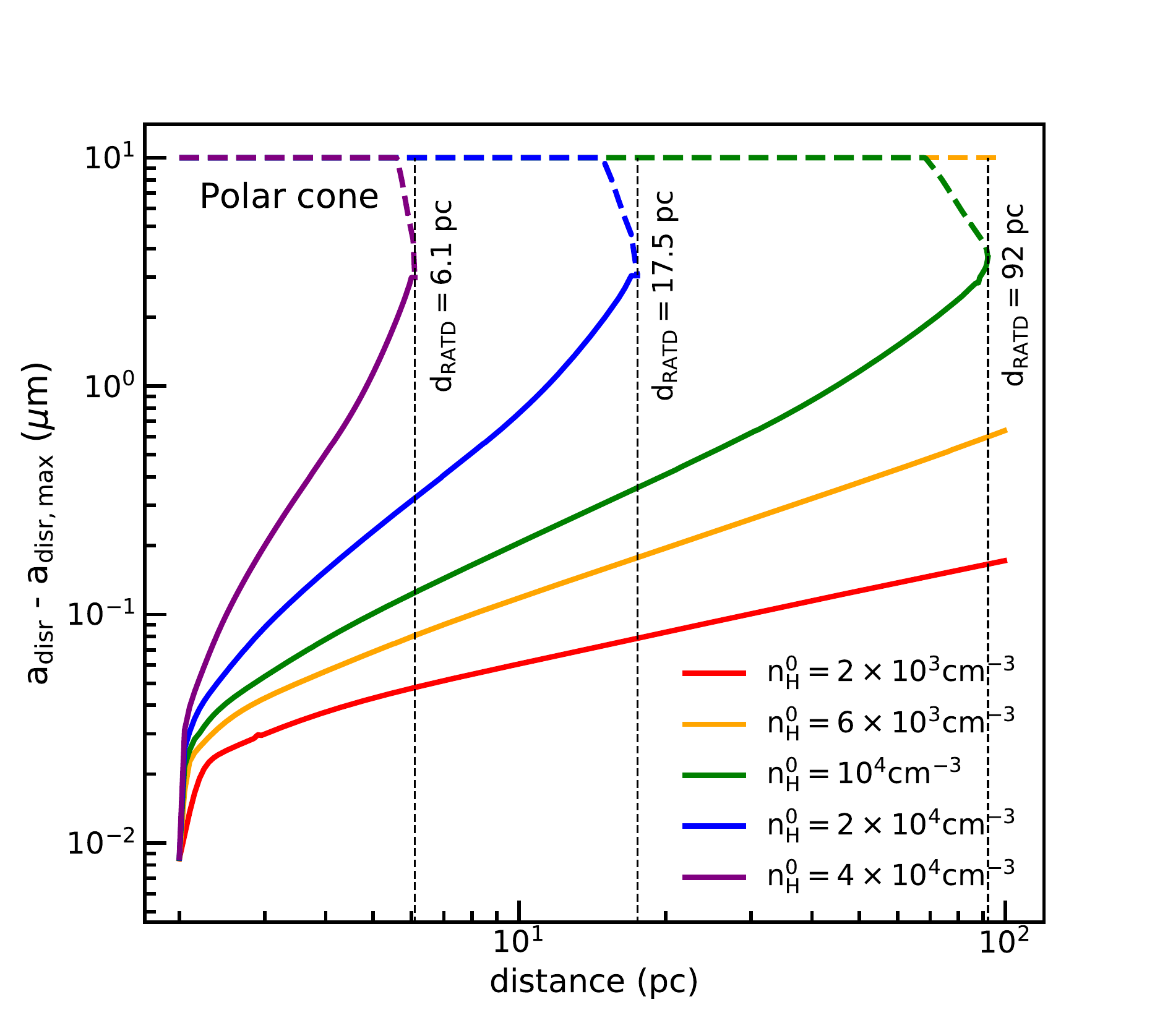}
\caption{Variation of the radiation field strength, $U$ (upper panel), and the range of grain sizes of $a_{\rm disr}-a_{\rm disr,max}$ which are disrupted by RATD (solid and dashed lines) (lower panel) with the cloud distances in the polar cone, assuming different initial gas density $n_{\rm H}^{0}$, $\gamma_{\rm polar} = -1$, and $S_{\rm max}=10^{8}\erg\cm^{-3}$.}
\label{fig:adisr_polar_nH}
\end{figure}
 
To model the RATD effect, we first calculate the radiation field strength $U$ as described in Section \ref{sec:radiative_transfer} and solve the equation of motion (Equation \ref{eq:eq_motion}) to get the terminal angular speed $\omega_{\rm RAT}$ for the range of grain size from $a_{\rm min}$ to $a_{\rm max}$. Then, by comparing $\omega_{\rm RAT}$ with the critical angular speed $\omega_{\rm disr}$ given by Equation (\ref{eq:w_disr}), one can determine the range of $a_{\rm disr}-a_{\rm disr, max}$ at which grains are disrupted into smaller sizes by RATD. 

 \subsection{Polar cone}\label{sec:a_disr_polar}
The upper panel of Figure \ref{fig:adisr_polar_nH} shows the variation of the radiation field strength $U$ in the polar cone as a function of distance for different values of initial gas density $n_{\rm H}^{0}$ at the sublimation front of $r_{\rm subli-polar} =2$ pc, assuming the density profile given by Equation (\ref{eq:profile_polar}) with $\gamma_{\rm polar} = -1$ and the maximum tensile strength of $S_{\rm max} = 10^{8}\erg\cm^{-3}$. The lower panel shows the corresponding range of grains which are disrupted by RATD, starting from the grain disruption size $a_{\rm disr}$ (solid lines) to the end at the maximum grain disruption size $a_{\rm disr, max}$ (dashed lines).   
 
For all considered gas density profiles, one can see that grains are not affected by RATD if they are very far from the center of AGN. It arises from the strong attenuation of AGN radiation such that it is not strong enough to spin up any grains to the disruption threshold. In this case, the grain disruption size is set to the maximum grain size. The distance where RATD ceases determines the boundary of the active region of RATD, denoted by $d_{\rm RATD}$ and marked by the vertical black dashed line. As the distance to AGN decreases, grains are spun up to faster rotation due to higher radiation flux, resulting in the expansion of the size range of disrupted grains. For example, with $n_{\rm H}^{0} = 10^{4}\cm^{-3}$, the disruption range expands continuously from $a_{\rm disr} = a_{\rm disr,max}$ at $d = 100$ pc (no disruption) to $1 - 10\mum$ at $\sim 40$ pc and $0.1 - 10\mum$ at $\sim 5$ pc.
 
For a higher value of $n_{\rm H}^{0}$, the radiation strength is decreased more drastically due to stronger dust reddening, decreasing the RATD efficiency (i.e., smaller active region of RATD). For instance, the value of $d_{\rm RATD}$ reduces continuously from $d_{\rm RATD} = 92$ pc to 17.5 pc and 6.1 pc if the initial gas density increases from $n_{\rm H}^{0} \leq 10^{4}\cm^{-3}$ to $2\times10^{4}\cm^{-3}$ and $4\times10^{4}\cm^{-3}$, respectively.
 
 \begin{figure}[htb!]
      \includegraphics[width=0.45\textwidth]{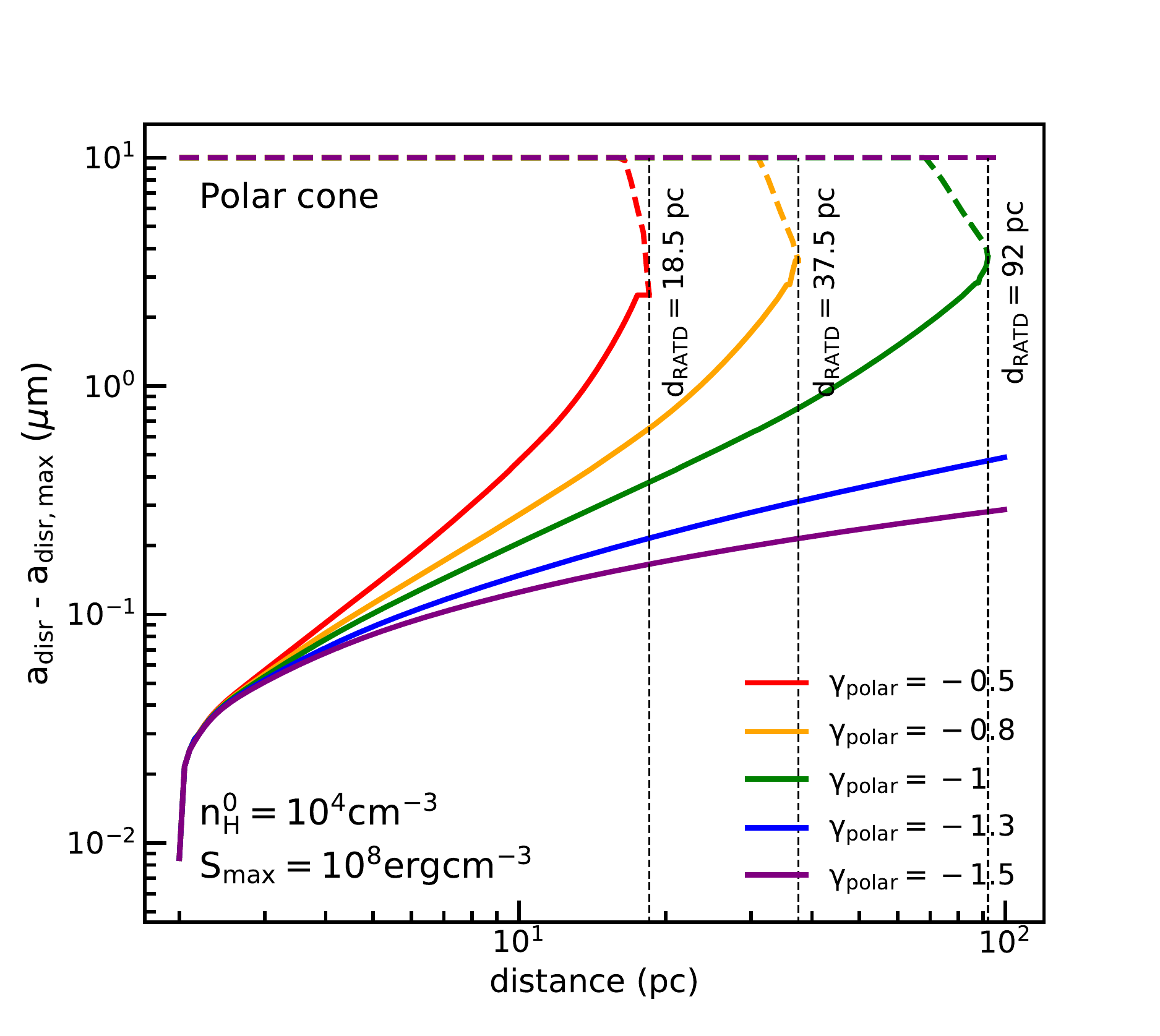} 
      \includegraphics[width=0.45\textwidth]{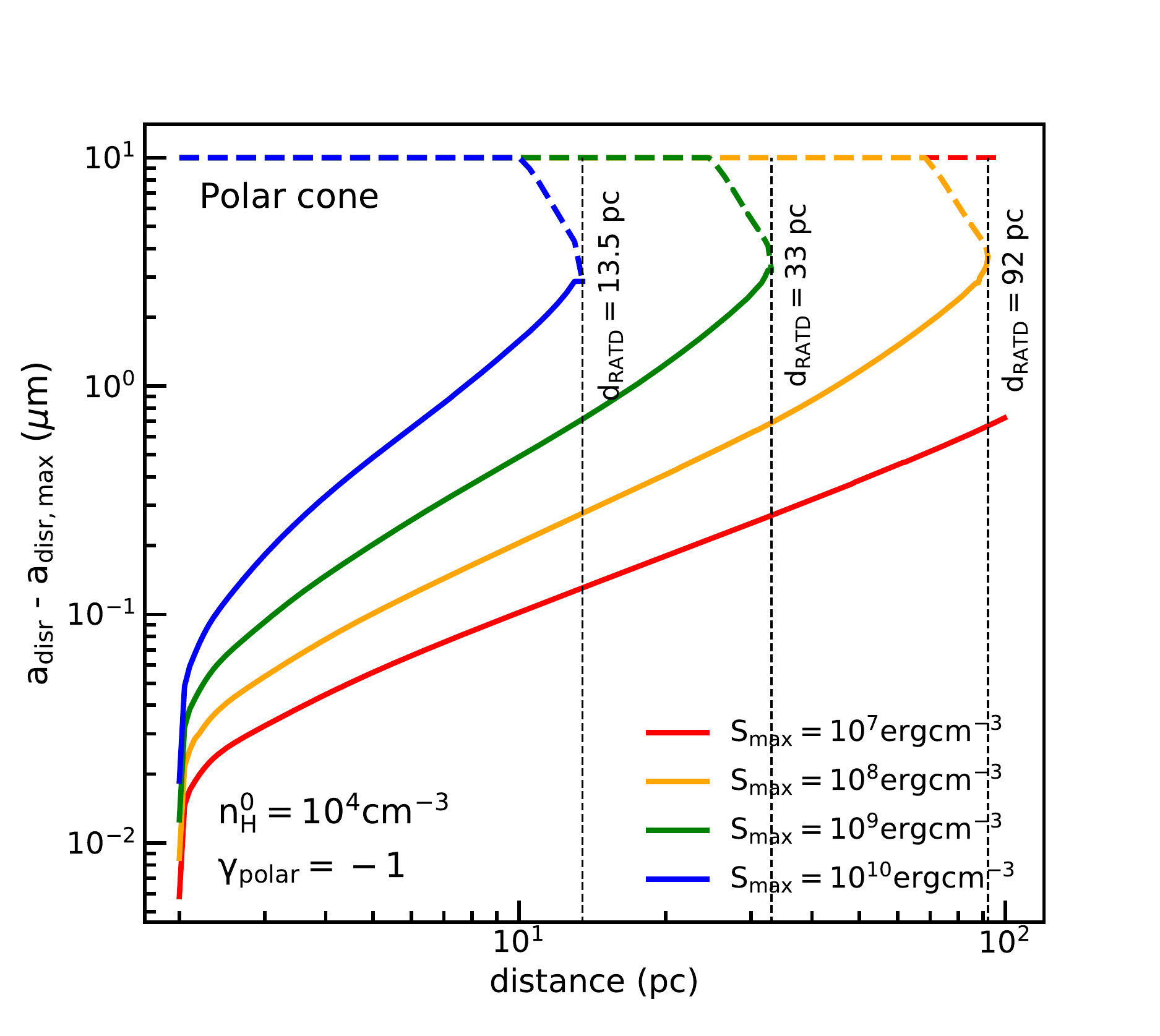}
\caption{The same as Figure \ref{fig:adisr_polar_nH} but for different power-law indexes $\gamma_{\rm polar}$, assuming $n_{\rm H}^{0}=10^{4}\cm^{-3}$ and $S_{\rm max}=10^{8}\erg\cm^{-3}$ (upper panel), and for different maximum tensile strengths $S_{\rm max}$, assuming $n_{\rm H}^{0} = 10^{4}\cm^{-3}$ and $\gamma_{\rm polar} = -1$ (lower panel).}
\label{fig:adisr_polar_gamma_Smax}
\end{figure}
 
The upper panel of Figure \ref{fig:adisr_polar_gamma_Smax} shows the range of disrupted grains within 100 pc in the polar cone for different values of the power-law index $\gamma_{\rm polar}$, assuming $n_{\rm H}^{0} = 10^{4} \cm^{-3}$ and $S_{\rm max}=10^{8}\erg\cm^{-3}$. For the same initial gas density, the slower decrease of $n_{\rm H}$ with distances, i.e., larger $\gamma_{\rm polar}$, induces stronger attenuation of AGN radiation and reduces the active region of RATD. For example, $d_{\rm RATD}$ reduces from $d = 92$ pc to 37.5 pc and 18.5 pc if $\gamma_{\rm polar}$ increases from $\leq -1$ to -0.8 and -0.5, respectively. 

The lower panel of Figure \ref{fig:adisr_polar_gamma_Smax} shows the same as the upper panel but for different tensile strengths, assuming $n_{\rm H}^{0}=10^{4}\cm^{-3}$ and $\gamma_{\rm polar}=-1$. One can see that with the same gas density profile, grains with higher $S_{\rm max}$ are less disrupted by RATD due to its higher disruption threshold $\omega_{\rm disr}$ (see Equation \ref{eq:w_disr}). For example, compact grains of $S_{\rm max}=10^{10}\erg\cm^{-3}$ and $S_{\rm max}=10^{9}\erg\cm^{-3}$ are only destroyed within 13.5 pc and 33 pc, while composite grains with $S_{\rm max} \leq 10^{8}\erg\cm^{-3}$ can be disrupted up to $d_{\rm RATD} \geq 92$ pc.

 \begin{figure}[htb!]
      \includegraphics[width=0.45\textwidth]{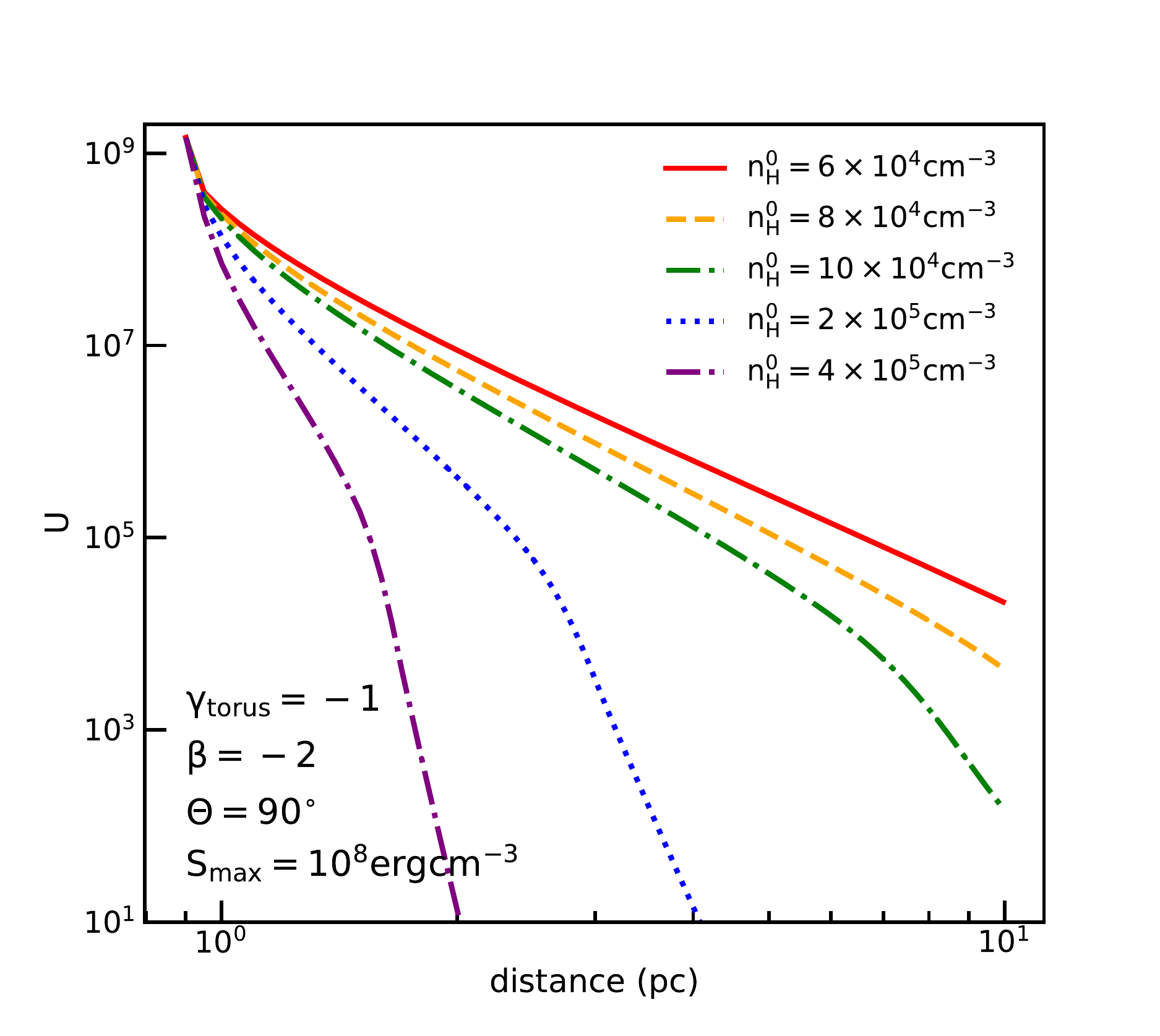}
          \includegraphics[width=0.45\textwidth]{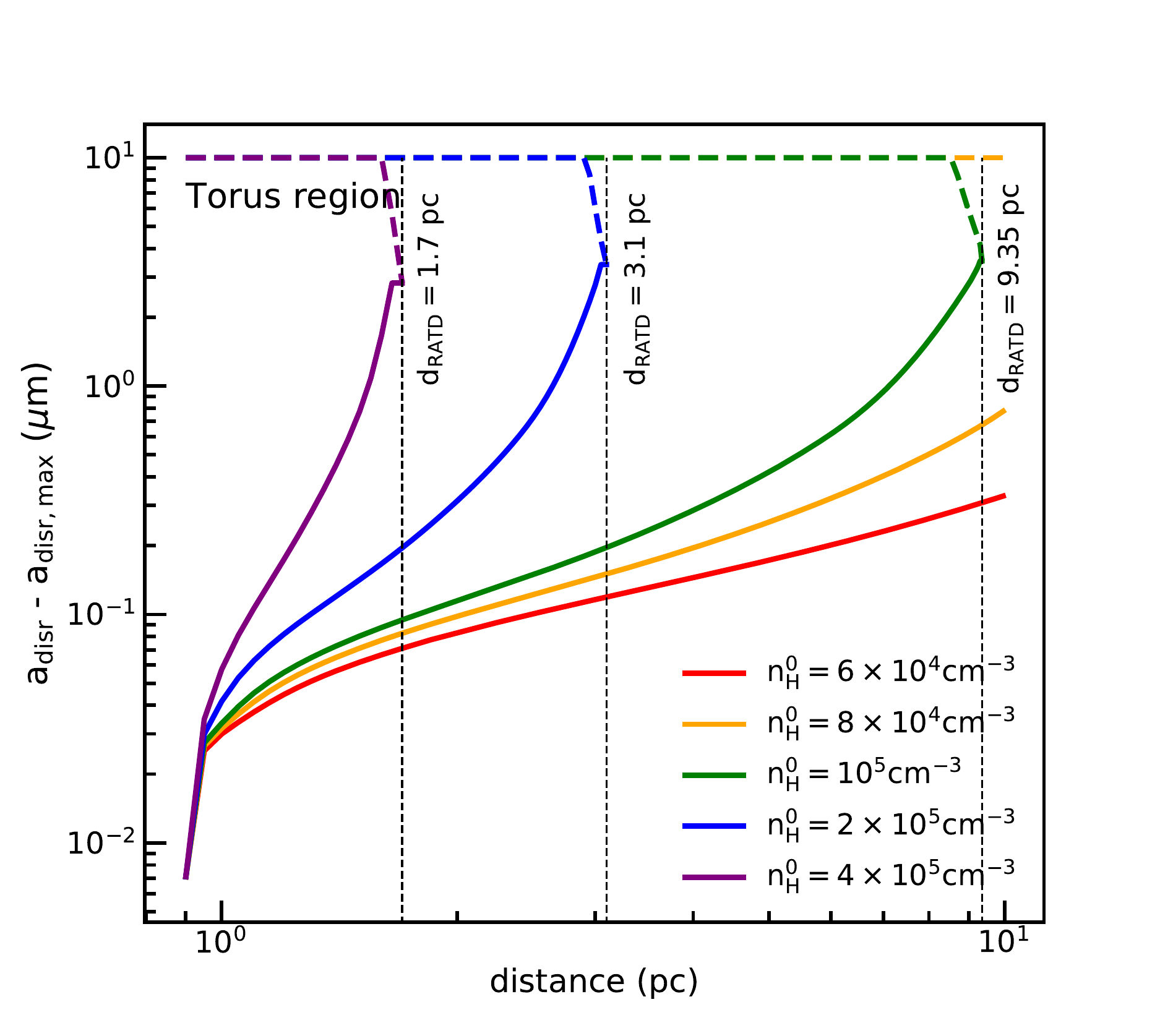}
\caption{Variation of the radiation field strength $U$ (upper panel) and the range of grain sizes which are disrupted by RATD (lower panel) with distance within 10 pc from the center AGN on the equatorial plane ($\Theta = 90^{\circ}$) for the torus, assuming different initial gas densities $n_{\rm H}^{0}$, $\gamma_{\rm torus} = -1$, $\beta = -2$ and $S_{\rm max}=10^{8}\erg\cm^{-3}$.}
\label{fig:adisr_torus_nH}
\end{figure} 
 \subsection{Torus region} \label{sec:a_disr_torus}

\begin{figure*}[t]
    \centering
    \includegraphics[width=0.45\textwidth]{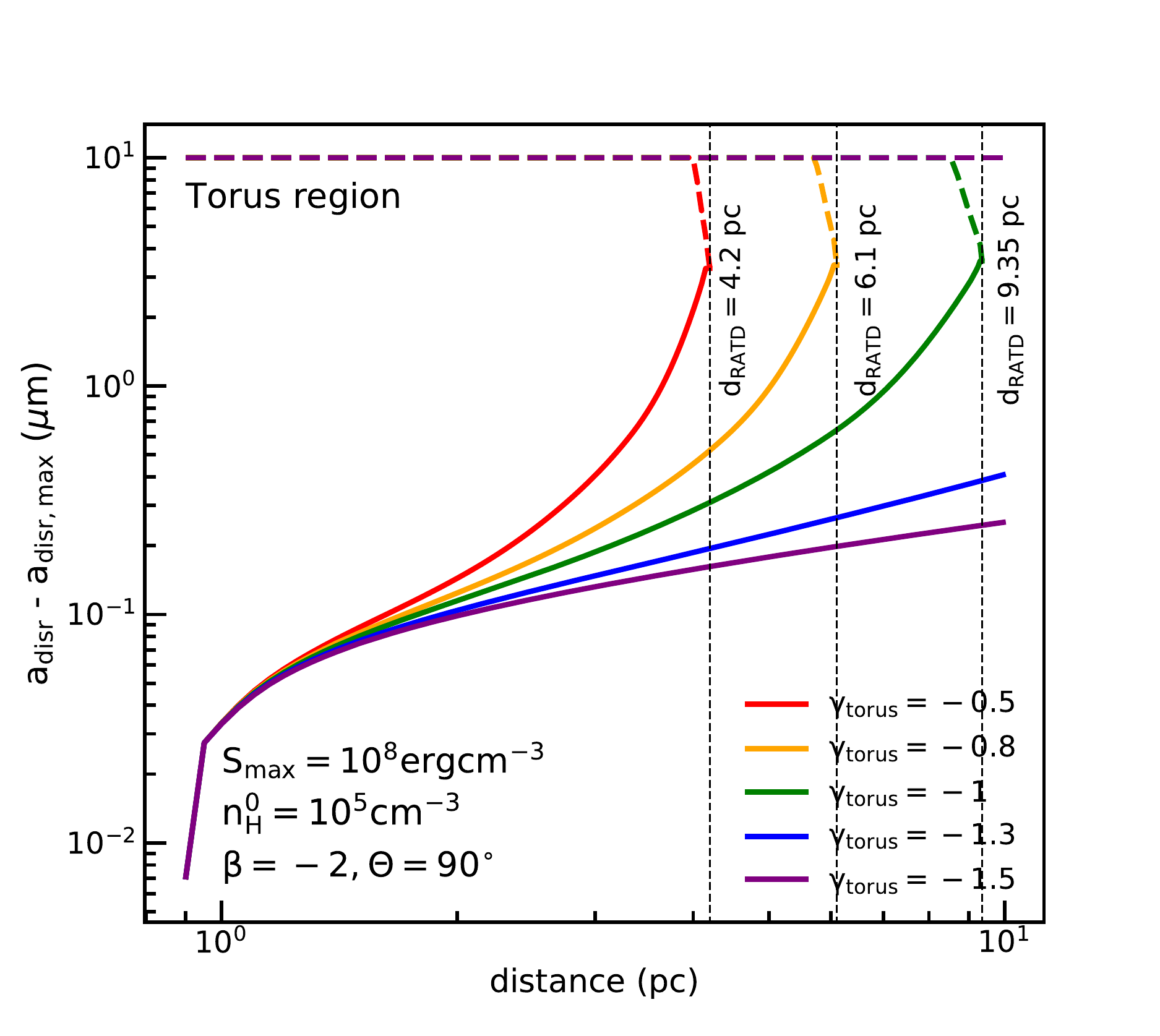}
    \includegraphics[width=0.45\textwidth]{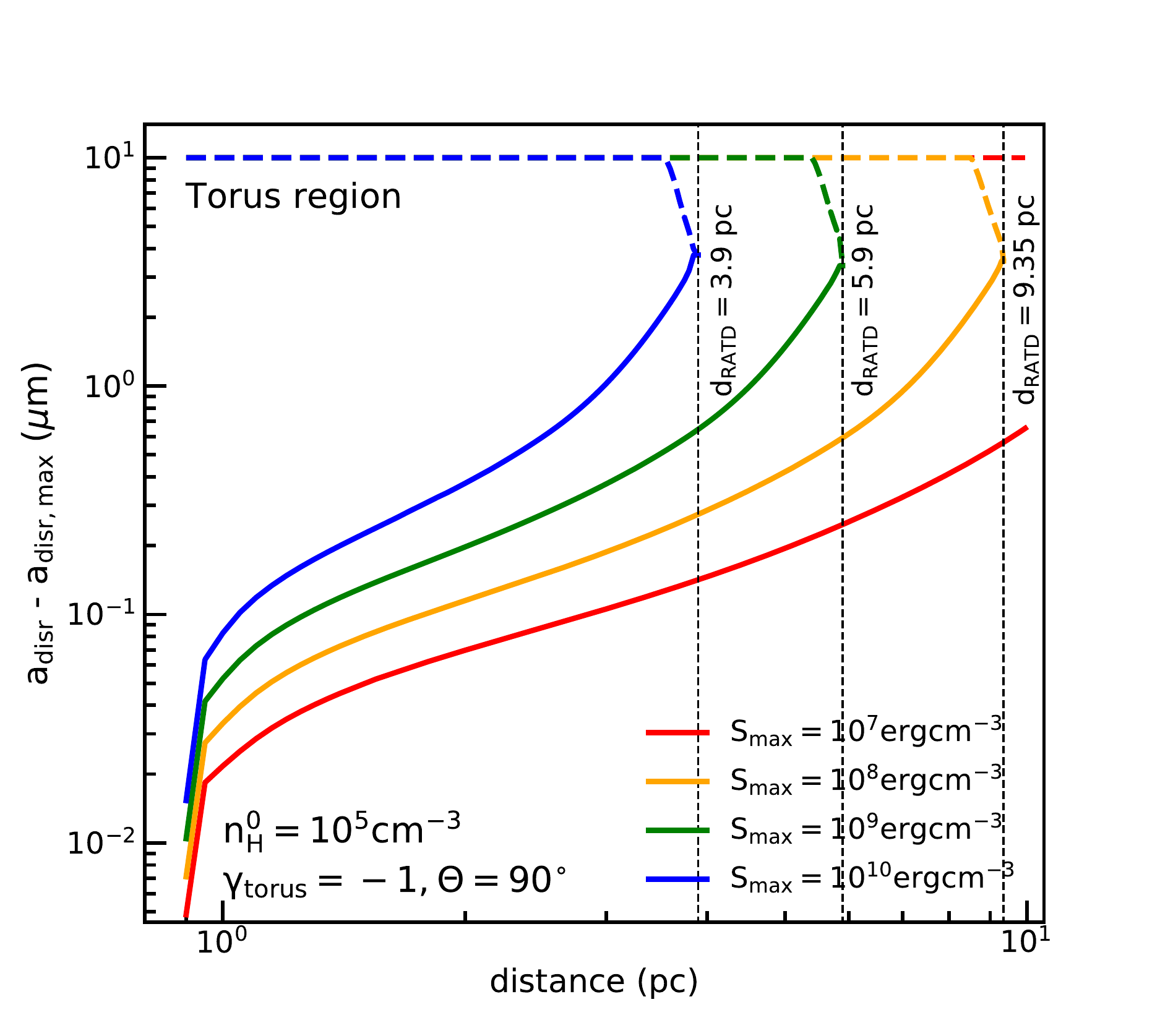}
    \includegraphics[width=0.45\textwidth]{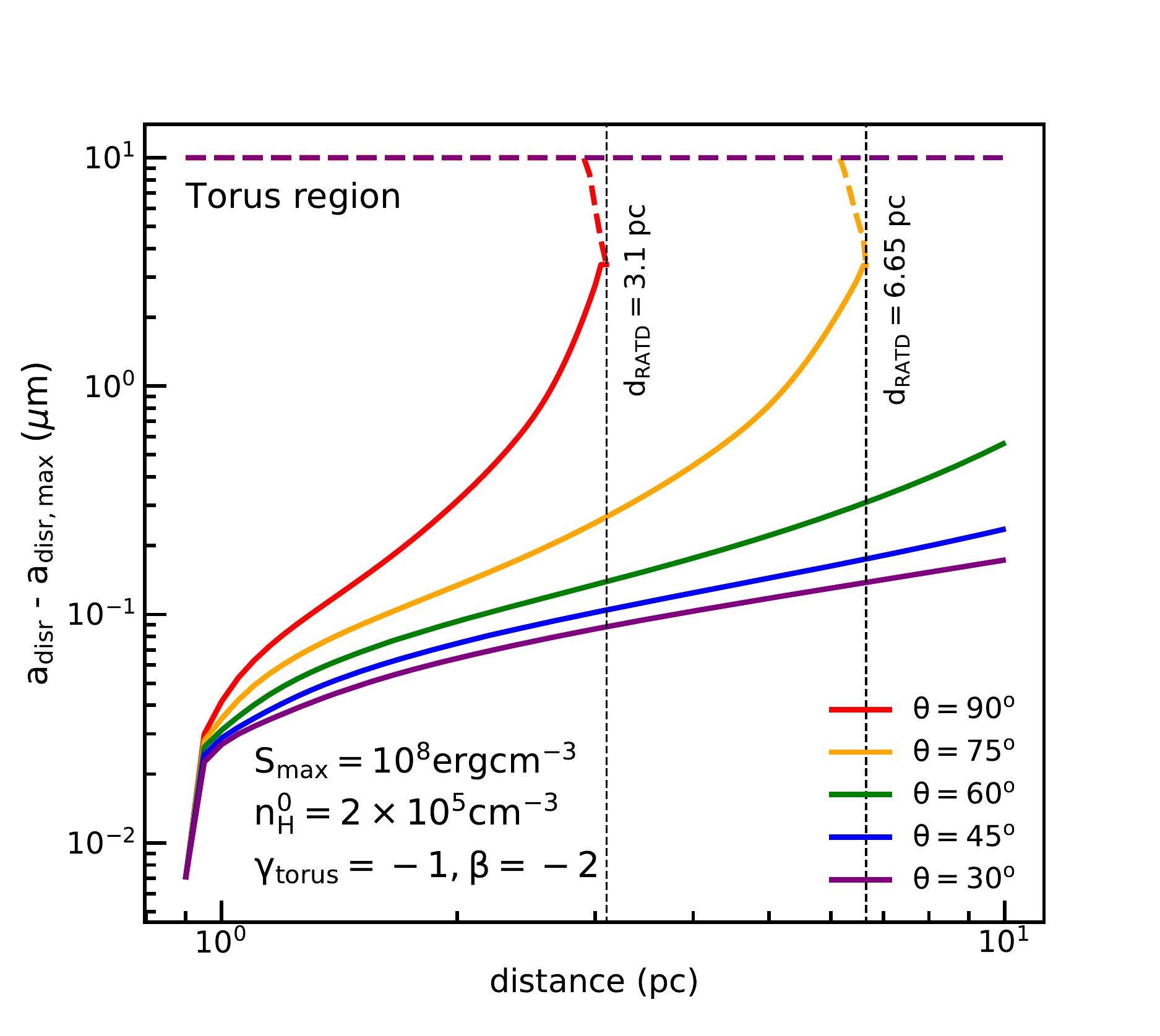}
    \includegraphics[width=0.45\textwidth]{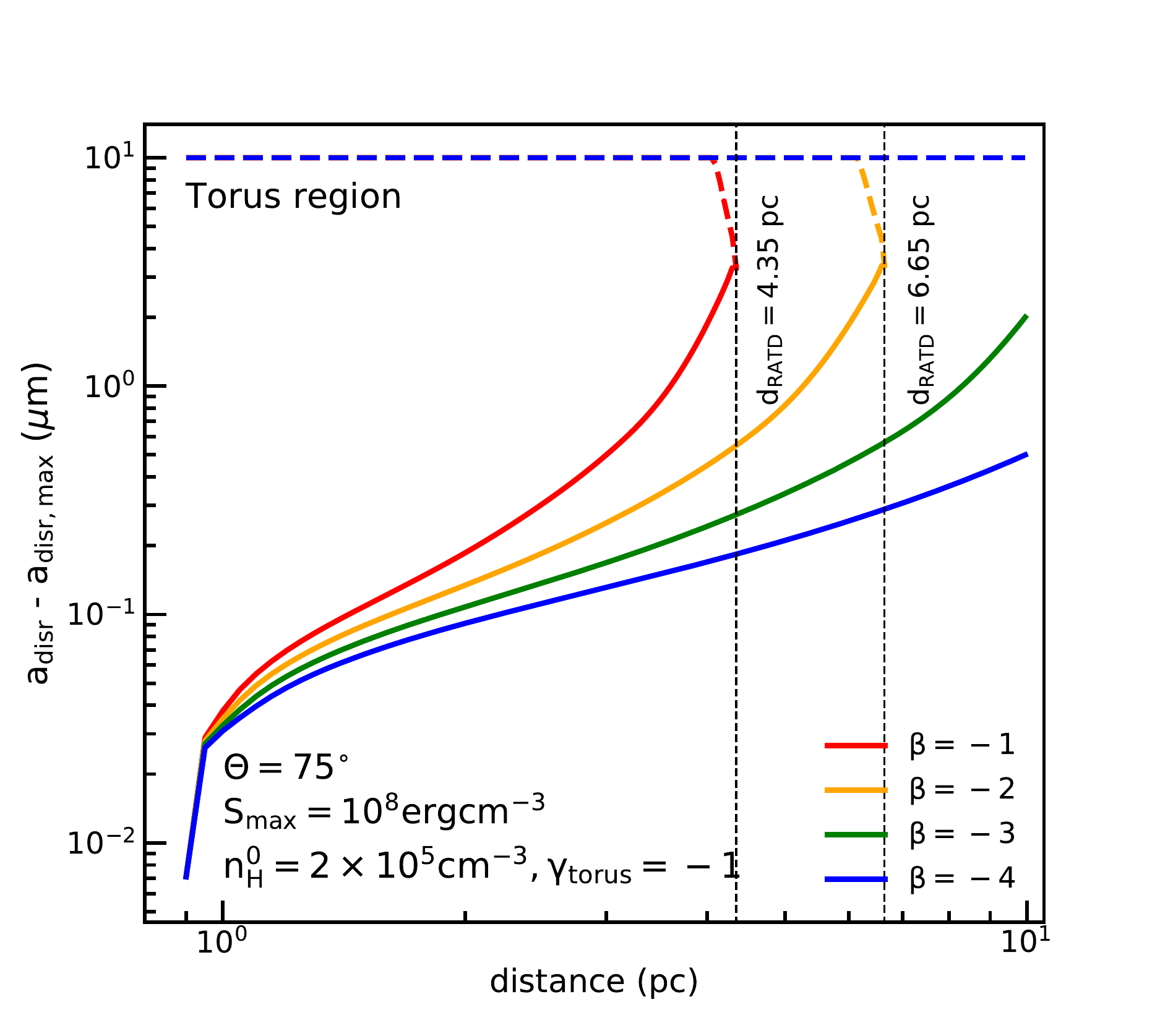}
\caption{Same as the lower panel of Figure \ref{fig:adisr_torus_nH} but for different parameters, including the radial power-law index $\gamma_{\rm torus}$ (upper left panel), the maximum tensile strength $S_{\rm max}$ (upper right panel), the observed angle $\Theta$ (lower left panel), and the polar index $\beta$ (lower right panel). Other parameters of the medium and grains are noted in each panel.}
\label{fig:adisr_torus_beta_Smax_theta_gamma}
\end{figure*}

Figure \ref{fig:adisr_torus_nH} shows the variation of the radiation field strength $U$ on the equatorial plane of the torus with distances (upper panel) and the corresponding size range of grains disrupted by RATD (lower panel). The number gas density in the torus is computed by Equation (\ref{eq:profile_torus}) with $\gamma_{\rm torus}= -1$, $\beta = -2$, $\Theta = 90^{\circ}$. The maximum tensile strength of grains is $S_{\rm max}=10^{8}\erg\cm^{-3}$. Similar to the polar cone, the radiation field strength is decreased stronger in the dense torus region (i.e., higher $n_{\rm H}^{0}$), which shrinks the active region of RATD. For example, the disruption region decreases from $\lesssim$ 9.35 pc to 3.1 pc and 1.7 pc when the initial gas density increases from $n_{\rm H}^{0} = 10^{5}\cm^{-3}$ to $2\times10^{5}\cm^{-3}$ and $4\times 10^{5}\cm^{-3}$, respectively.
 
Figure \ref{fig:adisr_torus_beta_Smax_theta_gamma} shows the dependence of the RATD efficiency with different parameters of the gas density profile and the maximum tensile strength. The upper left panel is for the varying power-index of the gas distribution in the radial direction $\gamma_{\rm torus}$. As increasing $\gamma_{\rm torus}$, the gas density drops slower resulting in the stronger dust reddening effect on the radiation field and the weaker effect of RATD in the torus region. For example, with $n_{\rm H}^{0} = 10^{5}\cm^{-3}$, the active region of RATD on the equatorial plane (i.e., $\Theta = 90^{\circ}$) reduces from 9.35 pc to 6.1 pc and 4.2 pc if the slope of the gas density profile becomes shallower from $\gamma_{\rm torus} = -1$ to -0.8 and -0.5, respectively. 

The upper right panel shows the results for the different values of $S_{\max}$. As increasing $S_{\rm max}$, the RATD is decreased due to larger disruption limit, which shrinks the disruption zone. For instance, with $n_{\rm H}^{0} = 10^{5}\cm^{-3}$ and for grains on the equatorial plane, $d_{\rm RATD}$ shrinks from $\geq 10$ pc to 3.9 pc with increasing the strength from $S_{\rm max} \leq 10^{8}\erg\cm^{-3}$ to $10^{10}\erg\cm^{-3}$.

The lower left panel shows the results for different observed angles $\Theta$. The active region of RATD expands when the observed direction is varied from the edge-on view ($\Theta = 90^{\circ}$) to the face-on view (e.g., $\Theta = 30^{\circ}$), due to the quick drop of gas density above the equatorial plane. The fast decrease of the gas density in the polar direction due to a lower value of $\beta$ also increases the active zone of RATD (see the lower right panel). For instance, with $n_{\rm H}^{0} = 2\times10^{5}\cm^{-3}$ and $\beta = -2$, RATD only can modify dust within $d \leq 3.1$ pc on the equatorial plane, but it can remove large grains within $d \leq 6.65$ pc along the direction of $\Theta =75^{\circ}$ and $d\geq 10$ pc for $\Theta > 60^{\circ}$. Moreover, along the direction of $\Theta = 75^{\circ}$, the active region of RATD can increase from 6.65 pc to $>$ 10 pc when the gas density drops with distances with $\beta \leq -3$.
 
 \begin{figure*}[t]
    \centering
    \includegraphics[width=0.45\textwidth]{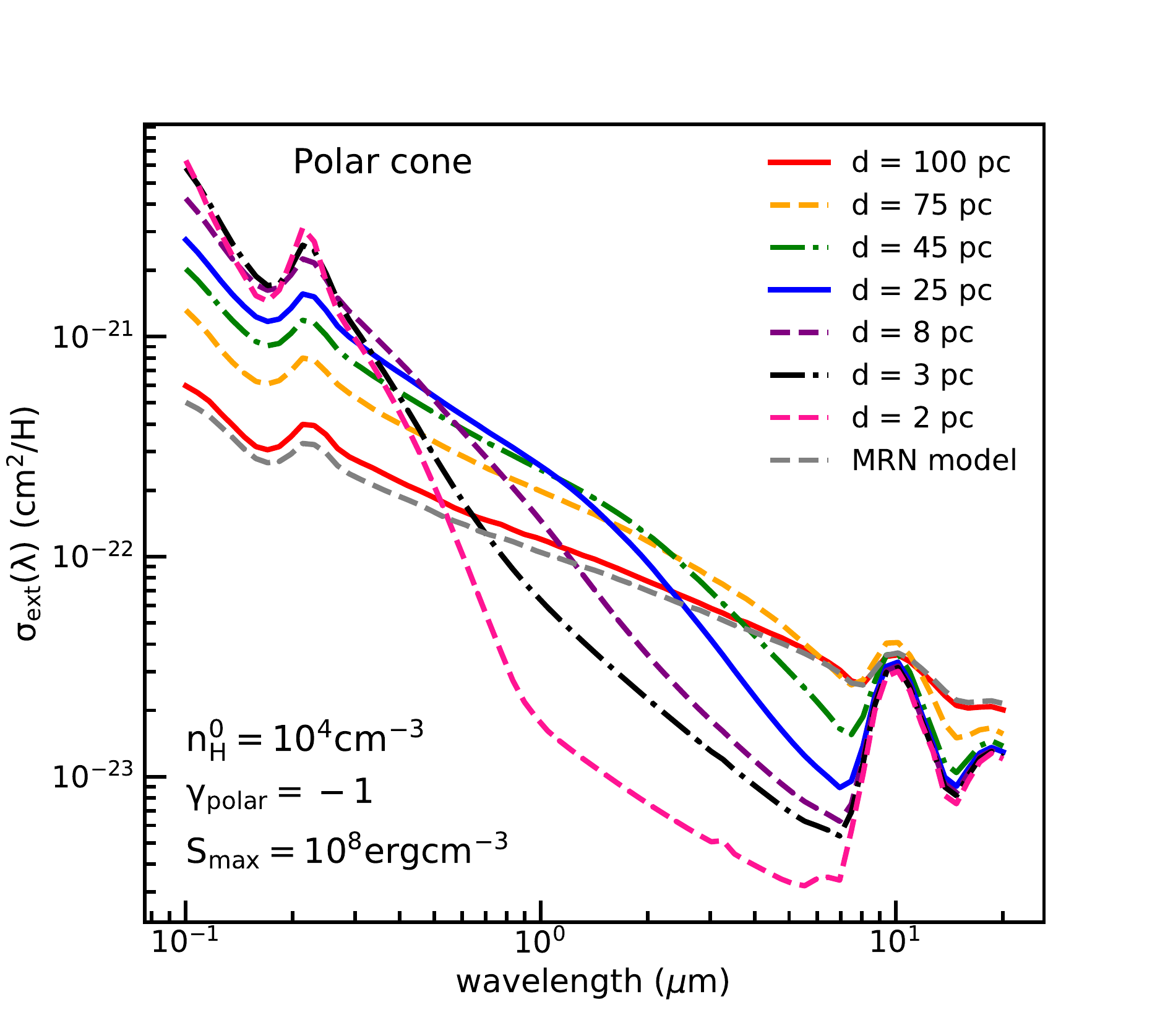}
    \includegraphics[width=0.45\textwidth]{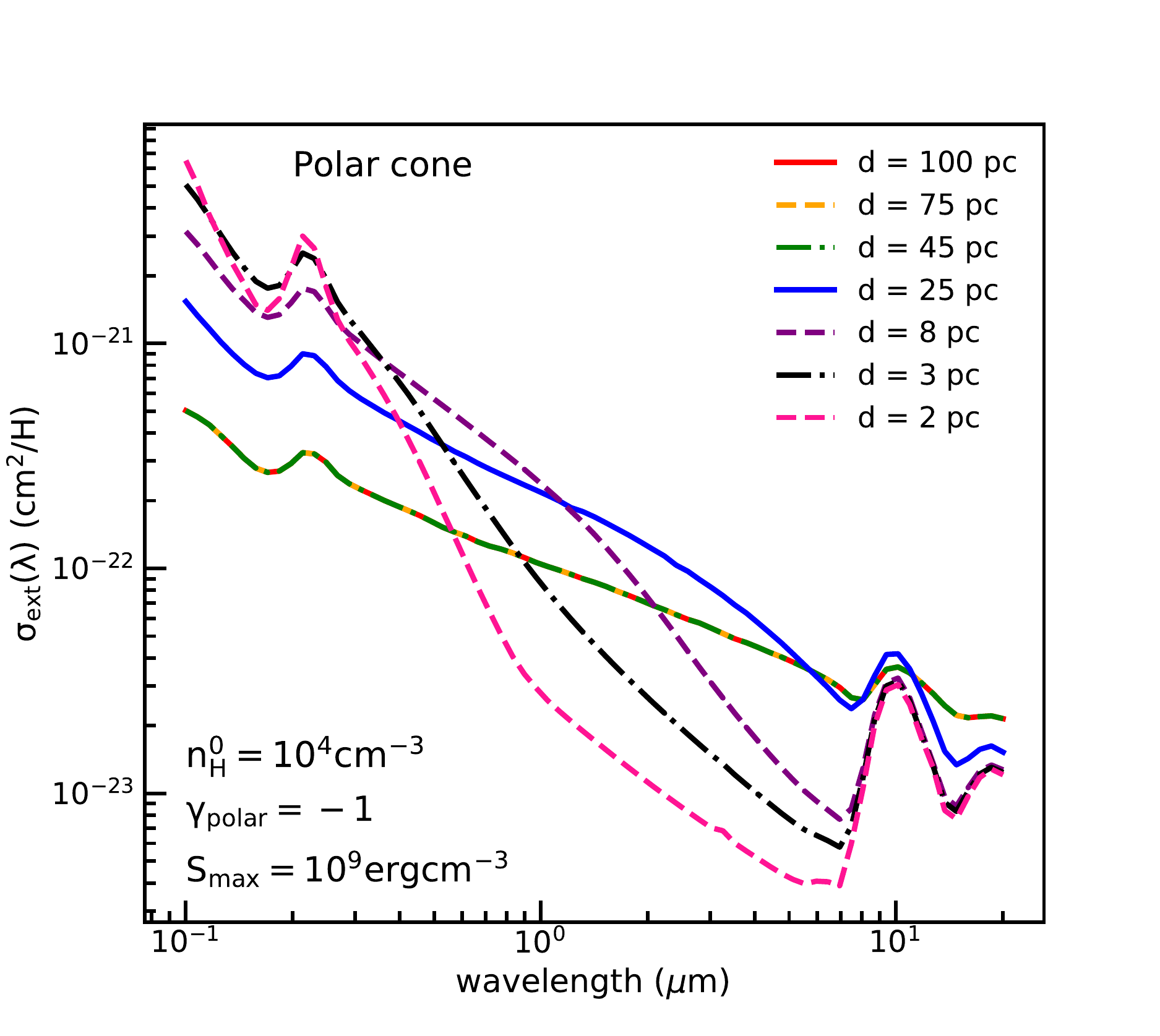}
\caption{Extinction cross-section per H, $\sigma_{\rm ext}(\lambda)$, caused by grains in a thin cell located at different distances for $S_{\rm max}=10^{8}\erg\cm^{-3}$ (left panel) and $S_{\rm max}=10^{9}\erg\cm^{-3}$ (right panel) for the torus. The gray dashed line in the left panel is produced by original grains that follows the MRN size distribution. As decreasing distances to the center of AGN, FUV-NUV extinction increases while optical-MIR extinction decreases owing to the strong conversion of large grains to smaller sizes by RATD.} 
\label{fig:A_lambda(d)}
\end{figure*}
 
\section{Modelling photometric parameters with RATD effect} \label{sec:Photometry}
\subsection{Extinction curves}\label{sec:Aext}
Now, we use the new grain size distribution obtained by RATD to model extinction curves toward AGN. To account for the fact that AGN may have a clumpy structure, we will model both the extinction cross section induced by grains in the dusty cell at different distances and the final extinction curve produced by all grains from the sublimation front.

From the numerical calculation of the optical depth described in Section \ref{sec:dust_redding}, the dust extinction at wavelength $\lambda$ induced by grains from the sublimation distance to $n^{\rm th}$ cell, $A(\lambda, n)$, is given by:
\bea 
A(\lambda, n) =  \sum_{\rm i = 1}^{n} 1.086 \Delta \tau_{\lambda,i} ~ \rm mag,
\label{eq:A_lambda}
\ena
where $\Delta \tau_{\lambda,i}$ is the optical depth produced by the $i^{\rm th}$ cell.

\subsubsection{Polar cone} \label{sec:Aext_polar} 
The left panel of Figure \ref{fig:A_lambda(d)} shows the extinction cross section per H, $\sigma_{\rm ext}(\lambda)$, produced by grains in a cell at different distances from $d=100$ pc to 2 pc in the polar region, assuming $n_{\rm H}^{0} = 10^{4}\cm^{-3}$ and $S_{\rm max}=10^{8}\erg\cm^{-3}$. Here, we assume that the cell is thin such that the grain disruption size does not change substantially in the cell. The gray dashed line is the wavelength-dependent extinction cross section produced by original grains that follows the MRN distribution. At a distance of $d = 100$ pc, carbonaceous micron-sized grains still be enhanced by RATD because of smaller dust mass density (see Equation \ref{eq:w_terminal}). Thus, one can obtain the higher extinction cross section by grains from FUV to MIR compared with ones produced by the original dust model. As the distance decreases to $d = 2$ pc, the efficiency of RATD increases due to the increase of radiation flux. As a result, $\sigma_{\rm ext}(\lambda)$ at optical-MIR continues to rise due to the enhancement of sub-micron grains, i.e., the curve at 75 pc, 45 pc, 45 pc, 25 pc, then decreases later when these grains are removed, i.e., the curve at 8 pc, 3 pc, and 2 pc (see the lower panel of Figure \ref{fig:adisr_polar_gamma_Smax}). NIR-MIR extinction cross section is reduced much stronger than optical range because of the quick removal of micron-sized grains by RATD. In contrast, $\sigma_{\rm ext}(\lambda)$ at FUV-NUV increases rapidly due to the significant enhancement of small grains in the polar cone. As a consequence, the steepness of FUV-MIR extinction cross section becomes steeper if the cell of grains locates closer to AGN.

In addition, at $d = 100$ pc, 75 pc, and 45 pc, the extinction cross section at $9.7\mum$ produced by the Si-O stretching mode of silicate grains is depleted due to the dominance of large grains in the polar cone. Moving closer to the center of AGN, the extinction at $9.7\mum$ is higher due to the increase of small polar dust.

The right panel of Figure \ref{fig:A_lambda(d)} shows similar results as the left one, but for grains with $S_{\rm max} = 10^{9}\erg\cm^{-3}$. The extinction cross section curve will be steeper toward FUV in the active region of RATD. However, the curve produced by grains with $S_{\rm max} = 10^{9}\erg\cm^{-3}$ only shows a significant rise in FUV-NUV range within 8 pc due to the narrower active region of RATD. Further than that, it becomes flattened due to the presence of large grains and ceases when polar dust is not modified by RATD, i.e., the curve at $d=45$ pc, 75 pc, and 100 pc. Besides, the high extinction at $9.7 \mum$ for $S_{\rm max} = 10^{8}\erg\cm^{-3}$ only happens within $d < 25$ pc, which is smaller than $d < 45$ pc in the left panel.

\begin{figure*}[t]
    \centering
    \includegraphics[width=0.45\textwidth]{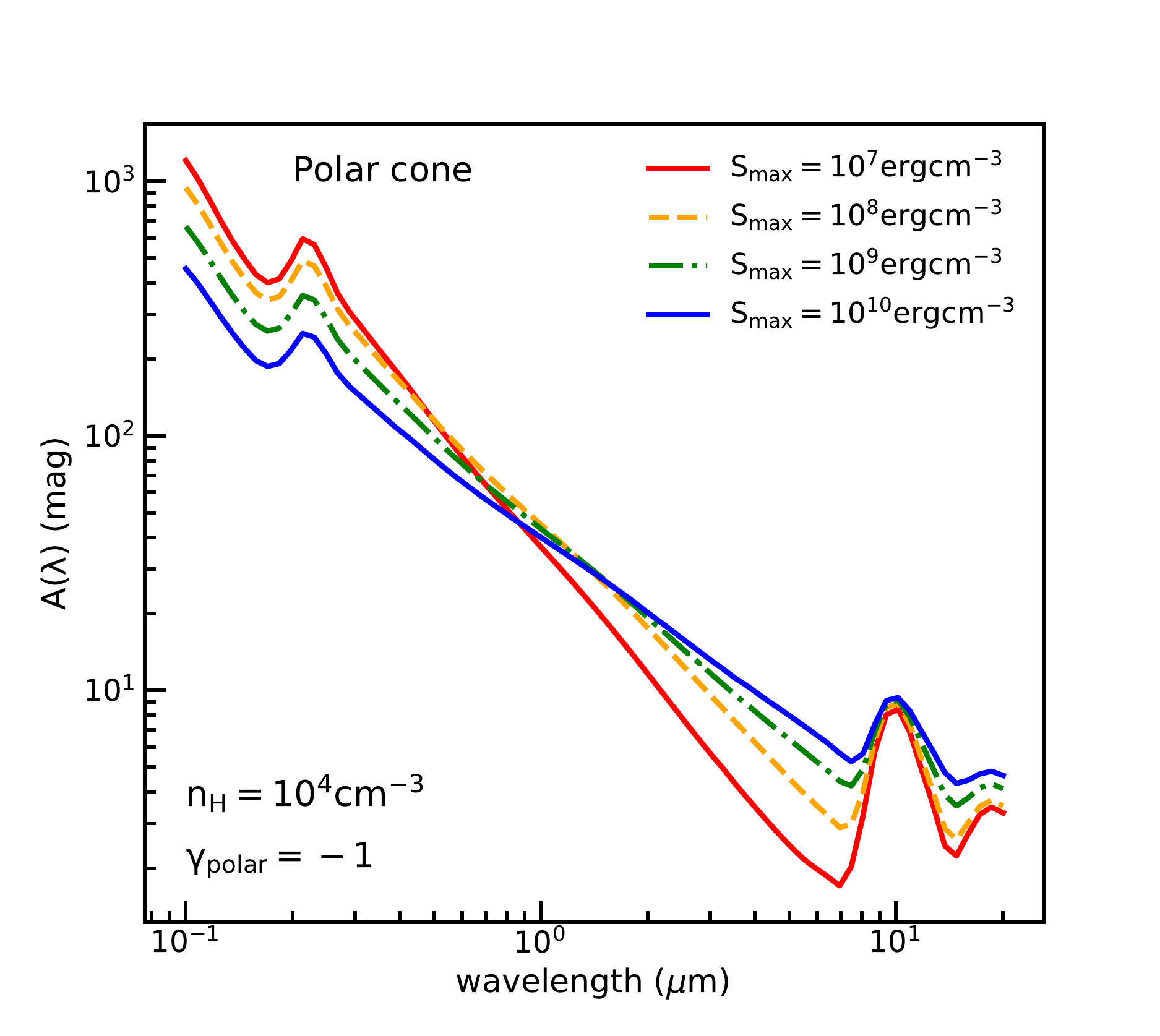} 
    \includegraphics[width=0.45\textwidth]{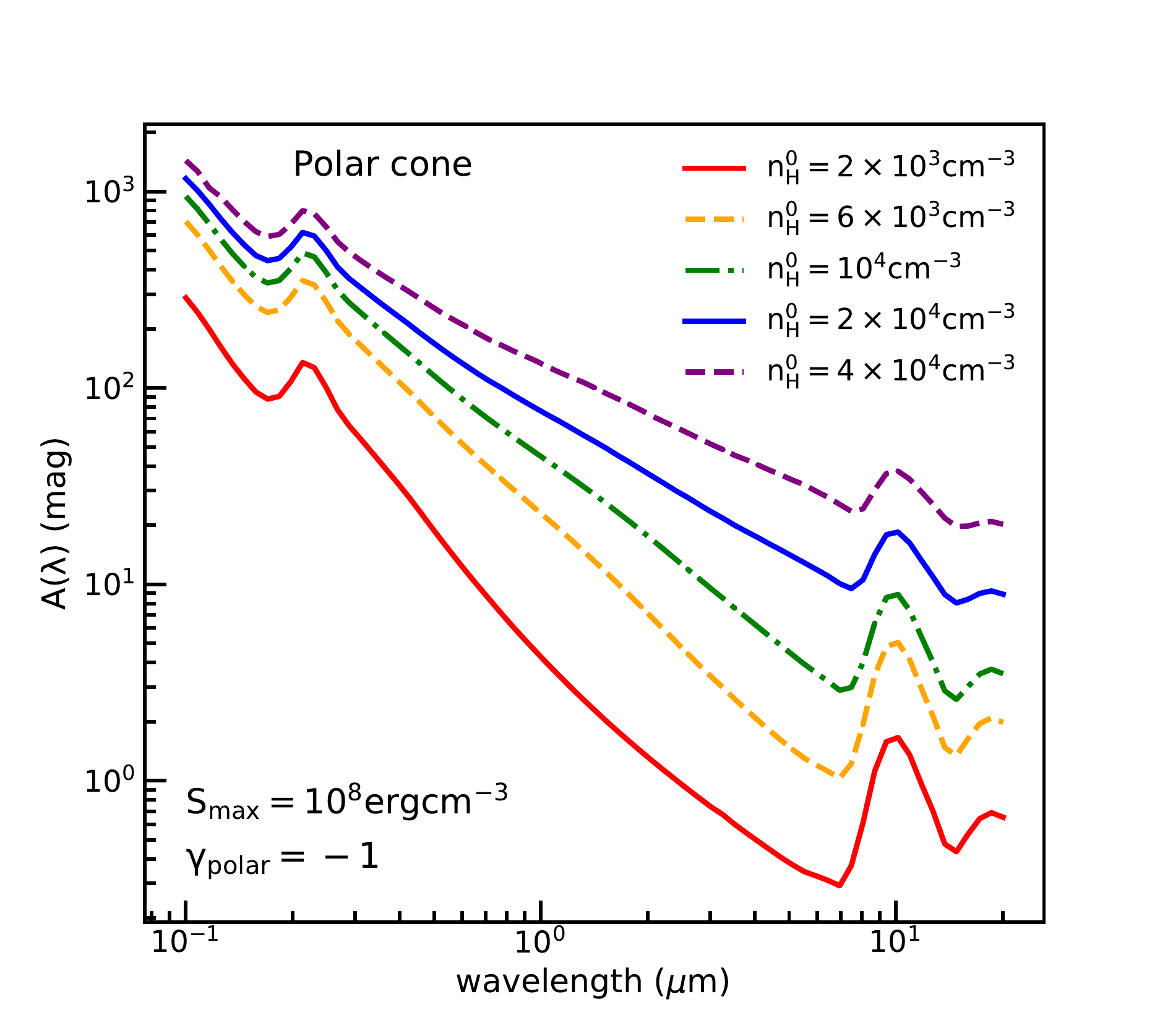} 
\caption{Final extinction curves caused by grains in the polar region within 100 pc for different values of $S_{\rm max}$, assuming $n_{\rm H}=10^{4}\cm^{-3}$ and $\gamma_{\rm polar} = -1$ (left panel), and for different values of $n_{\rm H}^{0}$, assuming $\gamma_{\rm polar} = -1$ and $S_{\rm max}=10^{8}\erg\cm^{-3}$ (right panel).}
\label{fig:A_lambda_polar}
\end{figure*}

The left panel of Figure \ref{fig:A_lambda_polar} shows the final extinction curves, $A_{\rm \lambda}$, produced by grains in the polar region within 100 pc from the sublimation front for different values of $S_{\rm max}$, assuming $n_{\rm H}^{0} = 10^{4}\cm^{-3}$ and $\gamma_{\rm  polar} = -1$. The extinction curve for high tensile strength of $S_{\rm max}=10^{10}\erg\cm^{-3}$ exhibits a prominent rise from MIR to FUV. This slope is produced by the decrease in optical-MIR extinction and the increase in FUV-NUV extinction due to the significant conversion of large polar dust to smaller sizes near the center of AGN (see the lower panel of Figure \ref{fig:adisr_polar_gamma_Smax} and Figure \ref{fig:A_lambda(d)}). Grains with $S_{\rm max} < 10^{10}\erg\cm^{-3}$ extinct FUV-NUV stronger but optical-MIR weaker. As a result, the slope of the extinction curve in FUV-MIR range will become steeper toward the blue. The $9.7 \mum$ extinction for lower $S_{\rm max}$ is also stronger due to higher abundance of small silicate grains by RATD.

The right panel of Figure \ref{fig:A_lambda_polar} shows similar results as the left one, but for different values of the initial gas density, assuming $\gamma_{\rm polar} = -1$ and $S_{\rm max} = 10^{8}\erg\cm^{-3}$. For a given $S_{\rm max}$, grains in the polar cone with lower density produce the clear far-UV rise extinction curve and the strong $10 \mum$ extinction feature due to the strong enhancement of small grains by RATD. However, its magnitude is smaller than ones produced in higher density polar cone because of the proportional of $A(\lambda)$ and $n_{\rm H}$ (see Equations \ref{eq:A_lambda} and \ref{eq:Aext}).

\begin{figure}[t]
        \includegraphics[width=0.45\textwidth]{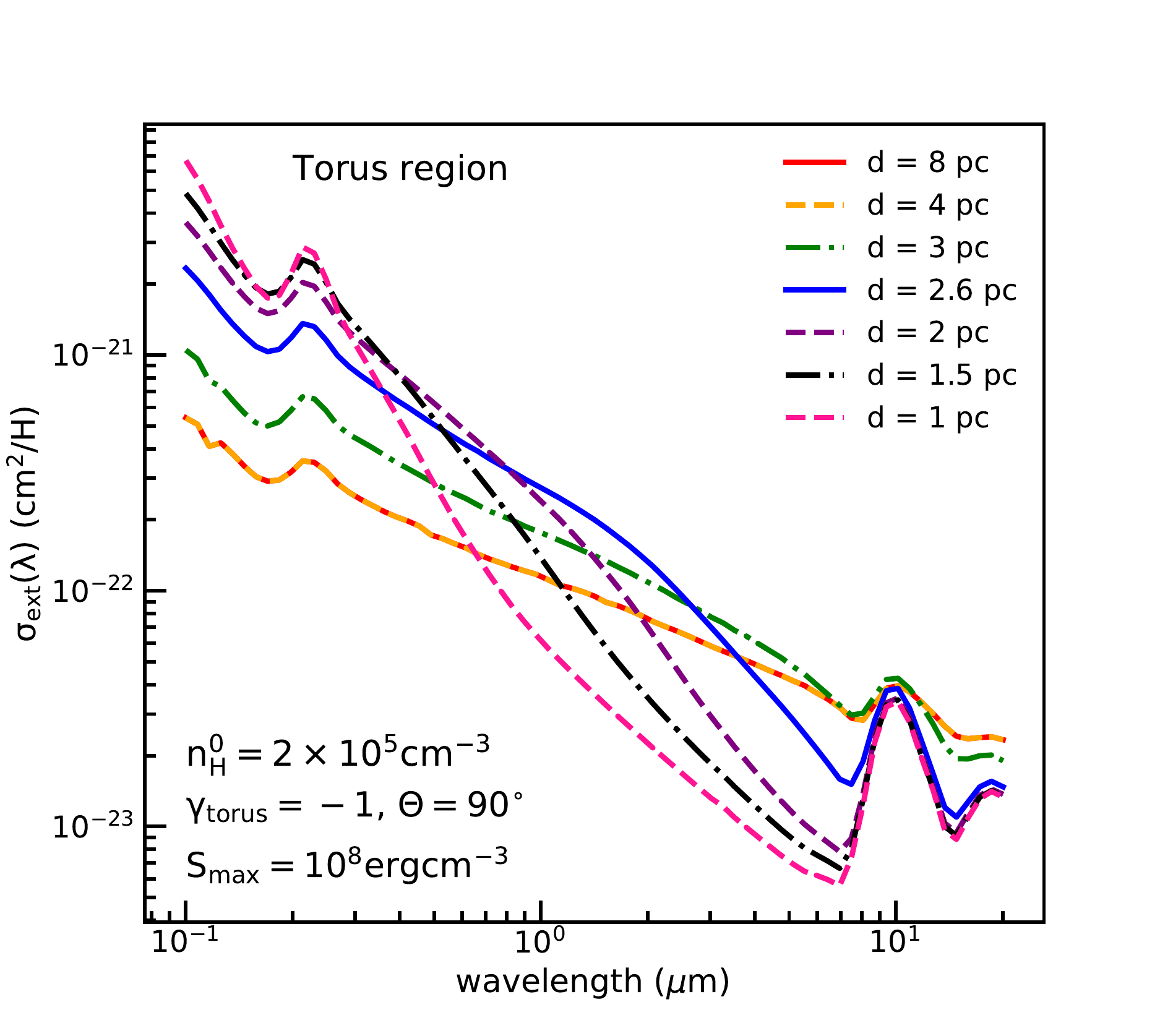}
     \caption{Extinction cross section, $\sigma_{\rm ext}(\lambda)$, caused by dust in the torus observed along the equatorial plane at different cell distances, assuming $S_{\rm max} = 10^{8}\erg\cm^{-3}$, $n_{\rm H}^{0} = 2\times10^{5}\cm^{-3}$, $\gamma_{\rm torus} = -1$, and $\beta = -2$. The extinction cross-section near AGN shows a steeper rise toward FUV than ones produced by dust at a distant cell.}
     \label{fig:normal_A_lambda_torus_Smax}
\end{figure}
 
\begin{figure*}
   \begin{minipage}[t]{0.47\textwidth}
     \centering
        \includegraphics[width=0.9\textwidth]{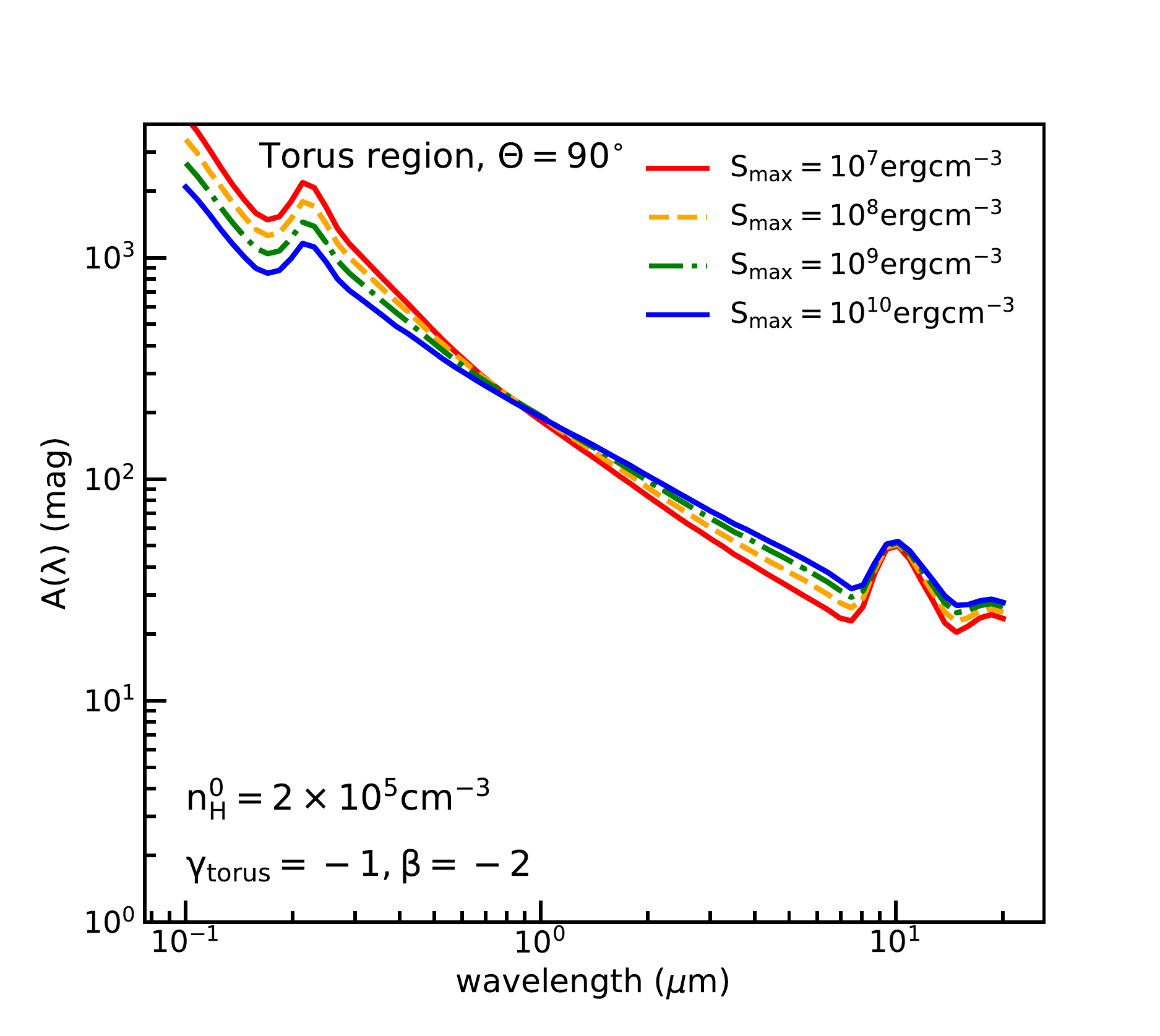}
        \includegraphics[width=0.9\textwidth]{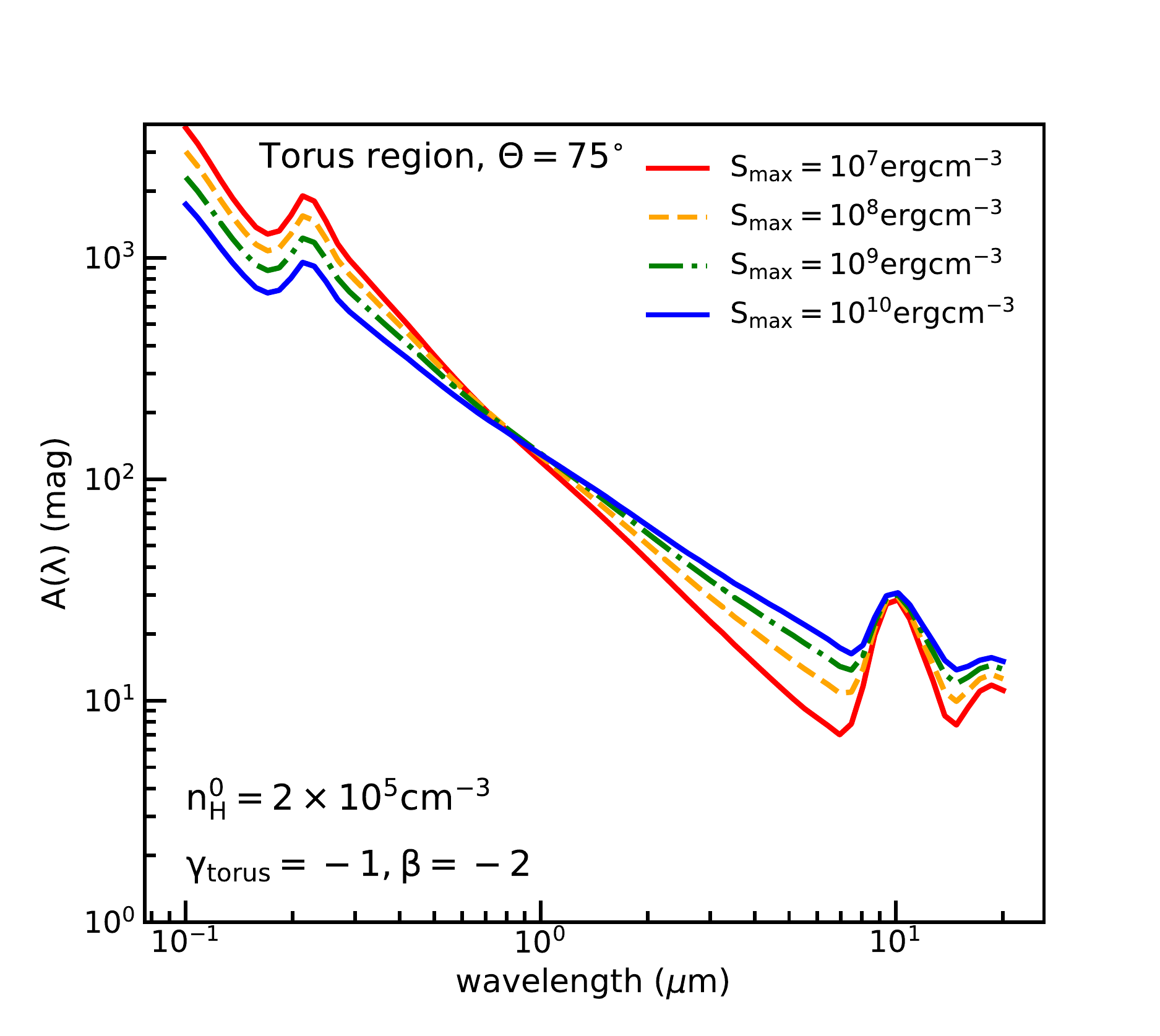}
        \includegraphics[width=0.9\textwidth]{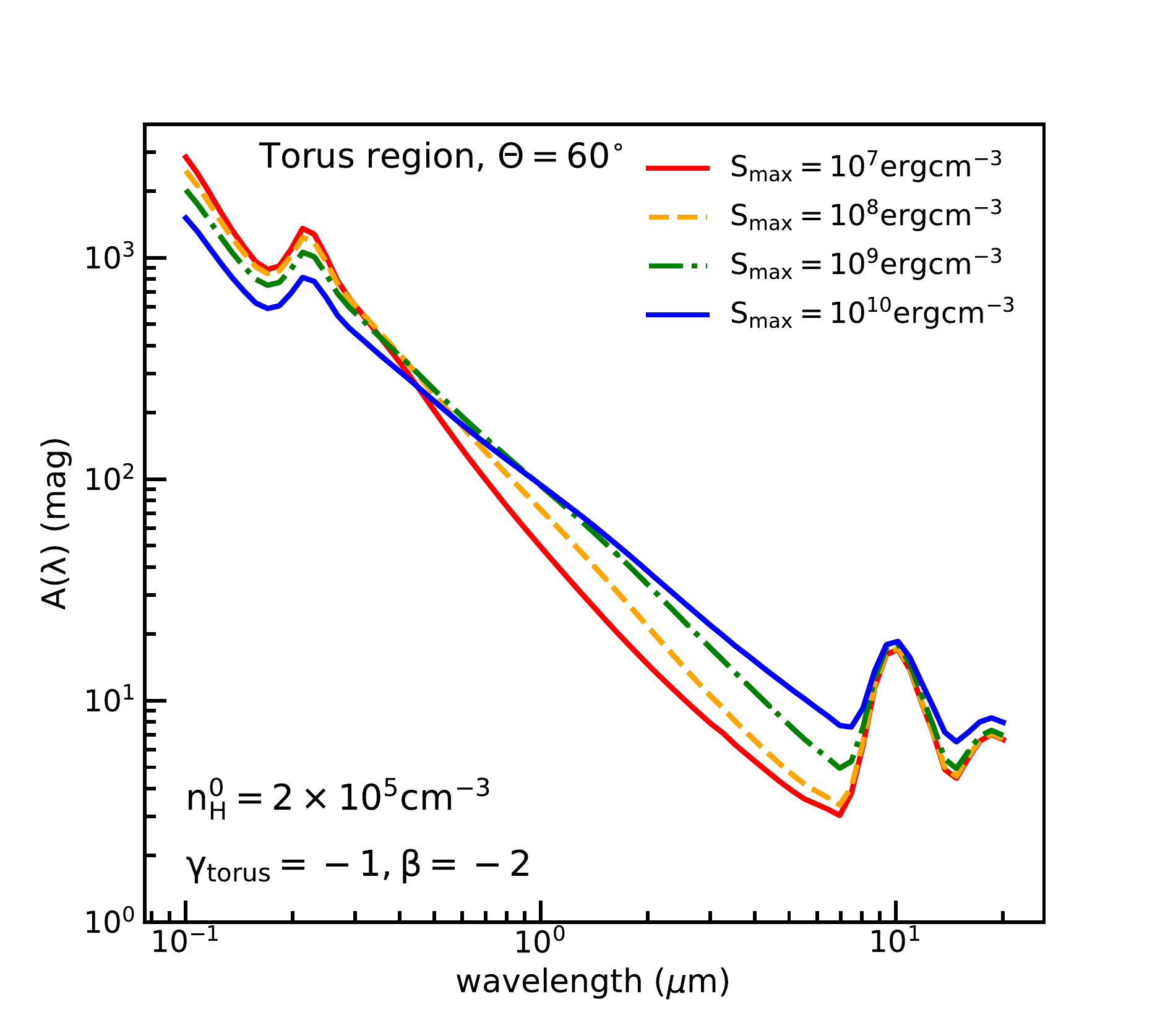}
     \caption{Total extinction curves produced by dust in the torus within 10 pc from the center of AGN for different values of $S_{\rm max}$, assuming $n_{\rm H}^{0} = 2\times10^{5}\cm^{-3}$, $\gamma_{\rm torus} = -1$ and $\beta = -2$. From top to bottom, the observed angle changes from $\Theta = 90^{\circ}$ to $\Theta = 75^{\circ}$ and $60^{\circ}$, respectively}. 
     \label{fig:A_lambda_torus_Smax}
   \end{minipage}\hfill
   \begin{minipage}[t]{0.47\textwidth}
     \centering
    \includegraphics[width=0.9\textwidth]{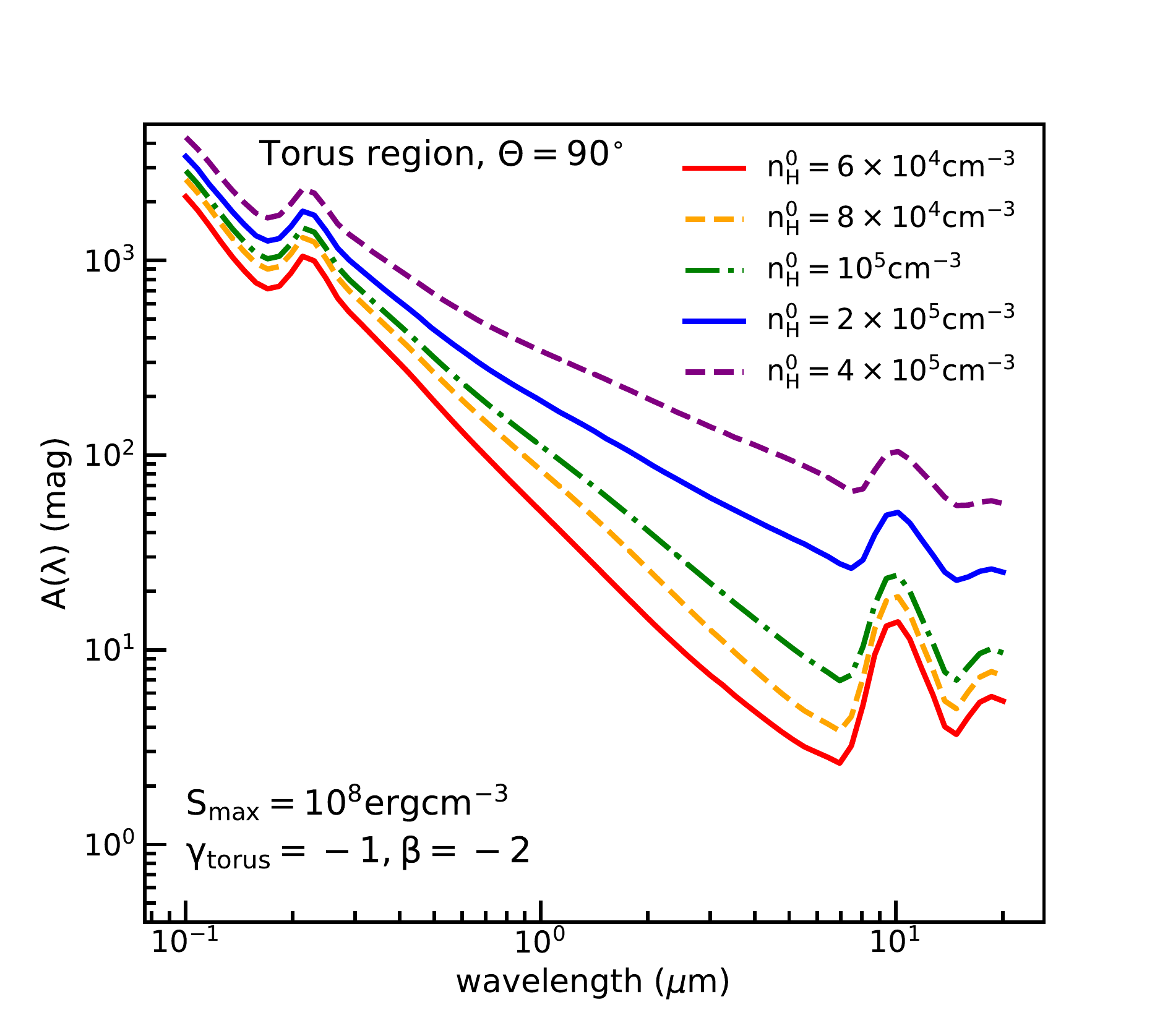}
    \includegraphics[width=0.9\textwidth]{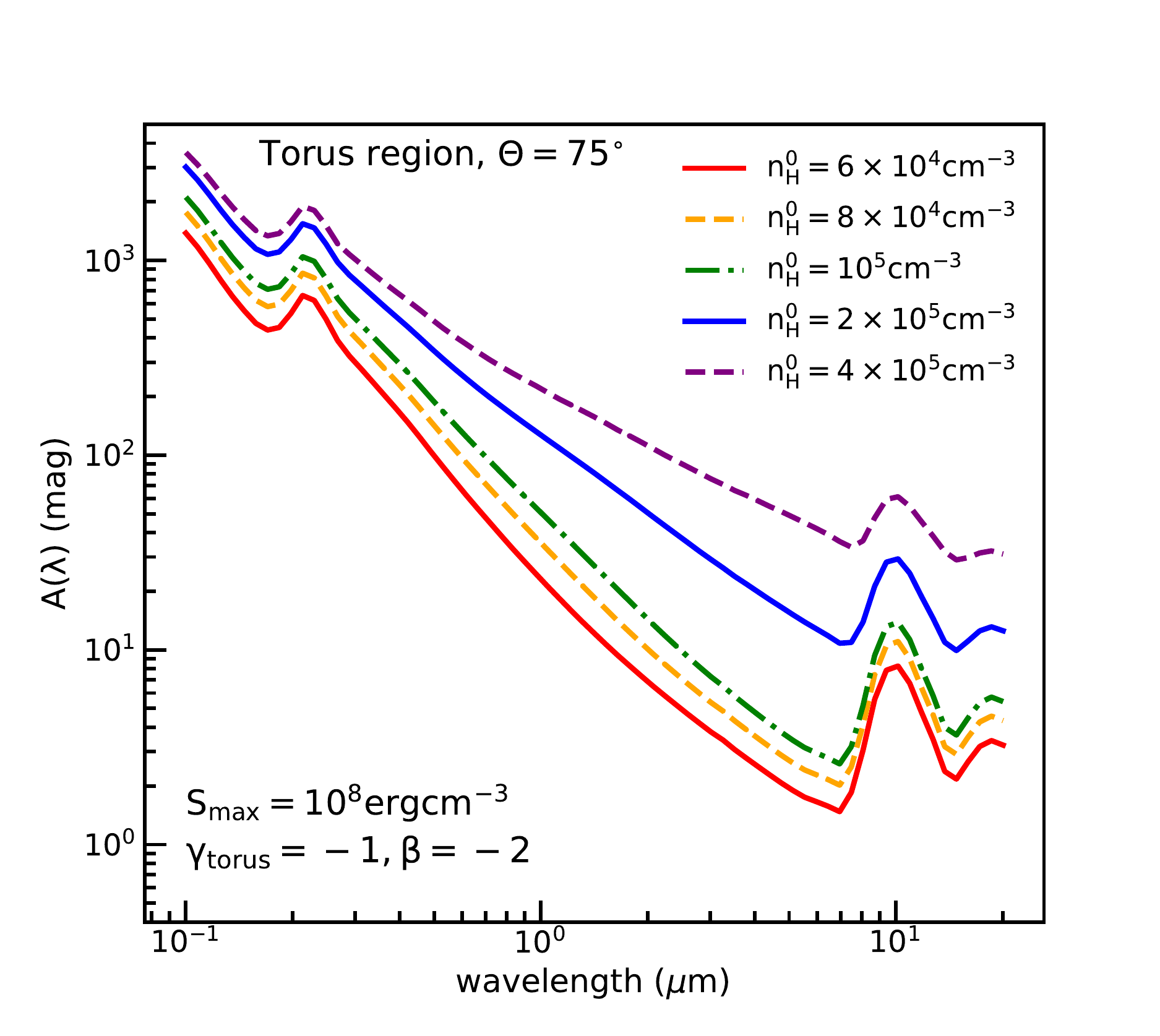}
    \includegraphics[width=0.9\textwidth]{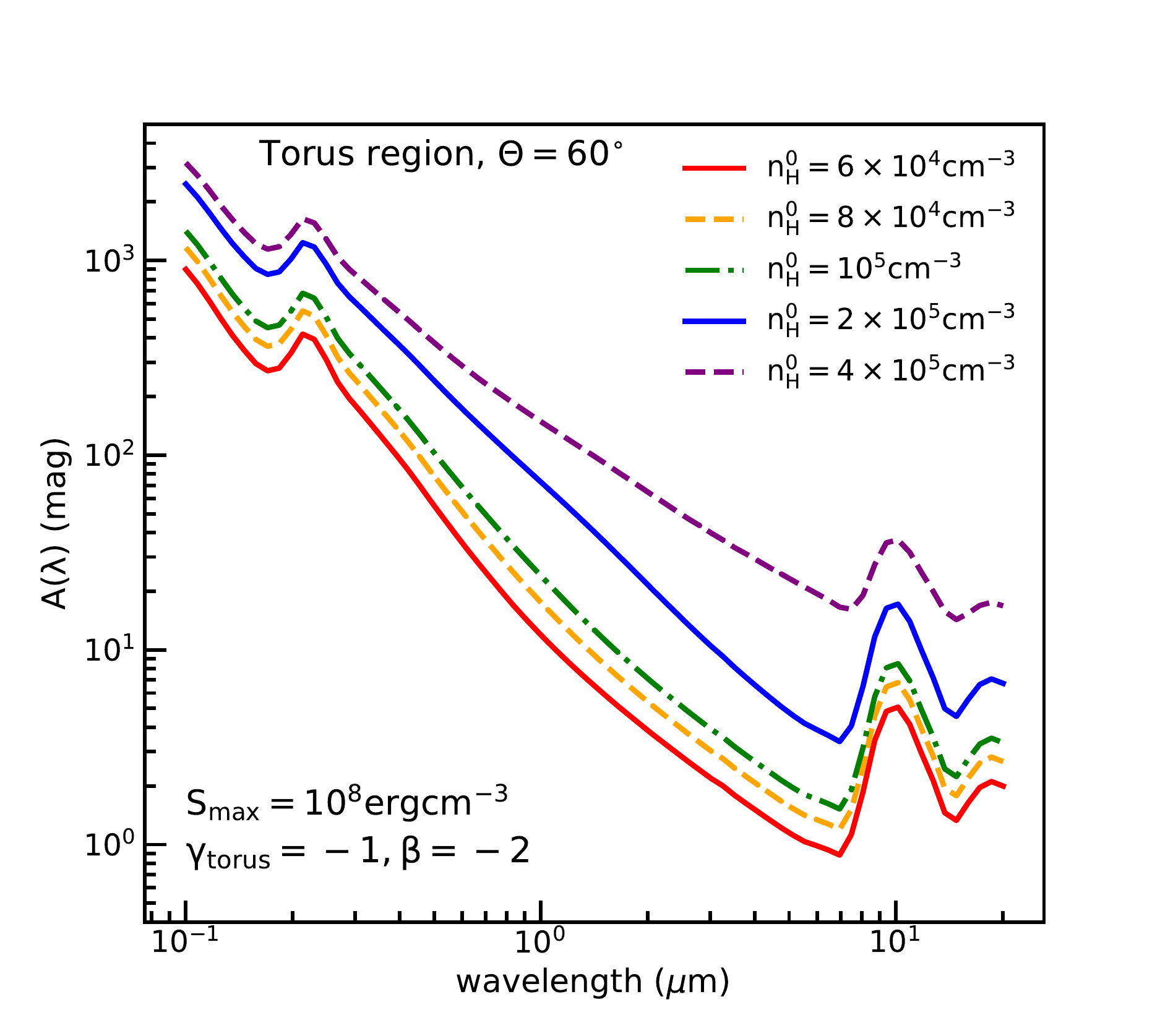}
   \caption{Same as Figure \ref{fig:A_lambda_torus_Smax} but for different values of $n_{\rm H}^{0}$, assuming $S_{\rm max} = 10^{8}\erg\cm^{-3}$, $\gamma_{\rm torus} = -1$, and $\beta=-2$ .} 
   \label{fig:A_lambda_torus_nH}
   \end{minipage}
\end{figure*} 

\subsubsection{Torus region} \label{sec:Aext_torus} 
Figure \ref{fig:normal_A_lambda_torus_Smax} shows the extinction cross section, $\sigma_{\rm ext}(\lambda)$, produced by grains at different distances in the midplane of the circumnuclear region, assuming $S_{\rm max} = 10^{8}\erg\cm^{-3}$, $n_{\rm H}^{0} = 10^{5}\cm^{-3}$, $\gamma_{\rm torus} = -1$, and $\beta = -2$. Similar to Figure \ref{fig:A_lambda(d)}, the extinction cross section curve will become steeper toward FUV with decreasing distances due to the predominance of small grains by RATD. The $9.7\mum$ absorption feature is also prominent for the cell near the center of AGN.

Figure \ref{fig:A_lambda_torus_Smax} shows total extinction curves $A_{\rm \lambda}$ produced by all grains from the sublimation front to the outer region of torus at 10 pc with different values of $S_{\rm max}$, assuming $n_{\rm H}^{0} = 10^{5}\cm^{-3}$, $\gamma_{\rm torus} = -1$, and $\beta = -2$. From top to bottom, the observed angle $\Theta$ changes from $90^{\circ}$ to $75^{\circ}$ and $60^{\circ}$, respectively. Similar to the left panel of Figure \ref{fig:A_lambda_polar}, for a fixed line of sight, as $S_{\rm max}$ decreases, optical-MIR extinction drops while FUV-NUV extinction rises, resulting in the steep far-UV rise extinction curve. The $9.7\mum$ feature is also extinct stronger due to the higher abundance of small grains. By changing the observed angle from the edge-on to the face-on view, the slope of FUV-MIR extinction curves of all considered cases of $S_{\rm max}$ become steeper toward shorter wavelengths, and the $9.7\mum$ feature becomes stronger. However, its magnitude slightly decreases due to the fast reduction of $n_{\rm H}$ above the equatorial plane.

Figure \ref{fig:A_lambda_torus_nH} shows the similar results as Figure \ref{fig:A_lambda_torus_Smax}, but for different values of $n_{\rm H}^{0}$, assuming $S_{\rm max} = 10^{8}\erg\cm^{-3}$. The extinction cure shows a clear rise from MIR to FUV, and the extinction feature at $10\mum$  becomes prominent for the lower gas density. The magnitude of the curve is smaller due to the proportional between $A(\lambda)$ and $n_{\rm H}$. These features are clearer when one observes AGN at the face-on direction due to the expansion of active region of RATD (see the lower left panel of Figure \ref{fig:adisr_torus_beta_Smax_theta_gamma}).

In conclusion, the modification of grains around AGN by RATD results in the far-UV rise extinction curve. The slope of the extinction curve depends on the RATD efficiency. Even when RATD is not effective at large distances from AGN, the net observed extinction curve is still steeper than the curve without RATD due to the enhancement of small grains near AGN center.  

\subsection{Color excess, $E(B-V)$, and total-to-selective extinction ratio, $R_{\rm V}$} \label{sec:Ebv_Rv}
 \begin{figure}[t]
    \includegraphics[width=0.45\textwidth]{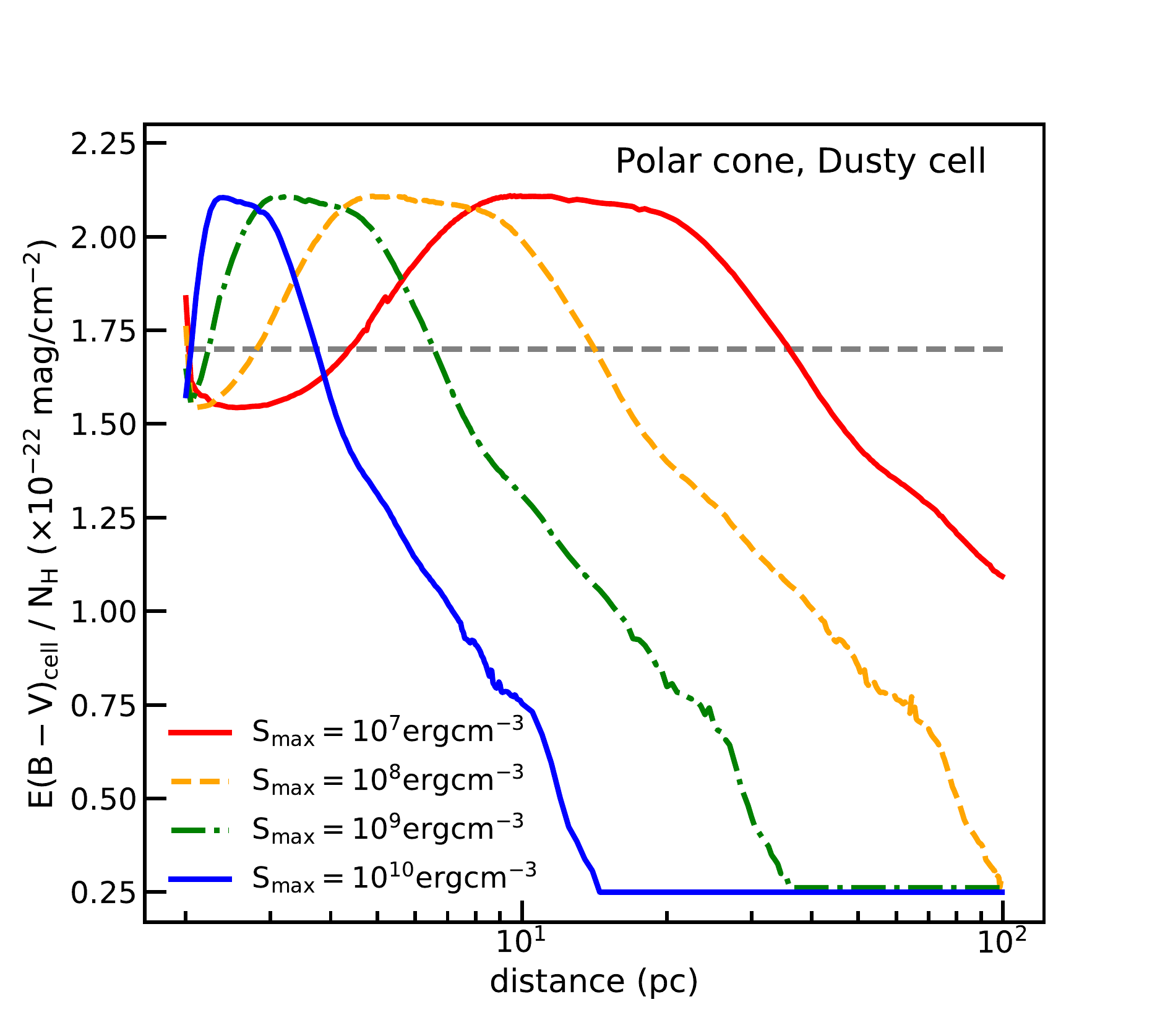}
        \includegraphics[width=0.45\textwidth]{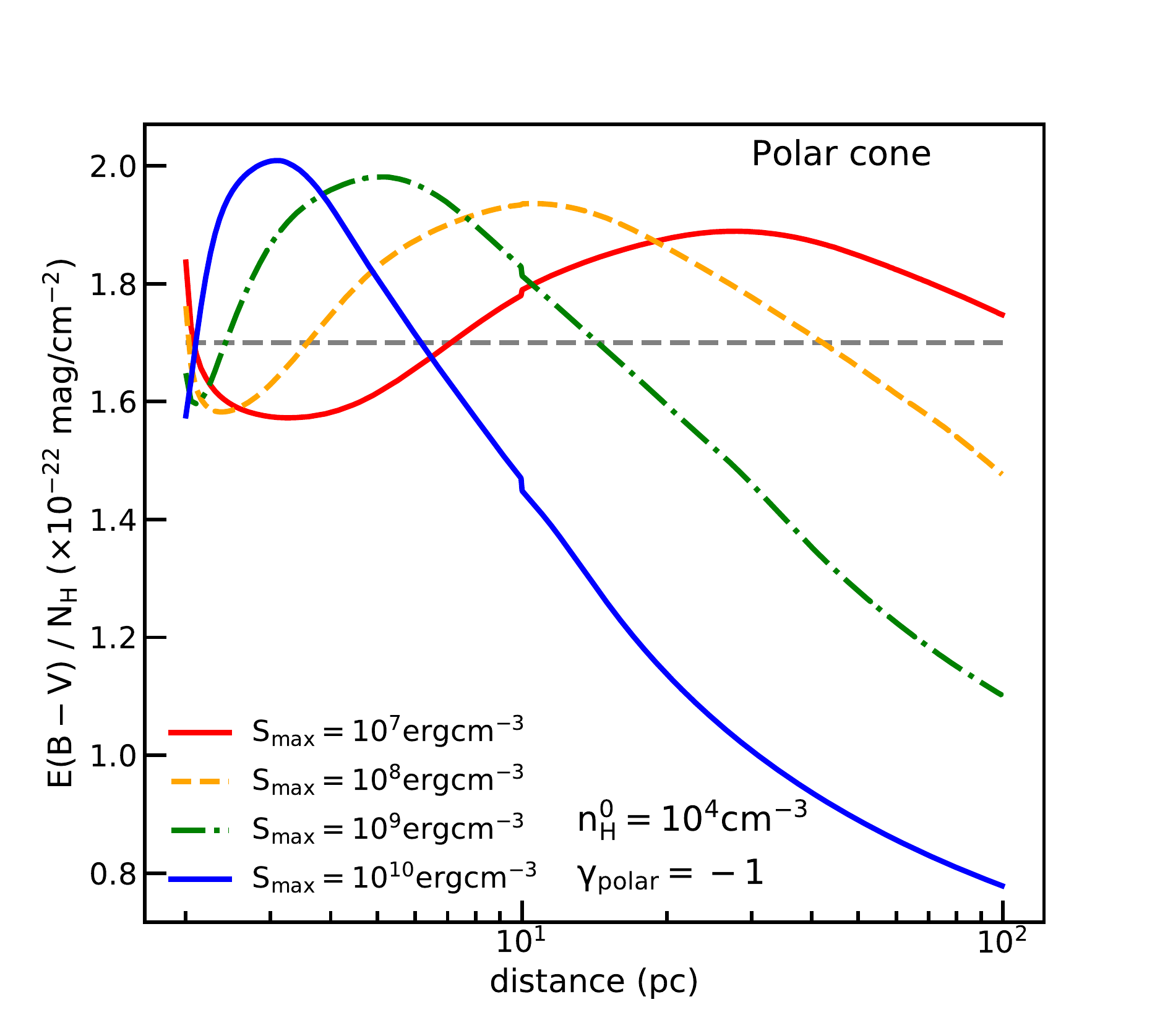}
    \caption{Variation of the color excess per H produced by grains in the cell of thickness $d_{\rm cell}$, $E(B-V)_{\rm cell}$ (upper panel) and $E(B-V)$ by all grains from the sublimation front to distance $d$ in the torus (lower panel), assuming different values of $S_{\rm max}$, $n_{\rm H}^{0} = 10^{4}\cm^{-3}$, and $\gamma_{\rm polar} = -1$. The horizontal black dashed line is the typical Galactic normalized color excess of $E(B-V)/N_{\rm H} = 1.7\times10^{-22}~\rm mag\cm^{2}$.}
\label{fig:Ebv(polar)}
\end{figure}

\begin{figure}[t]
    \includegraphics[width=0.45\textwidth]{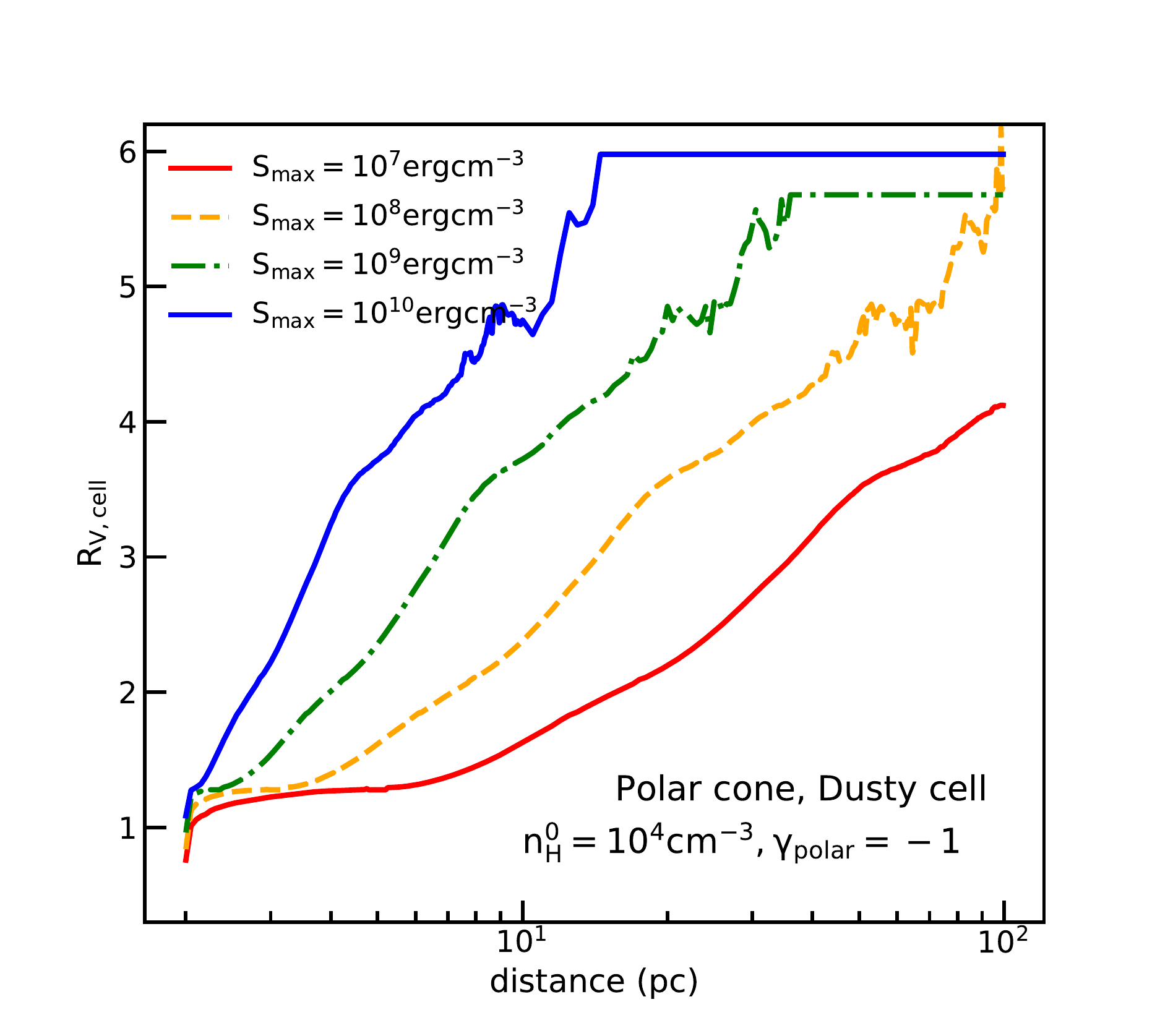}
    \includegraphics[width=0.45\textwidth]{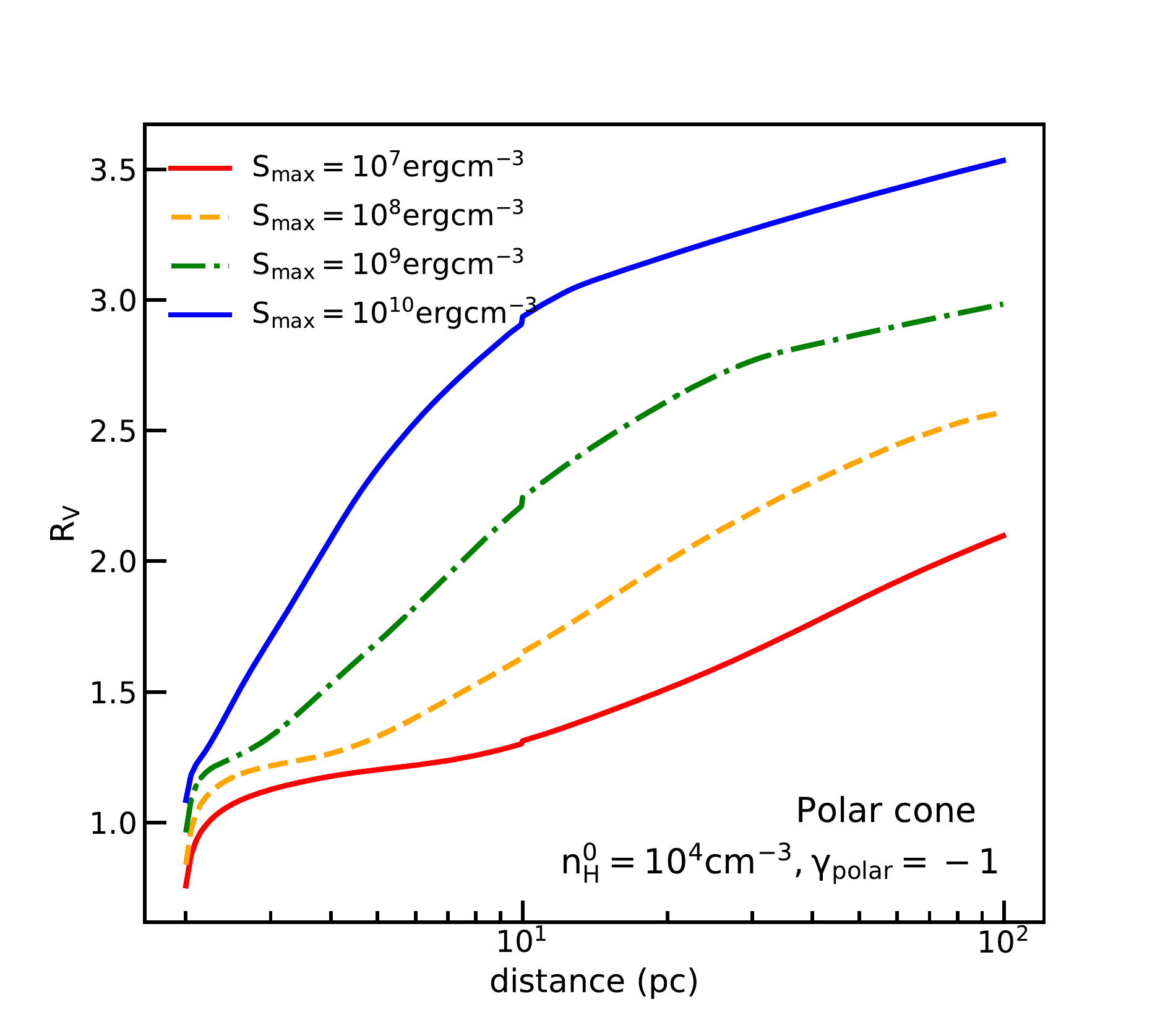}
\caption{Variation of the total-to-selective visual extinction ratio due to grains in the cell at different distances, $R_{\rm V cell}$ (upper panel), and $R_{\rm V}$ due to total grains from the sublimation front to distance $d$ (lower panel), for the torus, assuming different values of $S_{\rm max}$, $n_{\rm H}^{0} = 10^{4}\cm^{-3}$ and $\gamma_{\rm polar} = -1$.}
\label{fig:Rv(polar,Smax)}
\end{figure}

\begin{figure}[t]
    \includegraphics[width=0.45\textwidth]{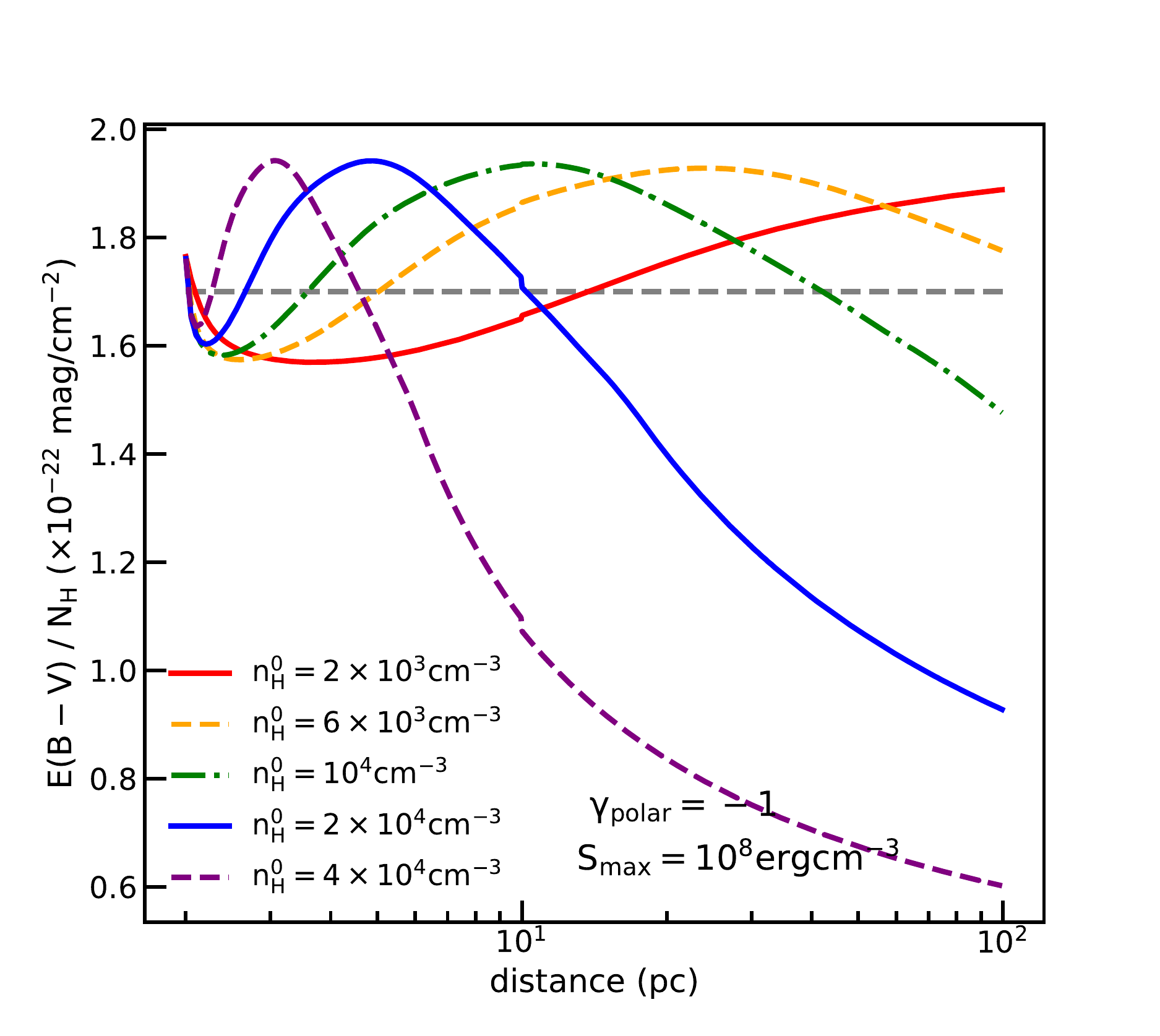}
    \includegraphics[width=0.45\textwidth]{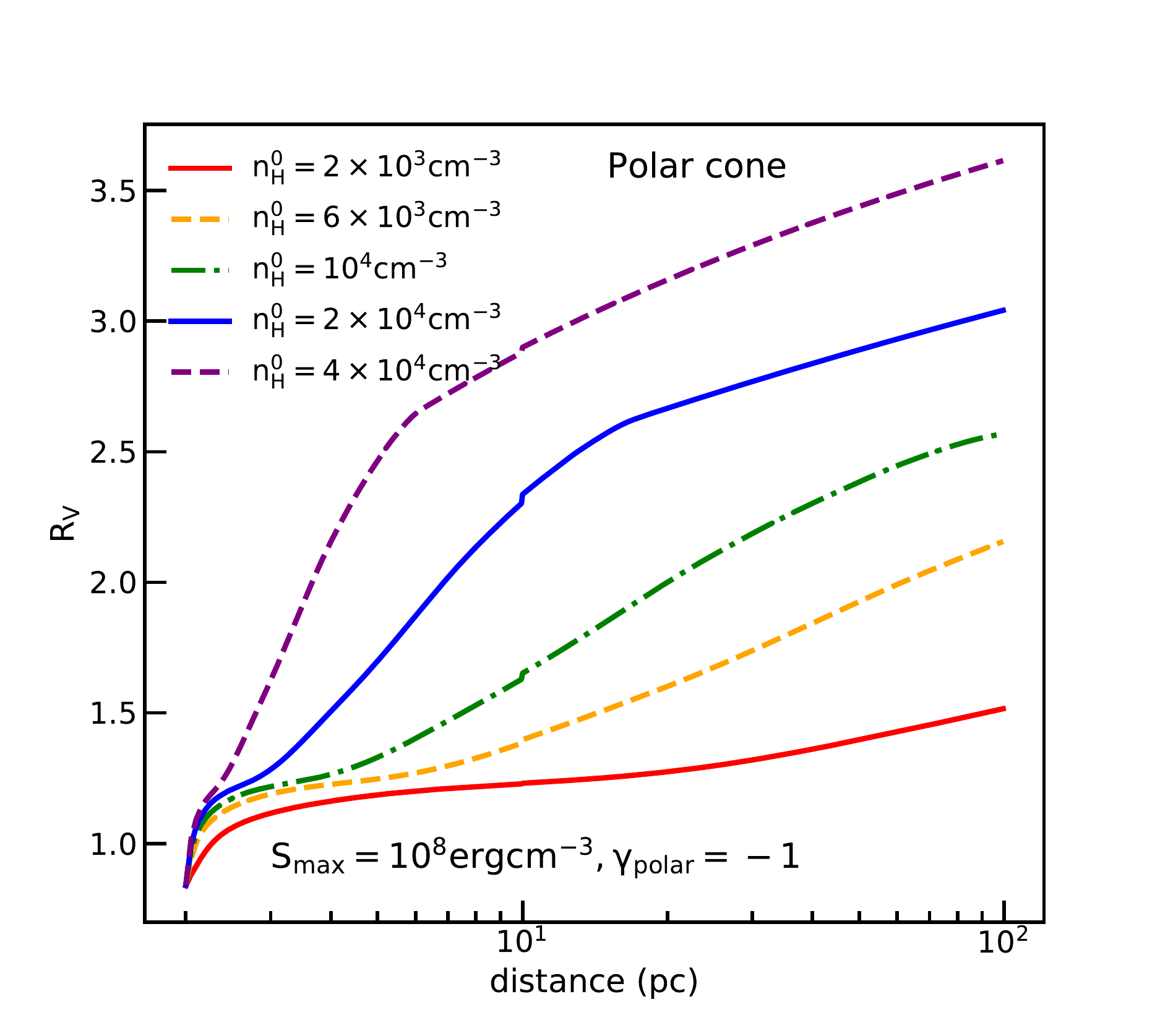}
\caption{Variation of $E(B-V)/N_{\rm H}$ and $R_{\rm V}$ due to all grains from the sublimation front as a function of distances $d$ in the polar cone with different values of $n_{\rm H}^{0}$, assuming $S_{\rm max} = 10^{8}\erg\cm^{-3}$ and $\gamma_{\rm polar} = -1$.} 
\label{fig:Rv(polar, nH)}
\end{figure}

Using the dust extinction $A(\lambda)$ obtained in Section \ref{sec:Aext}, we can calculate the color excess $E(B-V) = A_{\rm B} - A_{\rm V}$ with $A_{\rm B}$ and $A_{\rm V}$ being the extinction at blue and visible wavelength of $\lambda = 0.45 \mum$ and $0.55 \mum$, respectively. Then, we can get the total-to-selective visual extinction ratio $R_{\rm V} = A_{\rm V}/E(B-V)$.

\subsubsection{Polar cone}
The upper panel of Figure \ref{fig:Ebv(polar)} shows the variation of the color excess over the column density, $E(B-V)_{\rm cell}/N_{\rm H}$, for grains in the cell of thickness $d_{\rm cell}$ located at distances $d$ from 2 pc to 100 pc in the polar cone, assuming different values of $S_{\rm max}$, $n_{\rm H}^{0} = 10^{4}\cm^{-3}$ and $\gamma_{\rm polar} = -1$. At $d\sim 100$ pc, $E(B-V)_{\rm cell}/N_{\rm H}$ for $S_{\rm max}=10^{10}\erg\cm^{-3}$ is rather low of $\sim 0.25\times10^{-22}~\rm mag \cm^{2}$ due to the presence of large grains beyond the active region of RATD of $d_{\rm RATD} \geq 17$ pc (see the lower panel of Figure \ref{fig:adisr_polar_gamma_Smax}). Toward the AGN center, the value of $E(B-V)_{\rm cell}/N_{\rm H}$ increases due to the enhancement of small grains by RATD (see Figure \ref{fig:A_lambda(d)}), then it drops when small grains are further disrupted near the sublimation front (see the lower panel of Figure \ref{fig:adisr_polar_gamma_Smax}). By decreasing the maximum tensile strength, $E(B-V)_{\rm cell}/N_{\rm H}$ starts to rise at larger distances due to larger active region of RATD.

The lower panel of Figure \ref{fig:Ebv(polar)} shows similar results as the upper panel, but for $E(B-V)/N_{\rm H}$ produced by all grains from the sublimation front to distances $d$. In general, the value of $E(B-V)/N_{\rm H}$ increases with decreasing distances to the center region due to the increase of small grains by RATD and slightly decreases near the sublimation front when almost such small grains are destroyed. At $d \leq 10$ pc, the value of $E(B-V)/N_{\rm G}$ for lower $S_{\rm max}$ may be larger than the one for higher $S_{\rm max}$ due to the large active region of RATD.

The upper panel of Figure \ref{fig:Rv(polar,Smax)} shows the variation of the total-to-selective visual extinction ratio $R_{\rm Vcell}$ due to grains in the cell as a function of distances, assuming the same condition as Figure \ref{fig:Ebv(polar)}, i.e., different values of $S_{\rm max}$ and $n_{\rm H}^{0} = 10^{4}\cm^{-3}$. At large distances $d$ from the center of AGN, the disruption does not occur and $R_{\rm Vcell}$ is flat at $R_{\rm Vcell}\sim 6$. Note that the value $R_{\rm Vcell}$ here exceeds the typical value of $3.1$ for the standard ISM because very large grains of $a>1\mum$ are assumed to exist here. Then, toward the center region, the value of $R_{\rm Vcell}$ decreases continuously from $R_{\rm Vcell}= 6$ to $\sim 1$ due to increasing disruption of grains by RATD. At the same distance, $R_{\rm Vcell}$ is larger for a higher tensile strength due to weaker RATD. 
 
The lower panel of Figure \ref{fig:Rv(polar,Smax)} shows the similar results to the upper panel, but for the variation of $R_{\rm V}$ produced by all grains contained in the region between the sublimation front and distance $d$. At $d\sim 100$ pc, with $n_{\rm H}^{0} = 10^{4}\cm^{-3}$, one can see that grains with $S_{\rm max} \leq 10^{8}\erg\cm^{-3}$ produce $R_{\rm V} \leq 2.5$, much smaller than the Galactic standard value of $R_{\rm V} = 3.1$. It arises from the significant reduction of the optical-MIR extinction due to the strong removal of large grains by RATD near the center of AGN (see Figures \ref{fig:A_lambda(d)} and \ref{fig:A_lambda_polar}). Thus, the visual extinction $A_{\rm V}$ in this case is always smaller than $A_{\rm Vcell}$ produced in a cell at the same distance, resulting in a smaller value of $R_{\rm V}$. The value of $R_{\rm V}$ decreases with decreasing $d$, but at a much smaller rate compared to the upper panel. Besides, at the same cloud distance $d$, $R_{\rm V}$ is smaller for the lower maximum tensile strength because of the higher RATD efficiency. For example, at 100 pc, $R_{\rm V}$ decreases from $\sim 2.5$ to $\sim 2$ if the maximum tensile strength decreases from $10^{8}\erg\cm^{-3}$ to $10^{7}\erg\cm^{-3}$. However, in case of compact grains of $S_{\rm max} \geq 10^{9}\erg\cm^{-3}$, the observed value of $R_{\rm V}$ can increase to $\geq 3$ due to the dominance of sub-micron and micron grains in the polar direction, i.e., grains of $a \geq 0.1\mum$ start to survive at $d \geq$ 2 pc for $S_{\rm max} \geq 10^{9}\erg\cm^{-3}$ (see the lower panel of Figure \ref{fig:adisr_polar_gamma_Smax}).

The upper panel of Figure \ref{fig:Rv(polar, nH)} shows the variation of $E(B-V)/N_{\rm H}$ due to grains from the sublimation front to distance $d$ in the polar cone, assuming different gas density profiles and $S_{\rm max} = 10^{8}\erg\cm^{-3}$. For a fixed maximum tensile strength, the normalized color excess $E(B-V)/N_{\rm H}$ starts to change from the value produced by original dust grains at larger distances for smaller gas density due to the larger active region of RATD. On the other hand, the corresponding total-to-selective visual extinction ratio $R_{\rm V}$ (lower panel) drops faster with decreasing distances to AGN center and shows smaller values for lower gas density.

 \subsubsection{Torus region}
\begin{figure}[t]
    \includegraphics[width=0.45\textwidth]{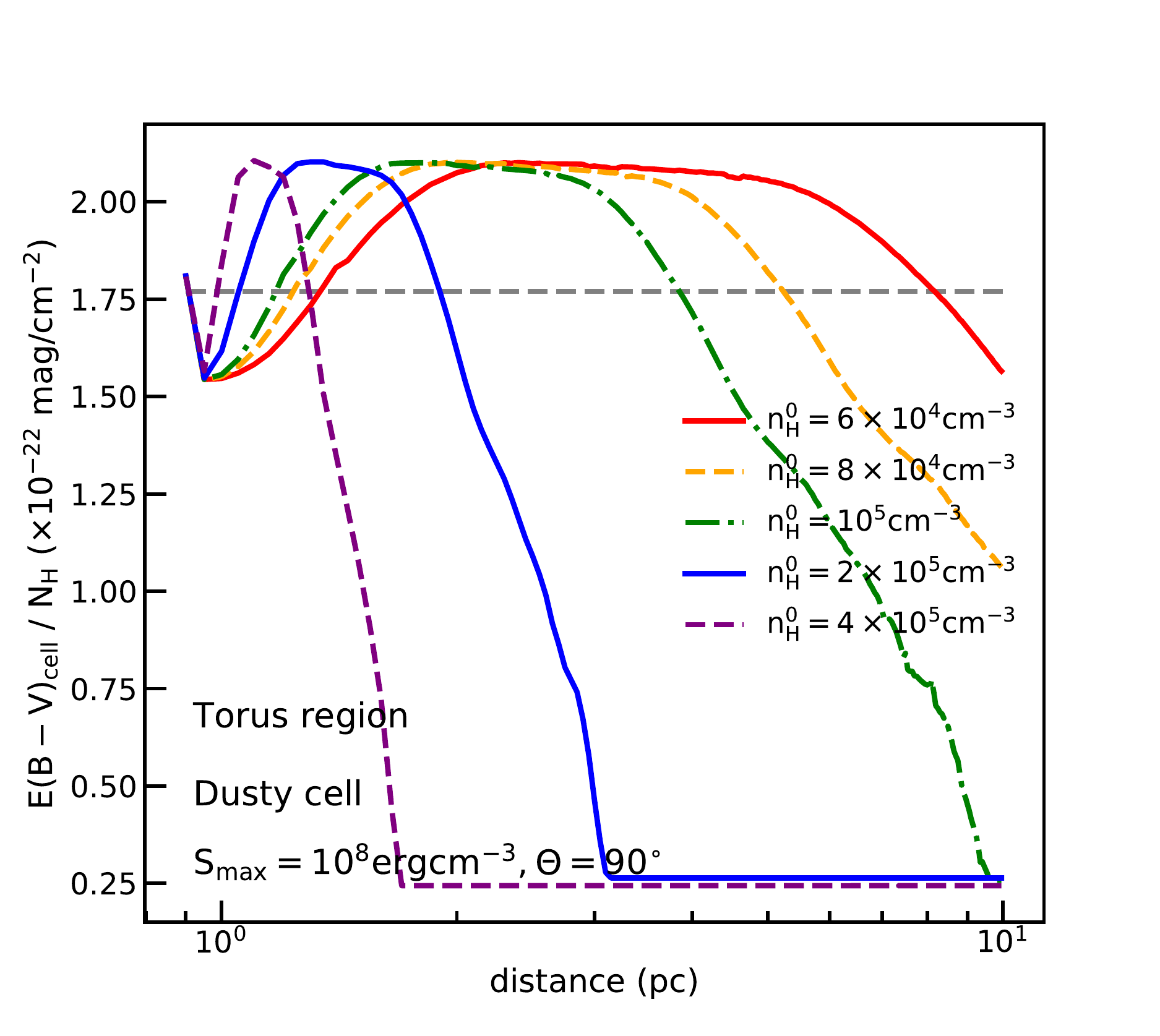}
    \includegraphics[width=0.45\textwidth]{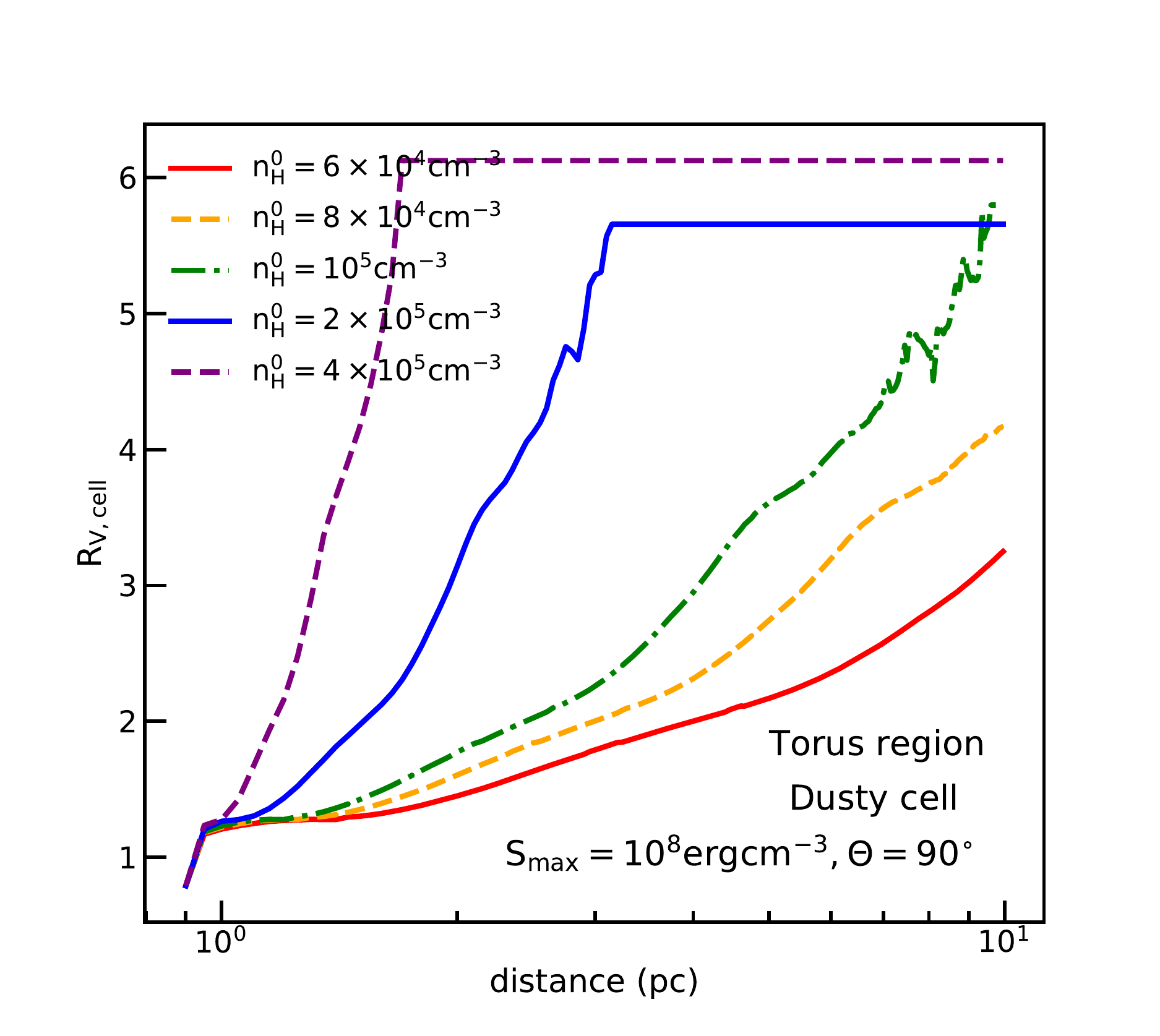}
\caption{Variation of $E(B-V)_{\rm cell}/N_{\rm H}$ (upper panel) and $R_{\rm Vcell}$ (lower panel) produced by grains in the torus at different cell's distance on the equatorial plane with different values of $n_{\rm H}^{0}$, assuming $S_{\rm max} = 10^{8}\erg\cm^{-3}$, $\gamma_{\rm torus} = -1$, $\beta = -2$ and $\Theta = 90^{\circ}$.}
\label{fig:Rv(cell, torus)}
\end{figure} 

 Figure \ref{fig:Rv(cell, torus)} shows the variation of $E(B-V)_{\rm cell}/N_{\rm H}$ (upper panel) and the corresponding $R_{\rm V cell}$ (lower panel) due to torus dust grains at different cell's distances on the equatorial plane, assuming different values of $n_{\rm H}^{0}$ and $S_{\rm max}  = 10^{8}\erg\cm^{-3}$. Similar to the polar cone, the value of $E(B-V)_{\rm cell}/N_{\rm H}$ and $R_{\rm Vcell}$ is different from the ones produced by original grains which follow the MRN distribution when RATD starts to modify the grain size distribution. $R_{\rm Vcell}$ decreases with decreasing distances to the center region and exhibits smaller values in lower gas density circumnuclear region. For example, grains in the cell at 10 pc give $R_{\rm Vcell} \sim 5.5$, 4, and 3 if the initial gas density of the torus decreases from $n_{\rm H}^{0} = 10^{5}\cm^{-3}$ to $8\times10^{4}\cm^{-3}$ and $6\times10^{4}\cm^{-3}$, respectively. 

\begin{figure}[t]
    \includegraphics[width=0.45\textwidth]{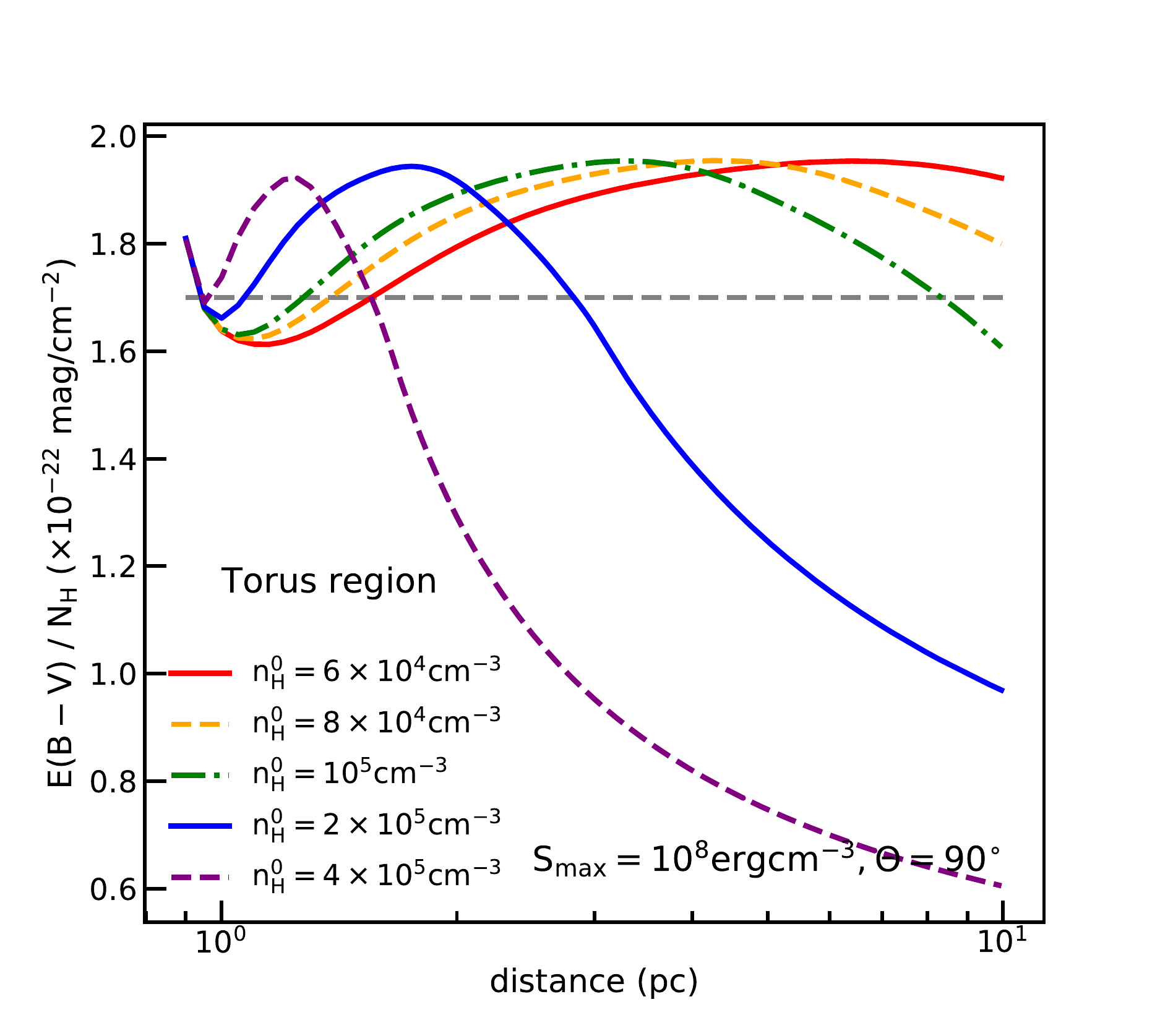}
    \includegraphics[width=0.45\textwidth]{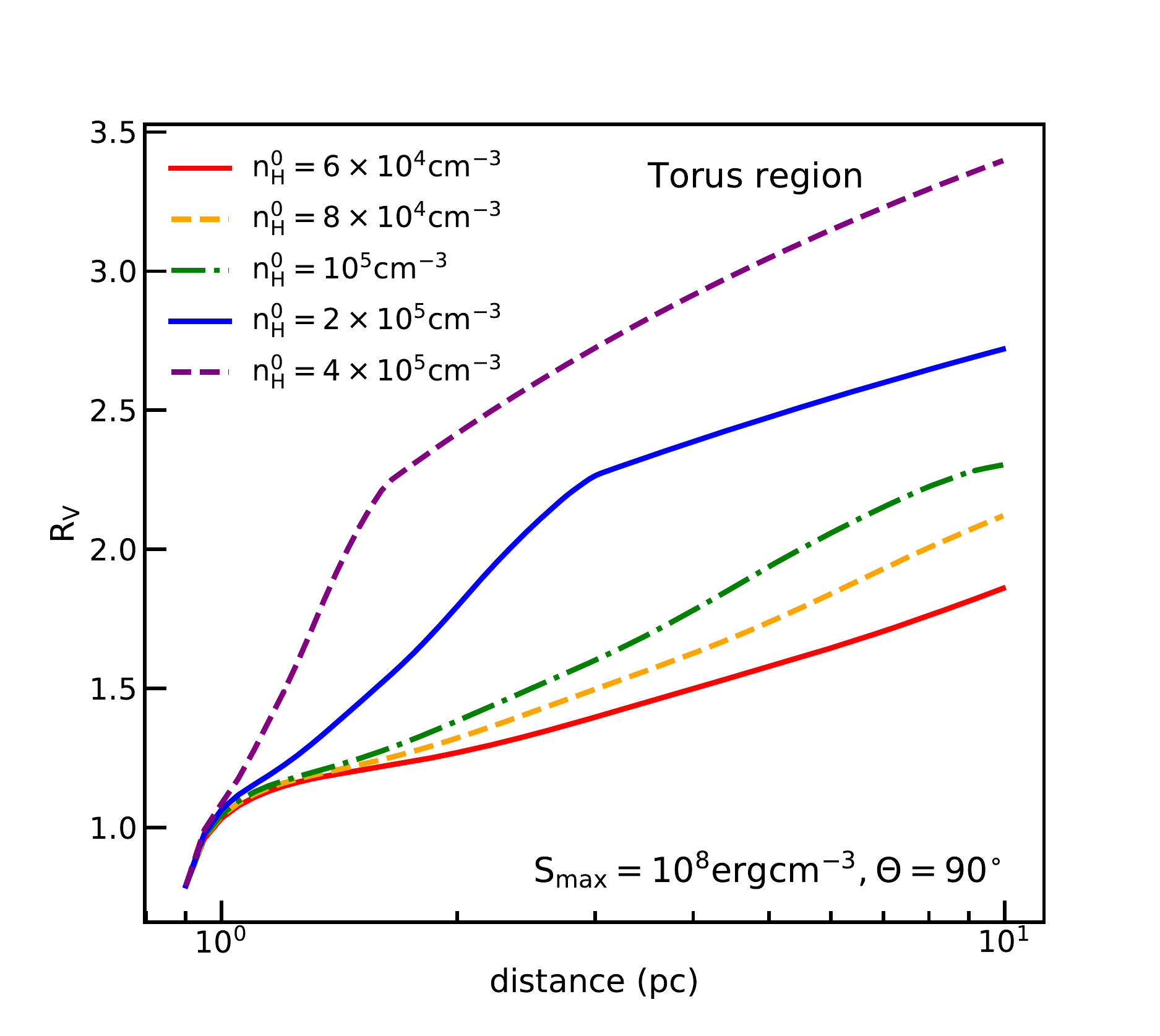}
\caption{Similar to Figure \ref{fig:Rv(polar, nH)} but for grains within 10 pc in the torus, assuming $S_{\rm max} = 10^{8}\erg\cm^{-3}$, $\gamma_{\rm torus} = -1$, $\beta = -2$ and $\Theta = 90^{\circ}$.}
\label{fig:Rv(torus)}
\end{figure}

Figure \ref{fig:Rv(torus)} shows the same results as Figure \ref{fig:Rv(cell, torus)}, but due to grains from the sublimation front to distance $d$. Similar to the results in the lower panel of Figure \ref{fig:Rv(polar, nH)}, one can get small values of $R_{\rm V} \sim 1-2.5$ if the torus size is small and the gas density is low. For example, for optically thin torus with $n_{\rm H}^{0} \leq 2\times10^{5}\cm^{-3}$, one can get $R_{\rm V} \sim 1.8-2.5$ with the torus size of radius up to $d \geq 10$ pc. In contrast, with high density of $n_{\rm H}^{0} = 4\times10^{5}\cm^{-3}$, one only can get $R_{\rm V} \leq 3$ if the torus size has the outer boundary at $d \sim 4$ pc. Further than that, $R_{\rm V}$ quickly rises to $R_{\rm V} \sim 3.5$ due to the presence of large grains.

Moreover, by changing the observed direction from the edge-on view to the face-on view, the total-to-selective visual extinction ratio produced in the isolated cell $R_{\rm Vcell}$ and in the cloud extending from the sublimation front $R_{\rm V}$ decreases. It arises from the stronger removal of large grains in the region above and below the equatorial plane of AGN by RATD (see the lower left panel of Figure \ref{fig:adisr_torus_beta_Smax_theta_gamma}). In contrast, the corresponding color excess $E(B-V)_{\rm cell}/N_{\rm H}$ and $E(B-V)/N_{\rm H}$ increase due to the enhancement of small grains by RATD.

\section{Discussion}\label{sec:dicuss}
In this section, we discuss the implications of our modeling results for photometric of AGN.

\subsection{Photometric observations toward AGN}
\subsubsection{Extinction curves with a steep far-UV rise}
Photometric observations toward individual AGN frequently report a 'red tail' in its spectral energy distribution (SED) (\citealt{Webster_1995}, \citealt{Brotherton_2001}, \citealt{Richards_2002}), which is proposed to originate from the intrinsic reddening caused by the central engine and the external dust in the host galaxy. By analyzing the spectra of 4576 quasars from the Sloan Digital Sky Survey (SDSS), \cite{Richards_2003} showed that the intrinsic slope of their SED is consistent with its general spectrum. Thus, the 'red tail' in the curves is originated from the extinction of grains in its host galaxy. By fitting the observational data with different dust models, they found that the extinction curves of 273 out of 4576 quasars ($6~\%$) are best described by the SMC-dust model, suggesting the dominance of small grains of size $a \leq 0.1\mum$ in the host galaxy. Similarly, the study of 9566 quasars from SDSS and expanding to longer wavelengths using a subset of 1866 SDSS-Two Micron All Sky, \cite{Hopkins_2004} came to the same conclusion.

In Section \ref{sec:Aext}, we show that RATD can convert large grains of $a\geq 0.1 \mum$ into small sizes up to 100 pc in the polar cone and 10 pc in the torus region around AGN. Consequently, optical-MIR dust extinction significantly decreases, while FUV-MIR extinction increases, resulting in a steep rise from MIR to far-UV range compared to the standard extinction curve of our galaxy.  Thus, the 'SMC'-like' extinction curve observed toward many individual AGN can be successfully explained by the effect of RATD on dust in AGN evironments.

Indeed, the extinction curves of AGN that show the steep far-UV rise are not completely described by the SMC-like dust model. Instead, its steepness varies between the Milky Way (MW), Large Magellanic Cloud (LMC), and SMC-like extinction curve. In detail, the rise of dust extinction from $\lambda ~ 3500-4000~\rm \AA$ to NUV observed toward NGC 3227 by \cite{Crenshaw_2001} is higher than the slope of the SMC-like curve. The observed extinction curve of Ark 564 shows a steeper slope than ones induced by interstellar grains, but shallower than the SMC-like curve and does not exhibit the $2175~\rm \AA$ bump (\citealt{Crenshaw_2002}). The well-determined extinction curve of B3 0754 +394 and Ton 951 studied by \cite{Gaskell_2007} can be described by the SMC and LMC-like dust models, respectively. However, the curve of Ton 951 shows a flat in FUV, suggesting the removal of very small grains. The dust reddening of Mrk 304 is also fitted well with LMC-like dust model, but its curve is not-well constrained (\citealt{Gaskell_2007}). 

Theoretically, the different steepness of the FUV-MIR extinction curve implies different grain size distributions. In Section \ref{sec:adisr}, we show that the modification of dust grains around AGN by RATD strongly depends on their internal structure and the properties of AGN environments. In particular, in both the polar cone and torus region (Section \ref{sec:adisr}), we show that large grains are easier to be disrupted to smaller sizes by RATD for lower gas density and lower maximum tensile strength of grains. Consequently, the outcome extinction curves in these cases will present the sharp rise toward far-UV compared with other cases (see Section \ref{sec:Aext}). Therefore, RATD appears to be a suitable explanation for the diversity of steepness of extinction curves observed toward many AGN. Moreover, if RATD can reproduce the 'redden' AGN, fitting extinction curves produced by RATD with observational data can help us constrain dust properties in AGN environments.

For example, the very steep far-UV rise extinction curve observed toward NGC 3227 (\citealt{Crenshaw_2001}) may arise from the high efficiency of RATD around the nuclei of the host galaxy. It thus implies that grains are likely to have a composite structure (i.e., low $S_{\rm max}$) instead of a compact one. In contrast, the intermediate slope between LMC and MW-like extinction curve observed toward Ark 564 (\citealt{Crenshaw_2002}), the SMC-like curve, and the LMC-like curve observed toward B3 0754+ 394 and Ton 951 (\citealt{Gaskell_2007}) suggests a moderate efficiency of RATD.
 
\subsubsection{Low dust reddening of AGN}
Several studies of the dust reddening toward AGN report a low value of $E(B-V)/N_{\rm H}$ and $A_{\rm V}/N_{\rm H}$ compared with the typical value in the interstellar medium. For example, most of 19 Seyfert galaxies studied by \cite{Maioline} exhibit a low value of $E(B-V)/N_{\rm H}$ by a factor of 3-100 compared with the standard value of $1.7\times10^{-22}~\rm mag \cm^{2}$ produced by interstellar dust grains. The low value of $A_{\rm V}/N_{\rm H}$ is also revealed through other evidences such as the weak broad line emission at $H\alpha$ and $H\beta$ in intermediate 1.8-1.9 type Seyfert, broad absorption at soft and hard X-ray range in Broad Absorption Line (BAL) QSO (\citealt{Reichert}, \citealt{Maioline}). This anomalous feature is attributed to the different properties of grains in the AGN environments. For example, \cite{Maioline} suggested that the low value of $E(B-V)/N_{\rm H}$ may arise from the dominance of large grains of $\geq 0.1\mum$. However, large grains produce a large visual extinction, which is inconsistent with the observed low value of $A_{\rm V}/N_{\rm H}$.
 
In Section \ref{sec:Aext}, we show that the RATD effect can reduce the optical to MIR extinction produced by polar and torus dust grains. Thus, by observing AGN in both the edge-on and face-on directions, one can obtain the low value of $A_{\rm V}/N_{\rm H}$ that is produced by the strong removal of large grains of $a\geq 0.1\mum$ by RATD in AGN environments.

However, the enhancement of small grains by RATD makes the extinction curve steeper, which increases the color excess $E(B-V)$ compared to the case of no dust disruption. Thus, the ratio $E(B-V)/N_{\rm H}$ can be higher or lower than the typical Galactic value of $ 1.7\times10^{-22}~\rm mag \cm^{2}$, depending on the destruction level of RATD in this region. In particular, one can observe the low value of $E(B-V)/N_{\rm H} \leq 1.7\times10^{-22}~\rm mag \cm^{2}$ if the sub-micron grains responsible for the extinction of optical radiation is less enhanced or totally destroyed by RATD (e.g., see Figure \ref{fig:Ebv(polar)} and the upper panel of Figure \ref{fig:Rv(polar, nH)} for polar cone and the upper panel of Figures \ref{fig:Rv(cell, torus)} and \ref{fig:Rv(torus)} for torus region). Therefore, the low value of $E(B-V)/N_{\rm H}$ still can be reproduced even if small grains are dominant. The high tensile strength of grains and high density in the surrounding region of AGN may be more preferred.

\subsection{Extinction curves without an UV 2175 \AA~ bump} 
In contrast to the SMC-like extinction curve, several studies based on analyzing the composite spectra quasar and individual AGN by \cite{Gaskell_2004} and \cite{Gaskell_2007} report a flat FUV-NUV extinction and a weak $2175~\rm \AA$ bump, that is believed to originate from $\pi-\pi*$ electronic transition of very small carbonaceous grains (Polycyclic Aromatic Hydrocarbons, PAHs). These features indicate the removal of small grains of $a\leq 0.01\mum$ which are responsible for the FUV-NUV extinction and suggest the preference of large grains of submicron and micron sizes around AGN.
 
In Section \ref{sec:Aext}, we assume that RATD only modifies the grain size, thus, FUV-NUV extinction cannot be reduced due to the enhancement of small grains in both the polar cone and torus region. Now, we consider the destruction mechanisms that can destroy the smallest grains (e.g., sputtering or sublimation). We adopt a space-varying minimum grain size of $a_{\rm min}(d)$ determined by the Coulomb explosion taken from \cite{Tazaki_20}, which is:
\begin{equation}
a_{\rm min}(d) = A d^{\alpha_{\rm amin}},
\end{equation}
where $A$ is the constant at which the minimum grain size at the sublimation front is $a_{\rm min, subli} = 0.1\mum$, and $\alpha_{\rm amin}$ is the power-law index of the profile.

\begin{figure*}[t]
    \centering
    \includegraphics[width=0.45\textwidth]{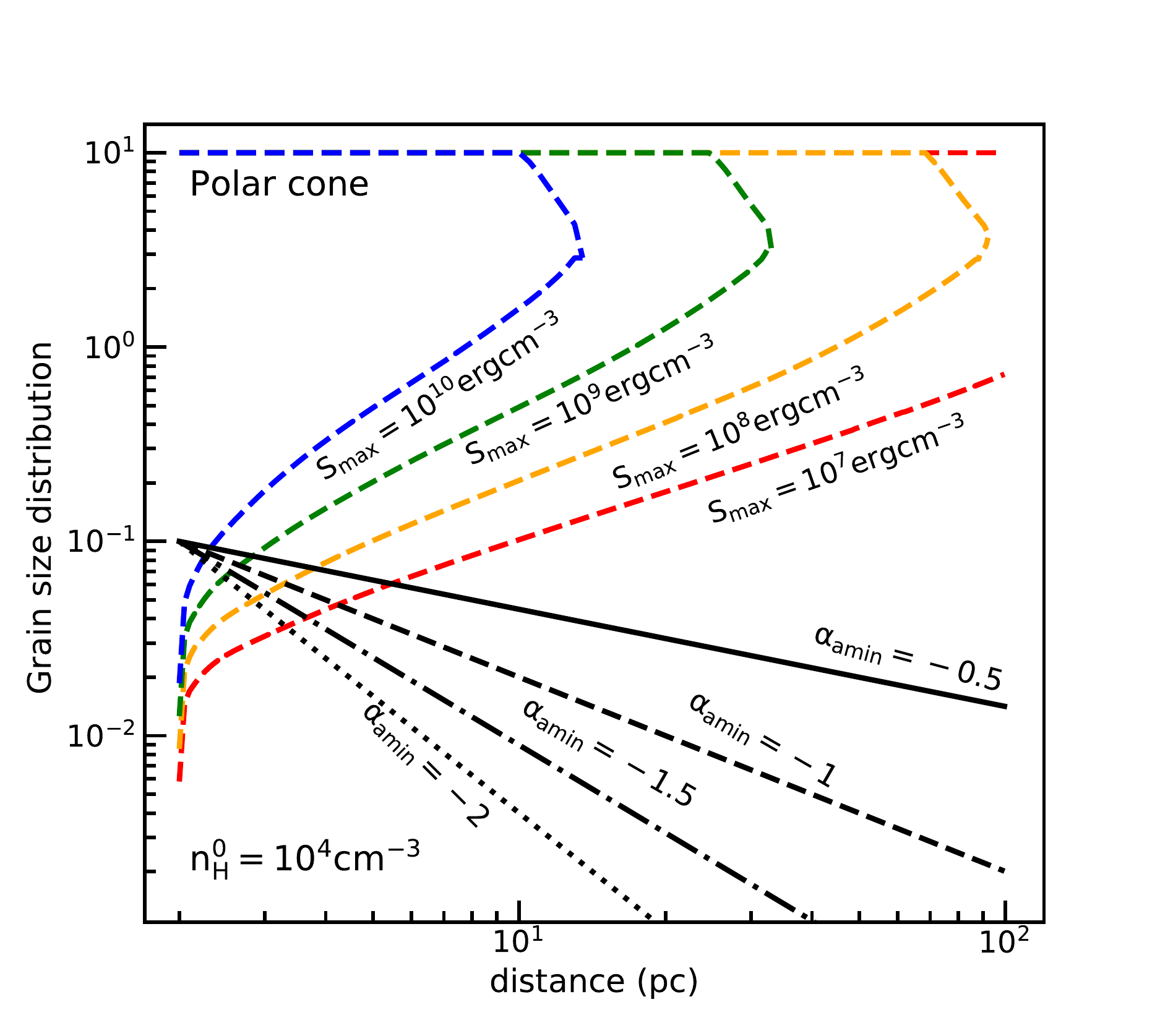}
    \includegraphics[width=0.45\textwidth]{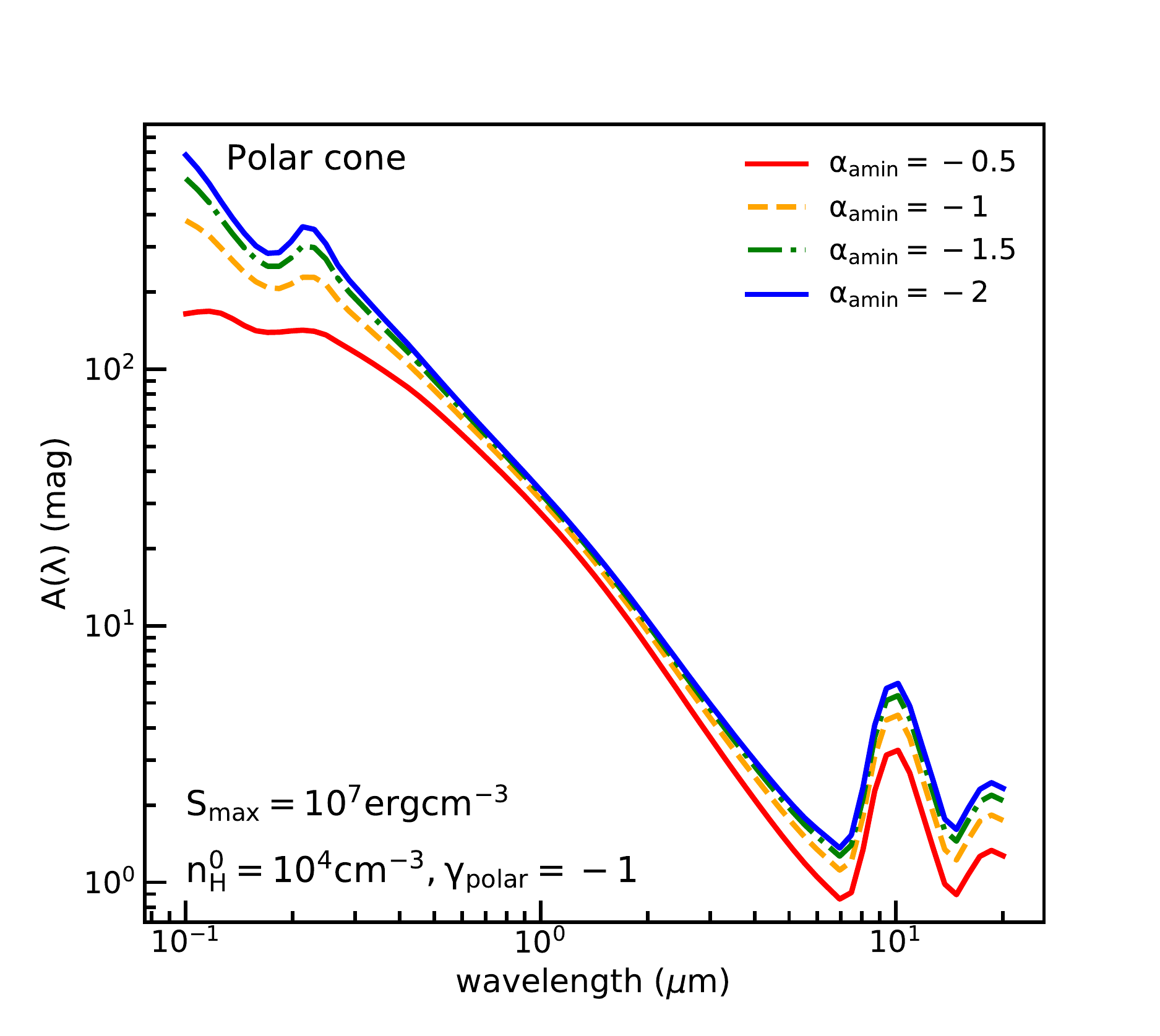}
    \includegraphics[width=0.45\textwidth]{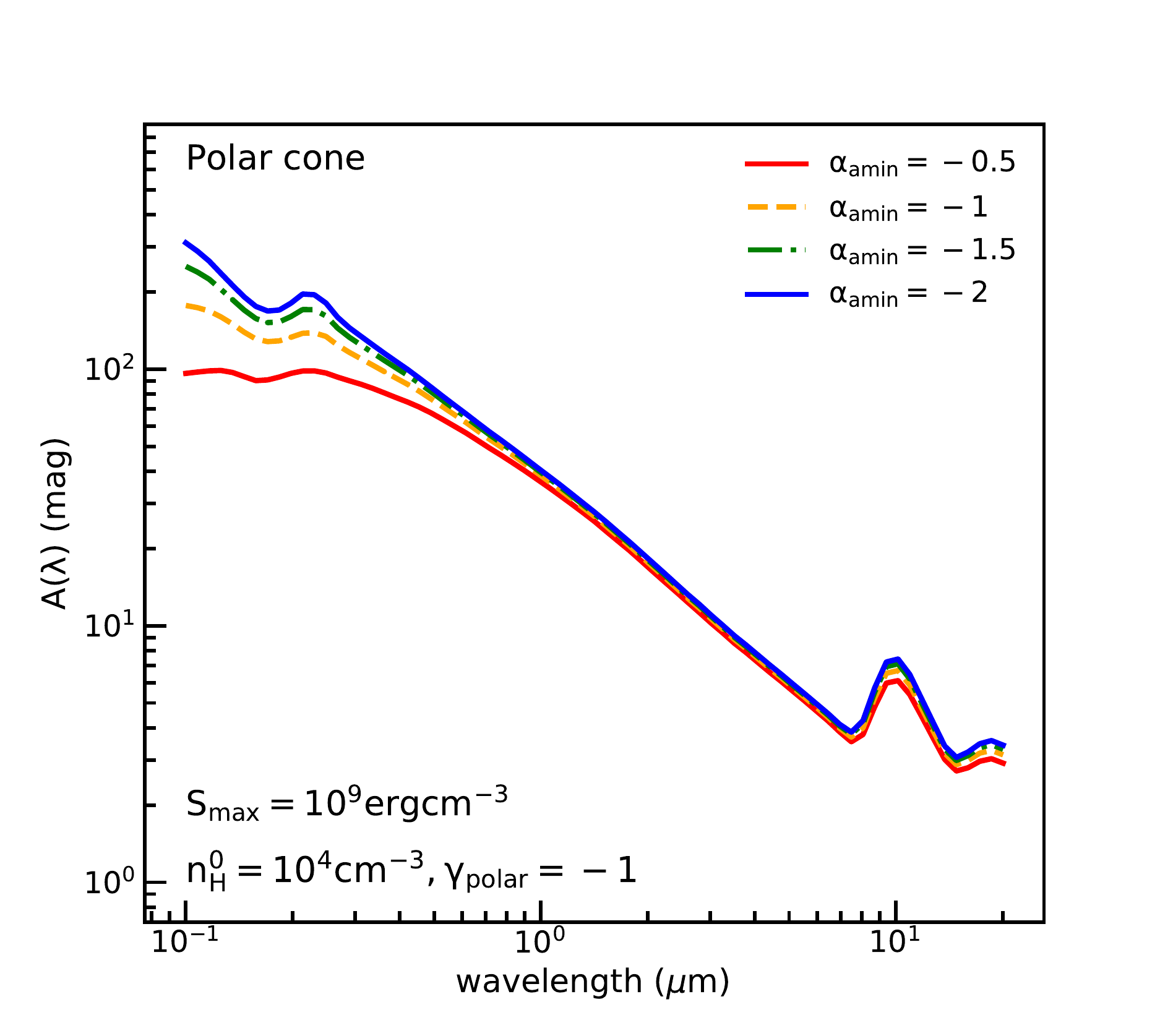}
    \includegraphics[width=0.45\textwidth]{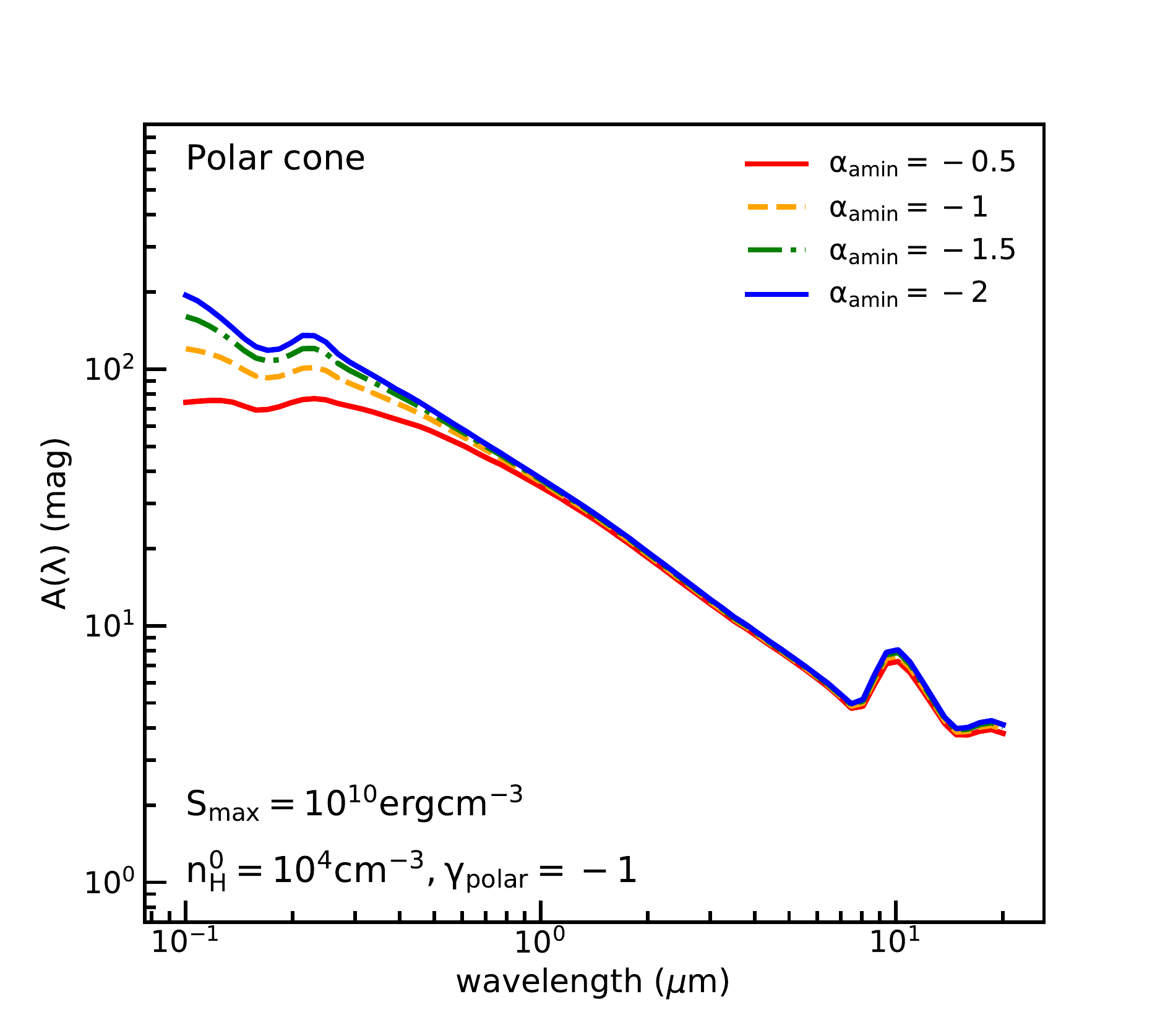}

\caption{Upper left panel: Variation of minimum grain size with different slope $\alpha_{\rm amin}$ and maximum grain size constrained by RATD with different values of $S_{\rm max}$ within 100 pc in the polar cone. Other panels: the corresponding final extinction curves of each case.}
\label{fig:a_min}
\end{figure*}

The upper left panel of Figure \ref{fig:a_min} shows the variation of the minimum grain size with different power-law indexes $\alpha_{\rm amin}$ from $-0.5$ to $-2$ (black line) within 100 pc in the polar cone. The minimum grain size decreases faster with increasing distances if the slope is steeper, i.e., $\alpha_{\rm amin} = -2$, implying the weak removal of small grains in AGN environment.

The upper right panel shows the final extinction curves produced by the grain size described in the upper left panel with different values of  $\alpha_{\rm amin}$, assuming $S_{\rm max}=10^{7}\erg\cm^{-3}$ and $n_{\rm H}^{0} =  10^{4}\cm^{-3}$. In the case that small grains of $a \leq 0.05\mum$ are only destroyed within $\sim 5$ pc from the center region, i.e., $\alpha_{\rm amin} = -2$, the final extinction curve still shows a steep rise toward FUV. By increasing the value of $\alpha_{\rm amin}$, small grains are removed to larger distances, resulting in the shallower slope of FUV-NUV extinction curve. For example, the curve will exhibit the flat FUV-NUV extinction if all small grains are removed up to the active of RATD of $d_{\rm RATD} \sim 50$ pc, i.e., $\alpha_{\rm amin} = -0.5$.

The lower left and right panels show the same results to the upper right panel, but for grains with $S_{\rm max} = 10^{9}\erg\cm^{-3}$ and $10^{10}\erg\cm^{-3}$, respectively. Similarly, the slope in FUV-NUV extinction becomes shallower if small polar dust of $\leq 0.05\mum$ is totally destroyed in the active region of RATD. The change from steep UV-rise to 'gray' extinction curve is faster for higher tensile strength of grains, and also higher gas density.
 
In conclusion, the strong removal of small grains of $a < 0.05\mum$ up to the active region of RATD induces the flatten in FUV-NUV extinction and the predominance of large grains as in the conclusion of \cite{Gaskell_2004} and \cite{Gaskell_2007}.
 
The destruction of small grains around AGN is first studied by \cite{Laor_93} for the polar cone and by \cite{Barvainis87} for the torus region. They found that near the center of AGN, grains are quickly sublimated to the gas phase due to the intense UV-optical radiation. \cite{Weingartner_2006} and recently \cite{TazakiRyo} study the effect of Coulomb explosion in AGN environments. They found that small grains are charged in the ionized polar cone then are destroyed up to $\sim 10$ pc, successfully reproducing the 'gray' extinction curve at this distance. This mechanism does not work well in the torus region due to optically thick environments. However, Coulomb explosion requires X-ray and extreme UV radiation, such as its efficiency is much weaker than RATD that can be effective with longer wavelength photons, i.e., optical-MIR, because such energetic photons are efficiently absorbed. Beside thermal sublimation and Colomb explosion, small grains around AGN also can be destroyed by non-thermal sputtering and MEchanical Torque Disruption (METD) (\citealt{Hoang_Lee_2020}). In the former mechanism, grains are accelerated to hypersonic speeds, i.e., $v \geq 500~\rm km ~s^{-1}$, by the strong radiation pressure from AGN center and be destroyed when moving into the ambient gas. The latter mechanism shares the same principle with RATD. However, in METD, grains are spun up by stochastic mechanical torques induced by grain-grain collision instead of the radiative torque in RATD. We expect that non-thermal sputtering can modify the distribution of small grain size beyond the active region of thermal sublimation or Coulomb explosion. METD can continue to modify it beyond the region where the speed of grains reduces to $v \leq 500 ~\rm km ~s^{-1}$ (\citealt{Hoang_Lee_2020}). A detailed study of the effect of METD and other dust destruction mechanisms in both the polar cone and torus of AGN is needed to explain the long-term puzzle of the 'flat' and 'SMC-like dust' extinction curve observed toward AGN.
 
\subsection{The depletion at $9.7\mum$ absorption feature} 
 The absorption/emission at $9.7\mum$ feature produced by the stretching mode of silicate oxide (Si-O) in the observed IR spectrum toward AGN is a tool for studying the properties and composition of grains in AGN environments. Theoretically, the unified model of AGN expects the observed absorption line at $9.7\mum$ in type 2 AGN due to the obscuration of cold dust in the torus. In contrast, one should obtain the emission line in type 1 due to the direct observed view to hot dust near the center of AGN. However, several observations toward type 1 AGN do not show the clear evidence for silicate feature in both absorption and emission (\citealt{Maioline}, \citealt{Clave20}, \citealt{Sturm_2005}). These anomalous features indicate the difference in physical properties and grain size distribution of grains. In particular, the concentration of silicate grains in clumpy structure (\citealt{Rowan}, \citealt{Nenkova2002}) or the predominance of micron-sized silicate grains are suggested in order to explain the depletion at $9.7\mum$ absorption in this region (\citealt{Laor_93}, \citealt{Maioline}). 

 According to \cite{Laor_93}, small grains of $a \leq 0.1\mum$ should be reduced to produce the dominance of large grains in the polar cone of AGN. So is the enhancement of small grains by RATD in conflict with the existence of large grains? The answer is that it depends on AGN environments. In Section \ref{sec:adisr}, we show that large grains of $a \geq 0.1\mum$ in both the polar cone and torus region are only destroyed near the center of AGN for higher gas density and higher internal strength of grains. Consequently, the depletion of $9.7\mum$ silicate absorption feature is still available in spite of accounting for the contribution of RATD in AGN environments (see Section \ref{sec:Aext}).  In other words, RATD still supports the depletion of $10\mum$ in observations. However, if the density around AGN is not dense and grains mostly have composite structure, the strong absorption at $9.7\mum$ is certainly prominent. The dependence of magnitude of $10\mum$ with the strength of RATD can be tested with observations using JWST or 10-m class ground-based telescope and be used as a tracer for probing the gas density and characteristic of AGN grains. 
 
 On the other hand, in reality, grains may concentrate in dense clumps instead of distributing uniform as our assumption. Depending on the location, the distance to AGN center, and the properties of clumps, grains may be less affected by an intense radiation field of AGN and may survive against RATD. As a result, one would observe the depletion of $9.7\mum$ feature in the MIR observed spectrum of type 1 AGN. Therefore, the detailed study on the role of RATD with a realistic clumpy distribution model (e.g., \citealt{Nenkova2002}, \citealt{Dullemond05}, \citealt{Honig06}, \citealt{Nenkova08a},  \citealt{Stalevski_2012}, \citealt{Sieben15}) and in polar cone (e.g., \citealt{Schart14}) is needed to see that if the anomalous feature at $9.7\mum$ be explained with RATD as in our study.

\subsection{Implications for grain structures in the AGN torus}
Our obtained results show that grains with a low tensile strength of $S_{\rm max}\sim 10^{7}\erg\cm^{-3}$ can be disrupted to larger distances than those with high tensile strength (see Section \ref{sec:adisr}).  From the comparison between our modeling and observational data, one can see the preference of compact grains in both the polar cone and torus region, i.e., higher $S_{\rm max}$, that is suitable for explaining the observed photometric features of AGN  However, we cannot rule out the presence of large fluffy grains in dense clumps. Moreover, the RATD mechanism implies an increase in the abundance of small grains with decreasing the radial distance to AGN. 

\section{Summary}\label{sec:sum}
We have studied the dust destruction by rotational disruption by RATs in the local environment of AGN and model the extinction curve and other photometric parameters, i.e., $R_{\rm V}$, and $E(B-V)$, of AGN. Our main findings are summarized as follows:

\begin{enumerate}

\item  Large grains of $a \geq 0.1\mum$ within $\sim$ 100 pc in the polar cone and $\sim$ 10 pc in the dusty torus can be rotationally disrupted by RATD.  The efficiency of RATD increases with increasing radiation field strength of AGN and decreasing the local gas density and the tensile strength of dust grains.  

\item  Assuming different density profiles and maximum tensile strengths, we calculate the extinction curve produced by polar and grains in the presence of RATD. We find that the final extinction curve produced in both the polar cone and torus region exhibits a steep rise toward far-UV due to the strong disruption of large grains to smaller sizes near the center of AGN. The far-UV rise in extinction is steeper for the lower gas density and lower tensile strength of grains. Thus, the RATD mechanism can explain the 'SMC-like' extinction curve observed toward many individual quasar and AGN.  

\item The absence of the $9.7\mum$ silicate absorption feature can be produced if the RATD effect is not too strong. The dependence of the depth of the $9.7\mum$ absorption feature with the strength of RATD can be used to probe the physical properties of dust and the effect of radiative feedback of AGN onto the surrounding environment.

\item Combining the RATD mechanism that affects large grains with other destruction mechanisms responsible for destroying very small grains of $\leq 0.05\mum$ is the key to explain the mixture between the 'gray' and the 'SMC-like' extinction curve observed toward individual AGN. The observed extinction curve will exhibit the flat FUV-NUV extinction if very small grains are strongly depleted up to the active region of RATD.
 
\item Modeling of the extinction curves for a dusty slab located at different distances $d$ in the polar cone and torus region reveals the strong variation of the curves with $d$. Both the total-to-selective ratio $R_{\rm V}$ observed in two regions decreases rapidly with decreasing $d$. This can be used to interpret observational data for AGN with a clumpy structure where each dusty slab can be considered a clump.

\end{enumerate}
The present paper focuses on grain disruption by RATs and its effect on AGN extinction. It is widely known that RATs can also induce grain alignment with the magnetic field (\citealt{Laza07}, \citealt{Hoang16}) that affects the AGN polarization. The optical-UV polarization is dominated by dust/electron scattering (e.g., \citealt{Marin_2}, \citealt{Marin_5}, \citealt{Marin20}), while MIR-submillimeter polarization is mainly produced by aligned dust grains (\citealt{Young95}, \citealt{Watanabe03}, \citealt{Marin_18}). The polarized emission from aligned dust grains in IR wavelengths thus becomes a powerful tool for studying magnetic fields in AGN environments (\citealt{Lopez2018}, \citealt{Lopez_20a}, \citealt{Lopez_20b}). This important issue will be studied in detail in our followup paper.

\acknowledgements
We are grateful to the referee for helpful comments that improve our manuscript. We thank Le Ngoc Tram for detailed comments on the early period of the draft. T.H. acknowledges the support by the National Research Foundation of Korea (NRF) grants funded by the Korea government (MSIT) through the Mid-career Research Program (2019R1A2C1087045). 

\bibliography{reference}
 
\end{document}